\input phyzzx.tex
%
%\draft
\tolerance=500000 \overfullrule=0pt

        \def\cmp{Comm. Math. Phys.}
     \def\ijmp{Int. J. Mod. Phys.}
    \def\jgp{J. Geom. Phys.} 
\def\jmp{J. Math. Phys.}

\def\mrl{Math. Res. Lett.}  \def\np{Nucl. Phys.}
\def\pl{Phys. Lett.}        
          
  \def\topo{Topology}  
\def\prsl{Proc. R. Soc. Lond.}

\def\mani{\cal{M}}  
\def\ad{\hbox{\rm ad}}
\def\Lie{\hbox{\bf g}}
\def\ind{{\hbox{\rm ind}}}

\def\ex{{\hbox{\rm e}}}  
\def\tr{{\hbox{\rm Tr}}}
\def\too{\longrightarrow}
\def\half{{1\over 2}} 

\def\to{\rightarrow}

\def\sqr#1#2{{\vcenter{\vbox{\hrule height.#2pt
        \hbox{\vrule width.#2pt height#1pt \kern#1pt
           \vrule width.#2pt}
        \hrule height.#2pt}}}}

\def\deriv{{\cal D}}
%
%en modo horizontal, \sqr64\sqr64 y \hskip -5pt \hskip -7pt
%

\def\raiz{\sqrt{2}}

\def\dalpha{{\dot\alpha}}
\def\dbeta{{\dot\beta}}
\def\dgamma{{\dot\gamma}}
\def\ddelta{{\dot\delta}}
\def\mapright#1{\smash{\mathop{\longrightarrow}\limits^{#1}}}

%
%alternativa: \buildrel ....... \over .....  ;ejemplos:
%\buildrel \alpha\beta \over \longrightarrow ,
%\buildrel \rm def \over = , etc.
%
%\input amssym.def
%\input amssym

%\def\complex{{\Bbb C}\thinspace}
%\def\Erre{\Bbb R}

\def\metrica#1#2{\eta_{\underline{#1}\underline{#2}}}
%
%
%$\Bbb C$, $\Bbb R$, $\Bbb Z$ 
%$$\Bbb C \Bbb R \Bbb Z$$
\def\Erre{\bf R}
\def\complex{\bf C}
\def\Bbb Z{\bf Z}
%
%%%%%%%%%%%%%%%%%%%%%%%%%%%%%%%%%%%%%%%%%%%%%%%%%%%%%%%%%%%%%%%%%%%%%%%%%%%%%%%
%                                                                             %
%%%%%%%%%%%%%%%%%%%%%%%%%%%%%%%%%%%%%%%%%%%%%%%%%%%%%%%%%%%%%%%%%%%%%%%%%%%%%%%
%                                                                           
%   
%
%\Pubnum={US-FT-4-97 \cr hep-th/9702106}
\pubnum={US-FT-4-97 \cr hep-th/9702106}
\date={February, 1997}
\pubtype={}
\titlepage

\title{MATHAI-QUILLEN FORMULATION OF TWISTED $N=4$ SUPERSYMMETRIC GAUGE
THEORIES IN FOUR DIMENSIONS} \author{J. M. F.
Labastida\foot{e-mail: labastida@gaes.usc.es}, Carlos Lozano
\foot{e-mail: lozano@fpaxp1.usc.es}}
\address{Departamento de F\'\i sica de Part\'\i culas\break Universidade de
Santiago de Compostela\break E-15706 Santiago de Compostela, Spain}  

\abstract{We present a detailed description of the three inequivalent twists
of $N=4$ supersymmetric gauge theories. The resulting topological quantum
field theories are reobtained in the framework of the Mathai-Quillen formalism
and the corresponding moduli spaces are analyzed. We study their geometric
features in each case. In one of the twists we make contact with the theory of
non-abelian monopoles in the adjoint representation of the gauge group. In
another twist we obtain a topological quantum field theory which is orientation
reversal invariant. For this theory we show how the functional integral
contributions to the vacuum expectation values leading to topological
invariants notably simplify.}

\endpage  
\pagenumber=1 
%  
%   
%
%%%%%%%%%%%%%%%%%%%%%%%%%%%%%%%%%%%%%%%%%%%%%%%%%%%%%%%%%%%%%%%%%%%%%%%%%%%%%%%%
\chapter{Introduction}

Topological quantum field theory \REF\tqft{E. Witten\journal\cmp
&117(88)353.} [\tqft] constitutes a very fruitful framework to apply and test 
different ideas emerged in the context of duality as a symmetry of extended
supersymmetric gauge theories. Two salient examples are the introduction of
Seiberg-Witten invariants in \REF\swequ{N. Seiberg and E. Witten, {\sl Nucl.
Phys.} {\bf B426} (1994) 19, Erratum {\bf B430} (1994) 485.}
\REF\swotro{N. Seiberg and E. Witten, {\sl Nucl. Phys.}
{\bf B431}  (1994) 484.}
\REF\monop{E. Witten\journal\mrl&1(94)769.} [\swequ,\swotro,\monop], and
the strong coupling test of $S$-duality carried out by Vafa and Witten in 
\REF\vafa{C. Vafa and E. Witten\journal\np&B431(94)3.} [\vafa] from the
analysis of  a twisted $N=4$ supersymmetric gauge theory. Subsequent
generalizations in the framework of Seiberg-Witten invariants have been
presented in \REF\marmon{J. M. F. Labastida and  M.
Mari\~no\journal\np&B448(95)373.} \REF\marpol{J. M. F. Labastida and M.
Mari\~no\journal\np&B456(95)633.}
\REF\korea{S. Hyun, J. Park and
J. S. Park,  {\sl Nucl. Phys.} {\bf B453} (1995) 199.}
\REF\zzeta{J. M. F. Labastida and M. Mari\~no, 
``Twisted $N=2$ Supersymmetry with
Central Charge  and Equivariant Cohomology", 
hep-th/9603169, to appear in {\sl Comm. Math. Phys}.}
\REF\marth{M. Mari\~no, ``The Geometry of Supersymmetric Gauge Theories 
in Four Dimensions", Ph. D. Thesis, hep-th/9701128.}
\REF\okte{C. Okonek and A. Teleman, {\sl Int. J. Math.}
{\bf 6} (1995) 893; {\sl \cmp} {\bf 180} (1996) 363; ``Recent developments in
Seiberg-Witten theory and  complex geometry", alg-geom/9612015.}
\REF\tele{A. Teleman, ``Non-abelian Seiberg-Witten theory and
projectively stable pairs",  alg-geom/9609020.}
\REF\pity{V. Pidstrigach and A. Tyurin, ``Localisation of the
Donaldson's invariants along Seiberg-Witten classes", dg-ga/9507004.}
\REF\oscar{S. Bradlow and O. Garc\'\i a-Prada, ``Non-abelian
monopoles and vortices", alg-geom/9602010.} [\marmon-\oscar]. However, no
further progress has been made on the role played by duality in twisted $N=4$
supersymmetric gauge theories. The main goal of this paper 
is to construct a sound
framework to pursue further developments on this issue.

The first analysis of twisted $N=4$ supersymmetric gauge theories was carried
out by Yamron in \REF\yamron{J. P. Yamron\journal\pl&B213(88)325.} [\yamron]
where he presented the structure of two of the possible non-equivalent
twists of these theories and pointed out the existence of a third one. This
third twist was first addressed by Marcus in \REF\marcus{N.
Marcus\journal\np&B452(95)331.} [\marcus].  These twists have not been fully
presented in these works. In the second section of this paper we will describe
in full detail the twisting procedure in each of the cases and we will present
complete off-shell topological actions for all the three cases. For the twist
treated by Vafa and Witten the construction completes the action presented in
[\vafa] while for the twist treated by Marcus it provides an off-shell
formulation which is equivalent to the one recently obtained in
\REF\blauthomp{M. Blau and G. Thompson, ``Aspects of $N_{T}\geq 2$ Topological
Gauge Theories and D-Branes", hep-th/9612143.} [\blauthomp].  In the case of
the other twist we make contact with the topological quantum field theory of
non-abelian monopoles introduced in [\marmon] for the case in which matter
fields are in the adjoint representation.

It is well known that topological quantum field theories obtained after
twisting $N=2$ supersymmetric gauge theories can be formulated in the
framework of the Mathai-Quillen formalism. One would expect that a similar
formulation should exist for the $N=4$ case.  Though it turns out that this
is so, there is an important issue that has to be addressed to clarify  what 
it is
meant by a Mathai-Quillen formulation in the latter case.
Twisted $N=2$ supersymmetric gauge theories have an off-shell formulation such
that the topological quantum field theory action can be expressed as a
$Q$-exact expression, being $Q$ the part of the $N=2$ supersymmetry which
remains after the twisting and is valid on curved space. Actually, this is true
only up to a $\theta$-term. However, due to the chiral anomaly inherent to the
$R$-symmetry of $N=2$ supersymmetric gauge theories,  observables are
independent of $\theta$-terms up to a rescaling. This allows to disregard
these terms and to just consider the $Q$-exact part of the action which
is precisely the one obtained in the Mathai-Quillen formalism.

In $N=4$ supersymmetric gauge theories $\theta$-terms are observable.
There is no chiral anomaly and these terms can not be shifted away as in the
$N=2$ case. This means that in the twisted theories one might have a dependence
on the coupling constants (in fact, this was one of the key observations in
[\vafa] to make a strong coupling test of $S$-duality). This being so we
first have to clarify what one expect to be the form of the twisted
theories in the framework of the Mathai-Quillen formalism. To do this let us
concentrate our attention on the part of the action  of a
twisted theory (originated from any gauge theory with extended supersymmetry)
involving  the gauge field strength,
$$
{\cal S}_{X} = -{1\over 4 e^2} \int_X \sqrt{g} d^4x \, \tr(F^{\mu\nu}F_{\mu\nu})
-{i\theta\over 16\pi^2} \int_X \tr(F\wedge F) + \dots,
\eqn\runrun
$$
where $X$ is an oriented Riemannian four-manifold and $g_{\mu\nu}$ a Riemannian
metric on it. We are using conventions such that,
$$
k={1\over 16\pi^2} \int_X \tr(F\wedge F)={1\over32\pi^2}\int_X
\sqrt{g}\tr( *  F_{\mu\nu}F^{\mu\nu})=
{1\over32\pi^2}\int_X
\sqrt{g}\tr\bigl\{\,( F^{+})^2-(F^{-})^2\bigr\},
\eqn\knumber
$$  
gives the instanton number. We also take the path integral to be $Z\sim
\int\ex^{\cal S}$. Using the decomposition of the field strength
$F$ into its self-dual and anti-selfdual parts,
$$
F^\pm_{\mu\nu} = {1\over 2}(F_{\mu\nu}\pm{1\over 2}\epsilon_{\mu\nu\rho\sigma}
F^{\rho\sigma}),
\eqn\coque
$$
\runrun\ can be written in the following two forms:
$$
\eqalign{
{\cal S}_X  = &-{1\over 2 e^2} \int_X \sqrt{g} d^4x \,
\tr(F^{+\mu\nu}F^+_{\mu\nu})
 -2\pi i\tau {1\over 16\pi^2} \int_X \tr(F\wedge F) + \dots \cr
     = & -{1\over 2 e^2} \int_X \sqrt{g} d^4x \, \tr(F^{-\mu\nu}F^-_{\mu\nu})-
      2\pi i\bar\tau {1\over 16\pi^2} \int_X \tr(F\wedge F) + \dots, \cr}
\eqn\tsi
$$
being,
$$
\tau = {\theta \over 2\pi} + {4\pi i\over e^2}.
\eqn\latau
$$

The twist consists of considering as the new rotation group an exotic subgroup
of the global group corresponding to the extended supersymmetry under
consideration. The global group of extended supersymmetry has the form 
$SU(2)_L\otimes SU(2)_R \otimes H$ where $SU(2)_L\otimes SU(2)_R$ constitutes
the rotation group and $H$ the internal or isospin group. For $N=2$, $H=U(2)$,
while for $N=4$,
$H=SU(4)$. In the twisting procedure one first selects one of the two
components of the rotation group and  then replaces it by the diagonal sum of
that component with a $SU(2)$ subgroup of the internal group. In the case of
$N=2$ this can be done in only one way while for $N=4$ there are three
possibilities. These will be fully described in the next section. What we
intend to discuss here is the difference between the two possible choices
which are present when picking up one of the components of the rotation group.
It turns out that choosing one of them, say, the left or twist $T$, one must
consider the first form of the action in \tsi\ since then, after working out
its off-shell formulation, it can be written as 
$$
{\cal S}^T_X =  {1\over 2 e^2} \int_X \sqrt{g} d^4x \, \{Q,\Lambda\} -
           2\pi i\tau {1\over 16\pi^2} \int_X \tr(F\wedge F),
\eqn\terouno
$$
for some $\Lambda$, while it one chooses the other one, the right or twist
$\tilde T$, one finds, 
$$
{\cal S}^{\tilde T}_X =  {1\over 2 e^2} \int_X \sqrt{g} d^4x \, \{\tilde
Q,\tilde\Lambda\} -
           2\pi i\bar\tau {1\over 16\pi^2} \int_X \tr(F\wedge F),
\eqn\terodos
$$
for some $\tilde\Lambda$ and some $\tilde Q$. These actions correspond to
an orientable four-manifold $X$ with a given orientation. The actions of the two
twists are related in the following way:
$$
{\cal S}^{T}_X  = {\cal S}^{\tilde T}_{\tilde X} \Big|_{\tau\rightarrow
-\bar\tau}, 
\eqn\titin
$$
where the four-manifolds $X$ and $\tilde X$ are related by a reversal of
orientation. 

For twisted theories originated from $N=2$ supersymmetric gauge theories,
observables do not depend on  $e$ because  it appears only in a term which
is $Q$-exact. They do not depend either on $\tau$, up to a rescaling, due to
the chiral anomaly. 
In the case of twisted theories originated from $N=4$
supersymmetric gauge theories, however, observables  are independent of $e$
but possess a dependence on $\tau$.
In both cases one needs to consider only one of the types of twist, say
$T$, since, according to \titin, the other just leads to the observables that
one would obtain considering $\tilde X$ instead of $X$. In the first case this
statement is exact and in the second case one must supplement it with the
replacement $\tau\rightarrow -\bar\tau$. Therefore one can say that up to a
reversal of orientation there is only one possible twist from $N=2$
supersymmetric gauge theories and three, as stated in [\yamron] and described
in detail in the next section, from  $N=4$ theories.

After these remarks on the twisting procedure we will state what is meant by a
Mathai-Quillen formulation of topological quantum field theories resulting
after  twisting  $N=4$ supersymmetric gauge theories. The Mathai-Quillen
formulation builds out of a moduli problem a representative of the Thom
class associated to the corresponding vector bundle. This representative can
always be written as an integral of the exponential of a $Q$-exact expression.
The three twists of $N=4$, after working out their off-shell formulation,
can be written as in \terouno. We will present for each case the moduli problem
which in the context of the Mathai-Quillen approach leads to the $Q$-exact part
of the action. In other words, we will find out the geometrical content
which is behind each of the three twists.

One of the three twists, the one first considered by Marcus in [\marcus],
possesses special features. It turns out that the  topological quantum
field theories resulting from the twist $T$ and from the twist
$\tilde T$ are the same:
$$
{\cal S}^T_X = {\cal S}^{\tilde T}_X.
\eqn\amphiuno
$$
We will call theories satisfying this property {\sl amphicheiral} topological
quantum field theories. The reason for this name is that for these theories,
after using \titin,
$$
{\cal S}^T_X = {\cal S}^T_{\tilde X}\Big|_{\tau\rightarrow -\bar\tau},
\eqn\amphidos
$$
in other words, for a fixed twist the observables of the theory on $X$ and
on $\tilde X$ are related after reversing the sign of their dependence on the
real part of $\tau$, \ie, of their dependence on the $\theta$-angle.
Amphicheiral topological quantum field theories 
seem to possess very special properties which make them rather simple.
An example of this type of theories, the one resulting from the third twist,
will be analyzed in sect. 4.

The paper is organized as follows. In sect. 2 we present a detailed
description of the three twists of $N=4$ supersymmetric gauge theories,
obtaining their off-shell formulation and their canonical form \terouno.
In sect. 3 the three moduli problems associated to each of the twists are
presented and the construction of the corresponding Thom class
representatives is carried out making contact with the actions obtained in
sect. 2. In sect. 4 we discuss the observables of these theories and the special
features of the amphicheiral topological quantum field theory which results in
the third the twist. Finally, in sect. 5 we state our conclusions.  An appendix
describes our conventions and collects a set of useful formulas used throughout
the paper.

\endpage

\chapter{Twisting of  $N=4$ Supersymmetric Gauge Theory}

In this chapter we will obtain the 
actions and BRST-like symmetries which result after twisting $N=4$
supersymmetric gauge theories. We will first introduce the $N=4$ physical
theory and then we will carry out in detail its three possible twists.  

\section{$N=4$ supersymmetric gauge theory}

We begin with the standard $N=4$ supersymmetric gauge theory on flat 
${\Erre}^4$. Our conventions regarding spinor notation are almost 
as in Wess and Bagger \REF\wessba{J. Wess,  J. Bagger, 
{\it Supersymmetry and Supergravity},
Princeton University Press, 1992.} 
[\wessba], with some differences that we conveniently 
compile in the appendix.
The field content of the model is the following: a gauge field
$A_{\alpha\dalpha}$, gauginos $\lambda_u{}^\alpha$ and
$\bar\lambda^u{}_{
\dalpha}$ transforming respectively in the representations ${\bf 4}$ and 
${\bf\bar 4}$ of  
$SU(4)_I$ ($SU(4)_I$ is the global isospin group of the theory, and
indices ($u,v,w,\ldots$) label its fundamental representation), and
scalars
$\phi_{uv}$ in the
${\bf 6}$ of
$SU(4)_I$. All the fields above take values in the adjoint representation
of some compact  Lie group $G$. Being in the 
representation ${\bf 6}$, the scalars $\phi_{uv}$ satisfy antisymmetry and
self-conjugacy constraints:
$$
\eqalign{&\phi_{uv}=-\phi_{vu}, \cr
&\phi^{uv}=(\phi_{uv})^{\dag}=\phi^{*}_{vu}=
-\half\epsilon^{uvwz}\phi_{wz};\quad \epsilon_{1234}=\epsilon^{1234}=+1.\cr}
\eqn\Valle
$$ 
%The second constraint (which expresses the
%selfconjugacy of the representation) is a consequence of the fact
%that the isospin
%group is just $SU(4)_I$ and not the full $U(4)_I$. 

The action for the model in Euclidean space is:
$$
\eqalign{ {\cal S}&= {1\over e^2}\int d^4 x\, \tr\, \bigl\{\,-{1\over
8}\nabla_{\!\alpha
\dalpha}\phi_{uv}\nabla^{\dalpha\alpha}\phi^{uv} 
-i\lambda_v{}^\alpha\nabla_{\!\alpha\dot\alpha}
\bar\lambda ^{v\dot\alpha} -{1\over4} F_{mn} F^{mn}\cr &-
{i\over\raiz}\,\lambda_u{}^\alpha [\lambda_{v\alpha},\phi^{uv}]
+{i\over\raiz}\,\bar\lambda^u{}_{\dot\alpha}[\bar\lambda^
{v\dot\alpha},\phi_{uv}]+{1\over 16}[\phi_{uv},\phi_{wz}]
[\phi^{uv},\phi^{wz}] \,\bigr\}\cr
&-{i\theta\over 32\pi^2}\int d^4 x\,\tr\,\bigl\{\, *  F_{mn}F^{mn}
\,\bigr\}.
\cr}
\eqn\Boris
$$  
We have introduced the covariant derivative $\nabla_{\!\alpha\dalpha}
=\sigma^m{}_{\alpha\dalpha}(\partial_m+i[A_m,~])$
(together with its corresponding field strength $F_{mn}=
\partial_m A_n -\partial_n A_m +i[A_m ,A_n]$) and the trace $\tr$ in the
adjoint representation, which we normalize as follows:
$\tr(T^a T^b)=\delta^{ab}$, being
$T^a$, $a=1,\dots,$dim$(G)$, the hermitian generators of the gauge group in
the adjoint representation.  The action \Boris\ is invariant under the
following four supersymmetries (in $SU(4)_I$ covariant notation):
$$
\eqalign{ 
&\delta A_{\alpha\dalpha} = -2i\bar\xi^u{}_\dalpha\lambda_{u
\alpha}+2i\bar\lambda^u{}_\dalpha\xi_{u\alpha },\cr 
&\delta\lambda_{u\alpha} =  -iF^{+}{}_{\!\alpha}{}^{\!\beta}\xi_{u\beta}+
i{\raiz}\bar\xi^{v\dot\alpha}\nabla_{\alpha\dot\alpha}
\phi_{vu}-i\xi_{w\alpha}[\phi_{uv},\phi^{vw}],\cr 
&\delta\phi_{uv}={\raiz}\bigl\{\xi_u{}^\alpha\lambda_{v\alpha}
-\xi_v{}^\alpha\lambda_{u\alpha} +
\epsilon_{uvwz}\bar\xi^w{}_{\dalpha}\bar\lambda^{z\dalpha}\bigr\},\cr}
\eqn\Vian
$$
where $F^{+}{}_{\!\alpha}{}^{\!\beta}=\sigma^{mn}{}_{\!\alpha}{}^{\!\beta}
F_{mn}$.
Notice that there are no auxiliary fields in the action \Boris.
Correspondingly, the transformations
\Vian\ close the supersymmetry algebra on-shell.   

As already discussed in the introduction, in $\Erre^4$, the global symmetry
group of $N=4$ supersymmetric theories is ${\cal H} =SU(2)_L\otimes
SU(2)_R\otimes SU(4)_I$, where  ${\cal K}= SU(2)_L\otimes SU(2)_R$ is the
rotation group $SO(4)$. The supersymmetry generators responsible
for the transformations \Vian\ are $Q^u{}_\alpha$ and $\bar Q_{u\dalpha}$
They  transform  as $({\bf 2},{\bf 1},{\bf \bar 4})\oplus({\bf
1},{\bf 2},{\bf 4})$ under ${\cal H}$.

From the point of view of $N=1$ superspace, the theory contains one $N=1$
vector  multiplet and three $N=1$ chiral multiplets. These supermultiplets
are represented  in $N=1$ superspace  by superfields
$V$ and $\Phi_s$ ($s=1,2,3$), which  satisfy the 
constraints $V=V^{\dag}$ and $\bar D_\dalpha \Phi_s=0$, being 
$\bar D_\dalpha$ a superspace covariant derivative. The physical
component fields of these superfields are:
$$
\eqalign{V &\longrightarrow\;   A_{\alpha\dalpha},\; 
\lambda_{4\alpha},\; \bar\lambda^4{}_{\dalpha},\cr 
\Phi_s, \Phi^{\dag s} &\longrightarrow\; B_s,\;
\lambda_{s\alpha},\; B^{\dag s},\;
\bar\lambda^s{}_{\dalpha}.\cr}
\eqn\ctres
$$ 
In terms of these fields, the $SU(4)_I$ tensors that we introduced above are
defined as follows: 
$$
\eqalign{&\{{\bf 4}\}  \to \lambda_u =
\{\lambda_1,\lambda_2,\lambda_3,\lambda_4\},\cr &\{{\bf 6}\} \to
\phi_{uv} \sim \{B_s,B^{\dag s}\},\cr  &\{{\bf \bar 4}\} \to
\bar\lambda^u = 
\{\bar\lambda^1,\bar\lambda^2,\bar\lambda^3,\bar\lambda^4\},\cr}
\eqn\cnt
$$ 
where by $\sim$ we mean precisely:
$$
\phi_{uv}=\!\pmatrix{\!0&-\!B^{\dag 3}&\!B^{\dag 2}&-\!B_1\cr 
                     \!B^{\dag 3}&0&\!-B^{\dag 1}&\!-B_2\cr
                     \!-B^{\dag 2}&\!B^{\dag 1}&\!0&\!-B_3\cr
                     \!B_1&        \!B_2&       \!B_3&\! 0\cr},\quad
\phi^{uv}=\!\pmatrix{\!0&\!B_3&\!-B_2&\!B^{\dag 1}\cr
                   \!-B_3&\!0&\!B_1&\!B^{\dag 2}\cr
                   \!B_2&\!-B_1&\!0&\!B^{\dag 3}\cr
                   \!-B^{\dag 1}&\!-B^{\dag 2}&\!-B^{\dag 3}&\!0\cr}.
\eqn\Inclan
$$

The action \Boris\  takes the following form in $N=1$ superspace:
$$
\eqalign{{\cal S} =& -{i\over 4\pi}\tau\int d^4 xd^2 \theta\, \tr (W^2) +
{i\over 4\pi}\bar\tau\int d^4 x d^2
\bar\theta\, \tr (W^{\dag 2}) \cr & 
+{1\over e^2}\sum_{s=1}^3 \int d^4 xd^2 \theta d^2
\bar\theta\, \tr(\Phi^{\dag s} \ex^V \Phi_s) \cr   &+{i\raiz\over e^2}
\int d^4xd^2\theta \, \tr\bigl\{\Phi_1[\Phi_2 ,\Phi_3]\bigr\} +
{i\raiz\over e^2}\int d^4 xd^2\bar\theta\,\tr\bigl\{\Phi^{\dag 1} 
[\Phi^{\dag 2},\Phi^{\dag 3}]\bigr\},
 \cr }
\eqn\cuno
$$ 
where $W_\alpha =-{1\over 16}\bar D^2 \ex^{-V}D_\alpha \ex^V$.

\section{Twisting $N=4$ Supersymmetry Gauge Theory}  The purpose of this
section is to analyze in detail the twists of $N=4$ supersymmetric gauge
theory. We  assume that the reader is familiar with the analogous (yet
simpler) procedure in $N=2$ theories 
\REF\wijmp{E. Witten\journal\jmp &35(94)5101.}
\REF\phyrep{D. Birmingham, M. Blau, M. Rakowski and G.
Thompson, {\sl Phys. Rep.} {\bf 209} (1991) 129.}
\REF\labmat{M. Alvarez, J. M. F. Labastida\journal\np&B437(95)356.}
\REF\labas{J. M. F. Labastida,
``Topological Quantum Field Theory: A Progress Report", Proceedings of the IV
Workshop on Differential Geometry and its Applications, 1995, edited by M.
Salgado and E. V\'azquez, CIEMAT, Anales de F\'\i sica Monograf\'\i as, {\bf
3}, 101-124, hep-th/9511037.} 
[\tqft,\phyrep,\wijmp,\labmat,\labas]. The aim of  the twist is
to extract from the supersymmetries of the theory under  consideration
one (or several)  scalar BRST-like symmetries which can be 
readily generalized to any arbitrary four manifold. To create a scalar 
supercharge out of spinor supercharges one has to modify somehow the
rotation group. The idea, as discussed in the introduction, is to replace 
one of the $SU(2)$ components of the rotation group ${\cal K}$
by its diagonal sum with an $SU(2)$ subgroup of the isospin group 
$SU(4)_I$. Depending on how we choose this subgroup, we will obtain
different theories after the twisting.  The
possible choices are found just analyzing how
the ${\bf 4}$  of $SU(4)_I$ splits in terms of
representations of the rotation group ${\cal K}$. There are just three
possibilities for a given choice of the $SU(2)$ component of ${\cal K}$: (1)
${\bf 4}\to({\bf 2},{\bf 1})\oplus ({\bf 2},{\bf 1})$, (2)
${\bf 4}\to({\bf 2},{\bf 1})\oplus ({\bf 1},{\bf 1})\oplus ({\bf 1},{\bf 1})$
and (3) ${\bf 4}\to ({\bf 2},{\bf 2})$, each of which leads to a different
topological quantum field theory.    
Choosing the other $SU(2)$ component of ${\cal K}$ one would obtain the other
three $\tilde T$ twists:
${\bf 4}\to({\bf 1},{\bf 2})\oplus ({\bf 1},{\bf 2})$, 
${\bf 4}\to({\bf 1},{\bf 2})\oplus ({\bf 1},{\bf 1})\oplus ({\bf 1},{\bf 1})$
and ${\bf 4}\to ({\bf 2},{\bf 2})$. As described in the introduction all these
twists are related to the other ones after a reversal of orientation of the
four-manifold $X$. Notice that in the third case both twists, $T$ and $\tilde
T$, involve the same splitting of the ${\bf 4}$ of $SU(4)_I$, anticipating its
amphicheiral character.

%%%%%%%%%%%%%%%%%  %%%%%%%%%%%%%%%%%%%%%%%%%%%%%%%%%%%%%%%%%%%%%%%%%%%%%%%%
%%VAFA-WITTEN%%%%
%%%%%%%%%%%%%%%%%  %%%%%%%%%%%%%%%%%%%%%%%%%%%%%%%%%%%%%%%%%%%%%%%%%%%%%%%%

\subsection{(1) ${\bf 4}\to({\bf 2},{\bf 1})\oplus ({\bf 2},{\bf 1})$
Vafa-Witten Theory}

This is the twist that has been considered by Vafa and Witten in
[\vafa].  After the twisting,  the 
symmetry group of the theory becomes ${\cal H'} =SU(2)'_L\otimes
SU(2)_R\otimes SU(2)_F$, where $SU(2)_F$ is a subgroup of $SU(4)_I$ that
commutes with the defining identification ${\bf 4}\to({\bf 2},{\bf
1})\oplus ({\bf 2},{\bf 1})$ and remains in the theory as a residual
isospin group. Under ${\cal H'}$, the supercharges split up as,
$$
Q^v{}_\alpha\to Q^i,\;Q^i{}_{\alpha\beta},\qquad \bar Q_{v\dalpha}\to
\bar Q^i_{\alpha\dalpha},
\eqn\elven
$$
where the index $i$  labels the fundamental
representation of
$SU(2)_F$. The twist has produced a scalar supercharge, the $SU(2)_F$
doublet $Q^i$. This scalar charge is defined in terms of the original
supercharges as follows:
$$
\eqalign{Q^{i=1}&\equiv Q^{v=1}_{\alpha =1} +Q^{v=2}_{\alpha=2},
\cr 
Q^{i=2}&\equiv Q^{v=3}_{\alpha =1} +Q^{v=4}_{\alpha=2}.
\cr }
\eqn\osiete
$$

The fields of the $N=4$ multiplet decompose under ${\cal H'}$ in the
following manner:
$$
\eqalign{
&A_{\alpha\dalpha} \too A_{\alpha\dalpha},\cr
&\lambda_{v\alpha}\too
\chi_{i\beta\alpha},~
\eta_{i},\cr
&\bar\lambda^v{}_{\dalpha} \too
\psi^{i\alpha}{}_{\dalpha},\cr
&\phi_{uv}\too \varphi_{ij},~G_{\alpha\beta}.\cr}
\eqn\poocho
$$
Notice that the fields $\chi^i_{\alpha\beta}$ and $G_{\alpha\beta}$ are
symmetric in their spinor indices and therefore can be regarded as
components of self-dual two-forms. $\varphi_{ij}$ is also symmetric in its
isospin indices and thus transforms in the representation ${\bf 3}$ of
$SU(2)_F$. Some of the definitions in
\poocho\ need clarification. Our choices for the anticommuting fields are, 
$$
\eqalign{&\chi_{i=1 (\alpha\beta)}=\cases{\chi_{i=1
(11)}=\lambda_{v=1,\alpha=1},\cr
\chi_{i=1(12)}=\half(\lambda_{v=1,\alpha=2}+\lambda_{v=2,\alpha=1}),
\cr\vdots\cr}\cr
&\eta_{i=1}=\lambda_{v=1,\alpha=2}-\lambda_{v=2,\alpha=1},\cr
&\psi^{i=1,\alpha=1,2}_\dalpha =\bar\lambda^{v=1,2}_\dalpha,
\cr}
\eqn\onueve
$$
while for the scalars $\phi_{uv}$:
$$
\eqalign{\varphi_{ij}=\pmatrix{\phi_{12}&\half(\phi_{14}-\phi_{23})\cr
\half(\phi_{14}-\phi_{23})&\phi_{34}\cr},&\cr
G_{\alpha\beta}&=\pmatrix{
\phi_{13}&\half(\phi_{14}+\phi_{23})\cr\half(\phi_{14}+
\phi_{23})&\phi_{24}\cr}.}
\eqn\noventa
$$
%Formulas like \onueve\ and \noventa , and like many other similar ones that 
%will be coming out soon must be taken with a grain of salt. They are supposed 
%to be not more than a tool for working out the twisted action. Once we have 
%written it down, we will not make further use of them. 
 
In terms of the twisted fields, the $N=4$ action \Boris\ takes the form 
(remember that we are still on flat $\Erre^4$):
$$
\eqalign{ 
{\cal S}^{(0)} &= {1\over e^2}\int d^4 x\, \tr\, \bigl\{\,{1\over
4}\nabla_{\alpha
\dalpha}\varphi_{ij}\nabla^{\dalpha\alpha}\varphi^{ij} -{1\over4}\nabla_
{\alpha\dalpha}G_{\beta\gamma}\nabla^{\dalpha\alpha}G^{\beta\gamma} 
-i\psi^{j\beta}{}_{\dot\alpha}\nabla^{\dot\alpha\alpha}
\chi_{j\alpha\beta}\cr& 
-{i\over 2}\psi^j_{\alpha\dot\alpha}\nabla^{\dot\alpha\alpha}
\eta_j
-{1\over4} F_{mn} F^{mn}
-{i\over\raiz}\,\chi_i{}^{\alpha\beta}[\chi_{j\alpha\beta},\varphi^{ij}]
+{i\over\raiz}\,\chi^{i\alpha}{}_\beta[\chi_{i\alpha\gamma},G^{\beta\gamma}]
\cr&-{i\over\raiz}\,\chi^i{}_{\alpha\beta}[\eta_i,G^{\alpha\beta}]-
{i\over{2\raiz}}\,\eta_i[\eta_j,\varphi^{ij}]
+{i\over\raiz}\,\psi^{i\alpha}{}_{\dot\alpha}[\psi_{i\beta}{}^
{\dot\alpha},G_\alpha{}^\beta]\cr&-
{i\over\raiz}\,\psi^{i\alpha}{}_{\dot\alpha}[\psi^j{}_{\alpha}{}^
{\dot\alpha},\varphi_{ij}]+{1\over 4}[\varphi_{ij},\varphi_{kl}]
[\varphi^{ij},\varphi^{kl}] 
-\half[\varphi_{ij},G_{\alpha\beta}][\varphi^{ij},G^{\alpha\beta}] 
\cr&+{1\over 4}[G_{\alpha\beta},G_{\gamma\delta}]
[G^{\alpha\beta},G^{\gamma\delta}]\,\bigr\}
-{i\theta\over 32\pi^2}\int d^4 x\,\tr\,\bigl\{\, *  F_{mn}F^{mn}
\,\bigr\}.
\cr}
\eqn\nuno
$$ 
 
The $Q^i$-transformations of the twisted theory can be readily obtained
from the corresponding $N=4$ supersymmetry transformations. These last
transformations are generated by $\xi_v{}^\alpha Q^v{}_\alpha + 
\bar\xi^v{}_\dalpha \bar Q_v{}^\dalpha$. According to our conventions, to
obtain the $Q^i$-transformations one must set $\bar\xi^v{}_\dalpha=0$ and
make the replacement:  
$$
\xi_{v\alpha}=\cases{\xi_{v=(1,2)\alpha}\to \epsilon_{i=
1}C_{\beta=(1,2)\alpha},\cr
\xi_{v=(3,4)\alpha}\to \epsilon_{i=
2}C_{\beta=(1,2)\alpha},\cr}
\eqn\ndos
$$
where $C_{\beta\alpha}$ (or $C_{\dbeta\dalpha}$, $C_{ij}$) is the
antisymmetric (invariant) tensor of
$SU(2)$ with the convention $C_{21}=C^{12}=+1$. The resulting
transformations are:
$$
\eqalign{
&\delta A_{\alpha\dalpha} = 2i\epsilon_j\psi^j_{\alpha\dalpha},\cr 
&\delta F^{+}_{\alpha\beta}=2\epsilon_i\nabla_{(\alpha}{}^{\dot\alpha}
\psi^i{}_{\beta)\dot\alpha},\cr
&\delta\psi^{i\alpha}{}_{\dot\alpha} = 
-i{\raiz}\epsilon_j\nabla^{\alpha}{}_{\dot\alpha}
\varphi^{ji} +
i{\raiz}\epsilon^i\nabla_{\beta\dot\alpha}
G^{\beta\alpha},\cr
&\delta\chi_{i\alpha\beta} = 
-i\epsilon_i F^{+}_{\alpha\beta}
-i\epsilon_i[G_{\gamma\alpha},G^{\gamma}{}_{\beta}]-
2i\epsilon_j[G_{\alpha\beta},\varphi^j{}_i],\cr 
&\delta\eta_i =2i\epsilon_k[\varphi_{ij},\varphi^{jk}],\cr
&\delta\varphi_{ij}=\raiz\epsilon_{(i}\eta_{j)},\cr
&\delta G_{\alpha\beta}=\raiz\epsilon^i \chi_{i\alpha\beta},\cr}
\eqn\ntres
$$
where, for example,
$\epsilon_{(i}\eta_{j)}=\half (\epsilon_{i}\eta_{j}+\epsilon_{j}\eta_{i})$.  
The transformations \ntres\ satisfy the on-shell algebra 
$[\delta_1,\delta_2]=0$
modulo a non-abelian gauge transformation generated by the scalars
$\varphi_{ij}$. For example, 
$
[\delta_1,\delta_2]G_{\alpha\beta}=
-4\raiz i\epsilon_1{}^{\!i}
\epsilon_2{}^{j}[\varphi_{ij}, G_{\alpha\beta}]$. 
In checking the algebra, use has to be made of the equations of motion for
the anticommuting fields $\psi^i_{\alpha\dalpha}$ and
$\chi^i_{\alpha\beta}$. In terms of the generators
$Q^i$, the algebra takes the form:
$$
\eqalign{\{Q^1,Q^1\}&= \delta_g (\varphi_{22}),\cr
\{Q^1,Q^2\}&= \delta_g (\varphi_{12}),\cr
\{Q^2,Q^2\}&= \delta_g (\varphi_{11}),\cr
}
\eqn\damned
$$
where by $\delta_g (\Phi)$ we denote the non-abelian gauge transformation
generated by, say, $\Phi$. As explained in [\yamron], it is possible to
realize the algebra off-shell by inserting the auxiliary
fields $N_{\alpha\beta}$ (symmetric in its spinor indices) and
$M_{\alpha\dalpha}$ in the transformations of 
$\psi^i_{\alpha\dalpha}$ and $\chi^i_{\alpha\beta}$. This is the opposite to the
situation one encounters in the associated physical $N=4$ theory, where an
off-shell formulation in terms of unconstrained fields is not possible. After
some suitable manipulations 
\REF\tesina{C. Lozano, {\it Teor\'\i as Supersim\'etricas y Teor\'\i as 
Topol\'ogicas}, Tesina, Universidade de Santiago de Compostela, June 1996.}
[\tesina], 
the off-shell formulation of the twisted theory takes the form:  
$$
\eqalign{ {\cal S}^{(1)}&={1\over e^2}\int d^4 x\, \tr
\bigl\{\,{1\over4}\nabla_{\!\alpha\dalpha}
\varphi_{ij}\nabla^{\dalpha\alpha}\varphi^{ij} +{i\over\raiz}
M^{\dalpha\alpha}\nabla_{\beta\dot\alpha}G^{\beta}{}_\alpha 
-i\psi^{j\beta}{}_{\dot\alpha}\nabla^{\dot\alpha\alpha}
\chi_{j\alpha\beta} 
\cr&-{i\over 2}\psi^j_{\alpha\dot\alpha}\nabla^{\dalpha\alpha}
\eta_j
+{i\over 2} N^{\alpha\beta} F^{+}_{\alpha\beta}
-{i\over\raiz}\,\chi_i{}^{\alpha\beta}[\chi_{j\alpha\beta},\varphi^{ij}]
+{i\over\raiz}\,\chi^{j\alpha}{}_\beta[\chi_{j\alpha\gamma},G^{\beta\gamma}]
\cr&-{i\over\raiz}\,\chi^j{}_{\alpha\beta}[\eta_j,G^{\alpha\beta}]-
{i\over{2\raiz}}\,\eta_i[\eta_j,\varphi^{ij}]
+{i\over\raiz}\,\psi^{j\alpha}{}_{\dot\alpha}[\psi_{j\beta}{}^
{\dot\alpha},G_\alpha{}^\beta]-{1\over4}M_{\alpha\dot\alpha}M^{\dalpha\alpha}
\cr&-
{i\over\raiz}\,\psi^{i\alpha}{}_{\dot\alpha}[\psi^j{}_{\alpha}{}^
{\dot\alpha},\varphi_{ij}]+
{1\over 4}[\varphi_{ij},\varphi_{kl}] [\varphi^{ij},\varphi^{kl}] 
-\half[\varphi_{ij},G_{\alpha\beta}][\varphi^{ij},G^{\alpha\beta}] 
\cr &+
{1\over4}N_{\alpha\beta}N^{\alpha\beta}+
{i\over 2}N_{\alpha\beta}
[G_\gamma{}^{\alpha},G^{\gamma\beta}]\,\bigr\}
-2\pi i\tau{1\over 32\pi^2}\int d^4 x\,\tr\,\bigl\{\, *  F_{mn}F^{mn}
\,\bigr\}.
\cr}
\eqn\nocho
$$ 

The corresponding off-shell transformations are:
$$
\eqalign{
&\delta A_{\alpha\dalpha} = 2i\epsilon_j \psi^j_{\alpha\dalpha},\cr 
&\delta F^{+}_{\alpha\beta}=2\epsilon_i\nabla_{(\alpha}{}^{\dot\alpha}
\psi^i{}_{\!\beta)\dot\alpha},\cr
&\delta\psi^i{}_{\!\alpha\dot\alpha} = 
-i{\raiz}\epsilon_j\nabla_{\alpha\dot\alpha}
\varphi^{ji} +
\epsilon^i M'_{\alpha\dot\alpha},\cr
&\delta\chi_{i\alpha\beta} = 
-2i\epsilon_j[G_{\alpha\beta},\varphi^j{}_{\!i}]+\epsilon_i
N'_{\alpha\beta},\cr
&\delta\eta_i =2i\epsilon_j[\varphi_{ik},\varphi^{jk}],\cr
&\delta\varphi_{ij}=\raiz\epsilon_{(i}\eta_{j)},\cr
&\delta G_{\alpha\beta}=\raiz\epsilon^j \chi_{j\alpha\beta},\cr
&\delta
M'_{\alpha\dot\alpha}=\epsilon^i\bigl\{\,-i\nabla_{\alpha\dot\alpha}
\eta_i+2\raiz i[\psi^j_{\alpha\dot\alpha},\varphi_{ij}]\,\bigr\},\cr
&\delta
N'_{\alpha\beta}=\epsilon^i\bigl\{\,
\raiz i[\eta_i ,G_{\alpha\beta}]-2\raiz
i[\chi_{j\alpha\beta},\varphi^j{}_{\!i}]\, \bigr\}.\cr }
\eqn\pnseis
$$ 
With the aid of the transformations \pnseis\ it is easy (but rather
lengthy) to show that the action \nocho\ can be written as a double
$Q$-commutator plus a $\tau$-dependent term, that is,
$$
\epsilon^2{\cal S}^{(1)}=\delta^2 \Lambda -\epsilon^2 2\pi i k \tau =
-\half\epsilon^2\{Q^i,[Q_i,\Lambda]\}-\epsilon^2 2\pi i k \tau,
\eqn\jackie
$$ 
(here $\delta \equiv \epsilon^i
[Q_i,\}$), with 
$$
\eqalign{\Lambda ={1\over e^2}\int d^4
x\,\bigl\{\,&{i\over{2\raiz}}F^{+}_{\alpha\beta}G
^{\alpha\beta}+{1\over{4\raiz}}N_{\alpha\beta}G^{\alpha\beta}+ {1\over
8}\psi_{j\alpha\dalpha}\psi^{j\dalpha\alpha}\cr &+
{i\over{6\raiz}}G_{\alpha\beta}[G^\beta{}_\gamma ,G^{\gamma\alpha}]-
{i\over{12\raiz}}\varphi_{ij}[\varphi^j{}_k ,\varphi^{ik}]\,\bigr\}.\cr}
\eqn\nnueve
$$

The next step is to couple the theory to an arbitrary background metric 
$g_{\mu\nu}$ of Euclidean signature. This can be done as follows: {\sl first},
covariantize the expression \nnueve\ and the transformations \pnseis ;
{\sl second}, define the new action to be $\delta^2 \Lambda_{cov}$. The
resulting action is:
%La ponemos o no la ponemos?
$$
\eqalign{ {\cal S}^{(1)}_{c}&={1\over e^2}\int_X  d^4 x\,\sqrt{g} \tr
\bigl\{\,{1\over4}\nabla_{\!\alpha\dalpha}
\varphi_{ij}\nabla^{\dalpha\alpha}\varphi^{ij} +{i\over\raiz}
M^{\dalpha\alpha}\deriv_{\beta\dot\alpha}G^{\beta}{}_\alpha 
-i\psi^{j\beta}{}_{\dot\alpha}\deriv^{\dot\alpha\alpha}
\chi_{j\alpha\beta} 
-\cr&-{i\over 2}\psi^j_{\alpha\dot\alpha}\nabla^{\dalpha\alpha}
\eta_j
+{i\over 2} N^{\alpha\beta} F^{+}_{\alpha\beta}
-{i\over\raiz}\,\chi_i{}^{\alpha\beta}[\chi_{j\alpha\beta},\varphi^{ij}]
+{i\over\raiz}\,\chi^{j\alpha}{}_\beta[\chi_{j\alpha\gamma},G^{\beta\gamma}]
-\cr&-{i\over\raiz}\,\chi^j{}_{\alpha\beta}[\eta_j,G^{\alpha\beta}]-
{i\over{2\raiz}}\,\eta_i[\eta_j,\varphi^{ij}]
+{i\over\raiz}\,\psi^{j\alpha}{}_{\dot\alpha}[\psi_{j\beta}{}^
{\dot\alpha},G_\alpha{}^\beta]-{1\over4}M_{\alpha\dalpha}M^{\dalpha\alpha}
\cr&-
{i\over\raiz}\,\psi^{i\alpha}{}_{\dot\alpha}[\psi^j{}_{\alpha}{}^
{\dot\alpha},\varphi_{ij}]
+{1\over 4}[\varphi_{ij},\varphi_{kl}] [\varphi^{ij},\varphi^{kl}] 
-\half[\varphi_{ij},G_{\alpha\beta}][\varphi^{ij},G^{\alpha\beta}] 
\cr &+
{1\over4}N_{\alpha\beta}N^{\alpha\beta}+
{i\over 2}N_{\alpha\beta}[G_\gamma{}^{\alpha},G^{\gamma\beta}]\,\bigr\}
-2\pi i\tau{1\over 32\pi^2}\int_X  d^4 x\,\sqrt{g}\tr\,\bigl\{\, * 
F_{\mu\nu}F^{\mu\nu}
\,\bigr\},
\cr}
\eqn\nnocho
$$ 
where we have introduced the full covariant derivative
$\deriv_{\alpha\dalpha}$. The action \nnocho\ is, by construction, invariant
under the appropriate covariantized version of the transformations \pnseis
.
%%%%%%%%%%%%%%%%%%%%%%%%%%%%%%%%%%%%%%%%%%%%%%%%%%%%%%%%%%%%%%%%%%%%%%%%%%%%%%

A sensible action for a so-called cohomological topological quantum field 
theory is expected to meet two basic requirements. First of all, it should  be
real, since we will eventually  interpret it as a real differential form defined
on a certain moduli space.  Likewise, it must display a non-trivial ghost number
symmetry which, from the  geometrical viewpoint, corresponds to the de Rham
grading on the moduli  space. These requirements are not fulfilled by the
action \nnocho. First of all, it is not manifestly real because it contains
fields in the fundamental representation of $SU(2)_F$, which are complex.
Second,  it is not possible to  assign a non-trivial ghost number to the fields
in \nnocho . 
%%%%%%%%%%%%%%%%%%%%%%%%%%%%%%%%%%%%%%%%%%%%%%%%%%%%%%%%%%%%%%%%%%%%%%%%%%%%%%

We solve these problems by breaking the $SU(2)_F$ internal symmetry group 
of the theory down to its $T_3$ subgroup. This
allows to introduce a non-anomalous ghost number in the theory
(basically twice the corresponding charge under $T_3$). With
respect to this ghost number, the field content of the theory can be
reorganized as follows (in the notation of reference [\vafa]): with
ghost number
$+2$, we have the scalar field
$\phi\equiv \varphi_{11}$; with ghost number $+1$, the anticommuting fields
$\psi_ {\alpha\dalpha}\equiv i\psi_{1\alpha\dalpha}$,
$\tilde\psi_{\alpha\beta}\equiv \chi_{1\alpha\beta}$ and $\zeta\equiv 
i\eta_1$; with ghost number $0$, the gauge connection
$A_{\alpha\dalpha}$, the scalar field $C\equiv i\varphi_{12}$, the
self-dual two-form  $B_{\alpha\beta}\equiv G_{\alpha\beta}$ and the
auxiliary fields
$H_{\alpha\beta}\equiv iN_{\alpha\beta}$ and 
$\tilde H_{\alpha\dalpha}\equiv M_{\alpha\dalpha}$; with ghost number $-1$,
 the anticommuting fields $\chi_
{\alpha\beta}\equiv i\chi_{2\alpha\beta}$,
$\tilde\chi_{\alpha\dalpha}\equiv \psi_{2\alpha\dalpha}$ and $\eta\equiv 
\eta_2$; and finally, with ghost number $-2$,  the scalar field
$\bar\phi\equiv
\varphi_{22}$.  
Notice that now we can consistently assume that all the fields above are real, 
in order to guarantee the reality of the topological action. 

In terms of these new fields, and after making the shifts:
$$
\eqalign{
\tilde H'_{\alpha\dalpha}&= \tilde
H_{\alpha\dalpha}+\raiz\,\nabla_{\!\alpha\dalpha}C,\cr
H'_{\alpha\beta}&= H_{\alpha\beta}+2i[B_{\alpha\beta},C],\cr}
\eqn\greisen
$$
the action \nnocho\ takes the form: 
$$
\eqalign{ {\cal S}^{(2)}_c&={1\over e^2}\int_X  d^4 x\,\sqrt{g}\, \tr
\bigl\{\,\half\deriv_{\alpha\dalpha}
\bar\phi\deriv^{\dalpha\alpha}\phi -{1\over4}
\tilde H'^{\dalpha\alpha}\bigl (\,\tilde H'_{\alpha\dalpha}-2\raiz \deriv
_{\alpha\dalpha}C-2\raiz i\,
\deriv_{\beta\dot\alpha}B^{\beta}{}_{\!\alpha}\,\bigr )\cr 
&-{1\over 4} H'^{\alpha\beta}\bigl (\, H'_{\alpha\beta}-2\,
F^{+}_{\alpha\beta}-2\,[B_{\gamma\alpha},B_\beta{}^{\!\gamma}]-4i\,
[B_{\alpha\beta},C]\,\bigr
)-i\psi^{\beta}{}_{\dot\alpha}\deriv^{\dot\alpha\alpha}
\chi_{\alpha\beta} 
\cr &-i\tilde\chi^{\beta}{}_{\dot\alpha}\deriv^{\dot\alpha\alpha}
\tilde\psi_{\alpha\beta} 
-{1\over 2}\tilde\chi_{\alpha\dot\alpha}\deriv^{\dalpha\alpha}
\zeta+{1\over 2}\psi_{\alpha\dot\alpha}\deriv^{\dalpha\alpha}
\eta-{i\over\raiz}\,\tilde\psi^{\alpha\beta}[\tilde\psi_{\alpha\beta},\bar
\phi]\cr&
+{i\over\raiz}\,\chi^{\alpha\beta}[\chi_{\alpha\beta},\phi]
-i\raiz\,\tilde\psi^{\alpha\beta}[\chi_{\alpha\beta},C]
-\raiz\,\tilde\psi^{\alpha}{}_\beta\,[\chi_{\alpha\gamma},B^{\beta\gamma}]
\cr&+{i\over\raiz}\,\chi_{\alpha\beta}[\,\zeta,B^{\alpha\beta}]
+{i\over\raiz}\,\tilde\psi_{\alpha\beta}[\,\eta,B^{\alpha\beta}]+
{i\over{2\raiz}}\,\zeta\,[\zeta,\bar\phi]-
{i\over{2\raiz}}\,\eta\,[\eta,\phi]\cr&-
{i\over{\raiz}}\,\zeta\,[\eta,C]
+\raiz\,\psi_{\alpha\dot\alpha}[\tilde\chi_{\beta}{}^
{\dot\alpha},B^{\alpha\beta}]-
{i\over\raiz}\,\tilde\chi^{\alpha}{}_{\dot\alpha}[\tilde\chi_{\alpha}{}^
{\dot\alpha},\phi]\cr&+
{i\over\raiz}\,\psi^{\alpha}{}_{\dot\alpha}[\psi_{\alpha}{}^
{\dot\alpha},\bar\phi]
-
i\raiz\,\psi^{\alpha}{}_{\dot\alpha}[\tilde\chi_{\alpha}{}^
{\dot\alpha},C]
-\half[\phi,\bar\phi]^2
+2[\phi,C][\bar\phi,C]\cr& 
-[\phi,B_{\alpha\beta}][\bar\phi,B^{\alpha\beta}] 
 \,\bigr\}
-2\pi i\tau{1\over 32\pi^2}\int_X  d^4 x\,\sqrt{g}\tr\,\bigl\{\, *  
F_{\mu\nu}F^{\mu\nu}
\,\bigr\}.
\cr}
\eqn\Phoebe
$$ 
The analysis of the bosonic part of a topological action is of great importance.
Apart from comparing to the corresponding theory on flat
${\Erre}^4$ (and possibly  unveiling some non-minimal curvature couplings which
are exclusive of the theory on general four-manifolds), it enables us to search
for vanishing theorems that can be used to constrain the space on which the
path integral localizes when passing to the weak coupling limit. After
integrating out the auxiliary fields in \Phoebe\ we find for the bosonic part
of the action not involving the scalars $\phi$ and $\bar\phi$ the following
expression:
$$
\eqalign{  \int_X  d^4 x\,\sqrt{g}\, &\tr
\bigl\{\,\half
\bigl (\,\deriv_{\alpha\dalpha}C+i\deriv_{\beta\dot\alpha}
B^{\beta}{}_{\!\alpha}\,\bigr )^2\cr 
&+{1\over4} \bigl (\, F^{+}_{\alpha\beta}+
[B_{\gamma\alpha},B_\beta{}^{\!\gamma}]+2i\,[B_{\alpha\beta},C]\,\bigr
)^2\,\bigr\}.\cr}
\eqn\one
$$
Expanding the squares in this expression  one obtains,
$$
\eqalign{  \int_X  &d^4 x\,\sqrt{g}\, \tr
\bigl\{\,-{1\over2}
\deriv_\mu C\deriv^\mu C
-\half\,(\,\deriv_{\beta\dot\alpha}B^{\beta}{}_{\!\alpha}
\deriv_{\gamma}{}^{\!\dalpha}B^{\gamma\alpha}-
F^{+\alpha\beta}[B_{\gamma\alpha},B_\beta{}^{\!\gamma}]\,)\cr 
 &-{1\over2}F^{+}_{\mu\nu}F^{+\mu\nu}+
[B^{+}_{\mu\nu},B^{+}_{\tau\lambda}][B^{+\mu\nu},B^{+\tau\lambda}]+
2[B^{+}_{\mu\nu},C][B^{+\mu\nu},C]\,\bigr\}.\cr }
\eqn\two
$$
where we have used $\deriv_{\alpha\dalpha}=\sigma^m{}_{\alpha\dalpha}\deriv_m$
and $F^{+}_{\alpha\beta}\equiv\sigma^{\mu\nu}_{\alpha\beta}F^{+}_{\mu\nu}$, 
$B_{\alpha\beta}\equiv\sigma^{\mu\nu}_{\alpha\beta}B^{+}_{\mu\nu}$. 
Let us now focus on the expression inside the parenthesis. Further
expansion leads to the identity:
$$
\eqalign{
&\int_X  d^4 x\,\sqrt{g}\,\tr\,\bigl\{\,-\half
\deriv_{\beta\dalpha}B^\beta{}_{\!\alpha}\deriv_
{\gamma}{}^{\!\dalpha}B^{\gamma\alpha}
+ \half F^{+\alpha\beta}[B_{\gamma\alpha},B_\beta{}^{\!\gamma}\,]\,\bigr\}=
\cr  
&\int_X  d^4 x\,\sqrt{g}\,\tr\,\bigl\{\,-\deriv_\mu
B^{+}_{\nu\lambda}\deriv^\mu B^{+\nu\lambda}-{\half}R
B^{+}_{\mu\nu}B^{+\mu\nu}+R_{\mu\nu\tau\lambda}
B^{+\mu\nu}B^{+\tau\lambda}\,\bigr\},\cr
}
\eqn\Beatrice
$$  
(using again  
$B_{\alpha\beta}\equiv\sigma^{\mu\nu}_{\alpha\beta}B^{+}_{\mu\nu}$). 
If we now express the Riemann tensor in \Beatrice\ in terms of its
irreducible components 
\REF\weinb{S. Weinberg, {\it Gravitation and Cosmology}, John Wiley,
1972.} [\weinb], 
$$
\eqalign{ R_{\mu\nu\tau\lambda}&=\half(g_{\mu\tau}R_{\nu\lambda}-
g_{\mu\lambda}R_{\nu\tau}-g_{\nu\tau}R_{\mu\lambda}+
g_{\nu\lambda}R_{\mu\tau})\cr
&-{R\over6}(g_{\mu\tau}g_{\nu\lambda}-g_{\nu\tau} g_{\mu\lambda})+
C_{\mu\nu\tau\lambda},\cr
}
\eqn\Riemann
$$ 
with $C_{\mu\nu\tau\lambda}$ the Weyl tensor, we finally obtain,
$$
\eqalign{
&\int_X  d^4 x\,\sqrt{g}\,\tr\,\bigl\{\,-\half
\deriv_{\beta\dalpha}B^\beta{}_{\!\alpha}\deriv_
{\gamma}{}^{\!\dalpha}B^{\gamma\alpha}
+ \half F^{+\alpha\beta}[B_{\gamma\alpha},B_\beta{}^{\!\gamma}\,]\,\bigr\}=
\cr  
&\int_X  d^4 x\,\sqrt{g}\,\tr\,\bigl\{\,-\deriv_\mu
B^{+}_{\nu\lambda}\deriv^\mu B^{+\nu\lambda}
-B^{+\mu\nu}\bigl (\,{1\over6}R\,(g_{\mu\tau}g_{\nu\lambda}-g_{\nu\tau}
g_{\mu\lambda})-C_{\mu\nu\tau\lambda}\,\bigr) B^{+\tau\lambda}\,\bigr\}.
\cr
}
\eqn\Turandot
$$  
Notice that all the terms in \one\ are negative definite 
except the terms contained in this
last equation which involve the scalar curvature and the Weyl tensor.

The associated fermionic symmetry splits up as well into BRST
($Q^{+}\equiv Q^1$) and  anti-BRST ($Q^{-}\equiv iQ^2$) parts. The
explicit formulas are:
$$
\eqalign{
[Q^{+}, A_{\alpha\dalpha}] &= -2\psi_{\alpha\dalpha},\cr 
\{Q^{+},\psi_{\alpha\dot\alpha}\} &= 
-{\raiz}\deriv_{\alpha\dot\alpha}\phi, \cr
[Q^{+},\phi]&=0,\cr
[Q^{+}, B_{\alpha\beta}]&=\raiz\tilde\psi_{\alpha\beta},\cr
\{Q^{+},\tilde\psi_{\alpha\beta}\}&=2i\,[B_{\alpha\beta},\phi],\cr
[Q^{+},C]&={1\over\raiz}\zeta,\cr
\{Q^{+},\zeta\,\} &=4i\,[C,\phi],\cr
[Q^{+},\bar\phi]&=\raiz\,\eta,\cr
\{Q^{+},\eta\,\} &=2i\,[\bar\phi,\phi],\cr
\{Q^{+},\tilde\chi_{\alpha\dalpha}\} &= 
\tilde H'_{\alpha\dalpha},\cr
[Q^{+},\tilde H'_{\alpha\dalpha}]&=
2\raiz i\,[\tilde\chi_{\alpha\dalpha},\phi],\cr\cr
\{Q^{+},\chi_{\alpha\beta}\} &= H'_{\alpha\beta},\cr
[Q^{+},H'_{\alpha\beta}]&=2\raiz i\,[\chi_{\alpha\beta},\phi],\cr 
\cr
}
\qquad\qquad
\eqalign{
[Q^{-}, A_{\alpha\dalpha}] &= -2\tilde\chi_{\alpha\dalpha},\cr 
\{Q^{-},\tilde\chi_{\alpha\dalpha}\} &= 
{\raiz}\deriv_{\alpha\dot\alpha}\bar\phi, \cr
[Q^{-},\bar\phi\,]&=0,\cr
[Q^{-}, B_{\alpha\beta}]&=-\raiz\,\chi_{\alpha\beta},\cr
\{Q^{-},\chi_{\alpha\beta}\}&=2i\,[B_{\alpha\beta},\bar\phi],\cr
[Q^{-},C]&=-{1\over\raiz}\eta,\cr
\{Q^{-},\eta\,\} &=4i\,[C,\bar\phi],\cr
[Q^{-},\phi]&=\raiz\,\zeta,\cr
\{Q^{-},\zeta\,\} &=-2i\,[\phi,\bar\phi],\cr
\{Q^{-},\psi_{\alpha\dalpha}\} &= 
-\tilde H'_{\alpha\dalpha}+2\raiz\,\deriv_{\alpha\dalpha}C,\cr
[Q^{-},\tilde H'_{\alpha\dalpha}]&=-2\,\deriv_{\alpha\dalpha}\eta +
2\raiz i\,[\psi_{\alpha\dalpha},\bar\phi\,]\cr&- 
4\raiz i\,[\,\tilde\chi_{\alpha\dalpha},C],\cr
\{Q^{-},\tilde\psi_{\alpha\beta}\} &= H'_{\alpha\beta}
-4i\,[B_{\alpha\beta},C\,],\cr
[Q^{-},H'_{\alpha\beta}]&=-2\raiz i\,[\tilde\psi_{\alpha\beta},
\bar\phi]-4\raiz i\,[\chi_{\alpha\beta},C]\cr&-2\raiz i\,
[B_{\alpha\beta},\eta],\cr 
}
\eqn\papagayo
$$  
satisfying the algebra,
$$
\eqalign{
\{Q^{+},Q^{+}\}&=\delta_g(\phi),\cr
\{Q^{-},Q^{-}\}&=\delta_g(-\bar\phi),\cr
\{Q^{+},Q^{-}\}&=\delta_g(C)
.\cr}
\eqn\leoon
$$

The $\tau$-independent part of the action \Phoebe\ can be written either  as a
BRST  ($Q^{+}$) commutator or as an anti-BRST ($Q^{-}$) commutator. Let us 
 focus on the former possibility. The appropriate ``gauge" fermion turns out
to be:
$$
\eqalign{
\Psi&=
{1\over e^2}\int_X  d^4 x\,\sqrt{g}\, \tr
\bigl\{\,-{1\over4}
\tilde \chi^{\dalpha\alpha}\bigl (\,\tilde H'_{\alpha\dalpha}-2\raiz\deriv
_{\alpha\dalpha}C-2\raiz i\,
\deriv_{\beta\dot\alpha}B^{\beta}{}_{\!\alpha}\,\bigr )\cr 
&-{1\over 4} \chi^{\alpha\beta}\bigl (\, H'_{\alpha\beta}-2\,
F^{+}_{\alpha\beta}-2\,[B_{\gamma\alpha},B_\beta{}^{\!\gamma}]-4i\,
[B_{\alpha\beta},C]\,\bigr
)\,\bigr\}\cr 
&+{1\over e^2}\int_X  d^4 x\,\sqrt{g}\, \tr
\bigl\{\,
{1\over{2\raiz}}\bar\phi\,\bigl
(\,\deriv_{\alpha\dalpha}\psi^{\dot\alpha\alpha}+i\raiz
\,[\tilde\psi_{\alpha\beta},B^{\alpha\beta}]
-i\raiz\,[\zeta,C]\,\bigr )\,\bigr\}\cr
&-{1\over e^2}\int_X  d^4 x\,\sqrt{g}\, \tr
\bigl\{\,{i\over4}\eta[\phi,\bar\phi]\,\bigr\}.\cr
}
\eqn\Mazinger
$$

For reasons of future convenience we will rewrite \Mazinger\ in vector 
indices. With the definitions, $X_{\alpha\dalpha}\buildrel {\sevenrm 
def}\over = \sigma^\mu_{\alpha\dalpha}X_\mu$, and, 
$Y_{\alpha\beta}\buildrel {\sevenrm
def}\over = \sigma^{\mu\nu}_{\alpha\beta}Y_{\mu\nu}$,
for any two given fields $X$ and $Y$, \Mazinger\ takes the form:
$$
\eqalign{
\Psi&=
{1\over e^2}\int_X  d^4 x\,\sqrt{g}\, \tr
\bigl\{\,\half
\tilde \chi^\mu\bigl (\,\tilde H'_\mu -2\raiz\deriv_\mu 
C+4\raiz\,
\deriv^\nu B^{+}_{\nu\mu}\,\bigr )\cr
&+\half\chi^{+\mu\nu}\bigl (\, H'^{+}_{\mu\nu}-2\,
F^{+}_{\mu\nu}-4i\,[B^{+}_{\mu\tau},B^{+\tau}{}_{\!\nu}]-4i
[B^{+}_{\mu\nu},C]\,\bigr
)\,\bigr\}\cr
&-{1\over e^2}\int_X  d^4 x\,\sqrt{g}\, \tr
\bigl\{\,
{1\over{2\raiz}}\bar\phi\,\bigl
(\,2\deriv_\mu\psi^\mu+2\raiz i
\,[\tilde\psi^+_{\mu\nu},B^{+\mu\nu}]
+\raiz i\,[\zeta,C]\,\bigr )\,\bigr\}\cr
&-{1\over e^2}\int_X  d^4 x\,\sqrt{g}\, \tr
\bigl\{\,{i\over4}\eta[\phi,\bar\phi]\,\bigr\}.\cr
}
\eqn\mermelada
$$

The gauge fermion, in turn, can itself be written as an anti-BRST
commutator \jackie :
$$
\eqalign{
\Psi=\bigr\{Q^{-},{1\over e^2}\int_X  d^4 x\,\sqrt{g}\, \tr\bigl (\, 
&-{1\over{2\raiz}}B^{\alpha\beta}\,\bigl (\,F^{+}_{\alpha\beta}-{1\over2}\,
H'_{\alpha\beta}+{1\over3}[\,B_{\alpha\gamma},B_{\beta}{}^{\!\gamma}\,]\,
\bigr )\cr &+{i\over{2\raiz}}C\,[\phi,\bar\phi\,]+{1\over4}\,
\psi_{\alpha\dalpha}\tilde\chi^{\dalpha\alpha}\,\bigr )\,\bigr\}.\cr
}
\eqn\Misha
$$

%%%%%%%%%%%%%%%%%%%%%%%%%%%%%%%%%  %%%%%%%%%%%%%%%%%%%%%%%%%%%%%%%%%%%%%%%%
%%ADJOINT%NON-ABELIAN%MONOPOLES%%
%%%%%%%%%%%%%%%%%%%%%%%%%%%%%%%%%  %%%%%%%%%%%%%%%%%%%%%%%%%%%%%%%%%%%%%%%%

\vskip 0.5cm

\subsection{(2) ${\bf 4}\too 
({\bf 2},{\bf 1})\oplus ({\bf 1},{\bf 1})\oplus ({\bf1},{\bf 1})$
Adjoint Non-Abelian Monopoles (Half-Twisted Theory)}

As explained in [\yamron], this amounts to a breaking of the $SU(4)_I$ isospin
group down to a subgroup $SU(2)_A\otimes SU(2)_F\otimes U(1)$ and then
a replacement of the $SU(2)_L$ factor of the rotation group by the
diagonal sum $SU(2)'_L$ of $SU(2)_L$ and $SU(2)_A$. The subgroup
$SU(2)_F\otimes U(1)$ remains in the theory as an internal symmetry group.
Hence, we observe that, as a by-product of the twisting procedure, it remains
in the theory a $U(1)$ symmetry which was not present in the
original
$N=4$ theory, and which becomes, as we shall see in a moment, the ghost
number symmetry associated to the topological theory. With respect to the
new symmetry group
${\cal H'} =SU(2)'_L\otimes SU(2)_R\otimes SU(2)_F\otimes U(1)$ the
supercharges $Q^v{}_{\!\alpha}$ split up into three supercharges: 
$Q_{(\beta\alpha)}\oplus Q\oplus Q^{i}{}_{\!\alpha}$, where the index $i$
labels the representation ${\bf 2}$ of $SU(2)_F$. In more
detail,
$$
Q^{v=1,2,3,4}_\alpha \to\cases{Q^{v=1,2}_\alpha \to
Q^{\beta}_{\!\alpha}\to\cases{Q=-Q^\alpha{}_{\!\alpha}\equiv
-Q^{v=2}{}_{\!2}-Q^{v=1}{}_{\!1},\cr
Q_{(\beta\alpha)}=-C_{\gamma(\beta}Q^\beta{}_{\alpha)},\cr}
\cr Q^{v=3}_{\!\alpha} \to Q^{i= 1}_\alpha,
\cr Q^{v=4}_{\!\alpha} \to Q^{i= 2}_\alpha.\cr}
\eqn\nuria
$$
The conjugate supercharges 
$\bar Q_{v\dot\alpha}$ split up accordingly into a vector isosinglet and a
right-handed spinor isodoublet supercharge, $\bar Q_{\alpha\dot\alpha}
\otimes \bar Q_{i\dot\alpha}$. 

The fields of the $N=4$ multiplet give rise, after the twisting, to the 
following topological multiplet (in the notation of reference [\yamron]):
$$
\eqalign{
&A_{\alpha\dalpha} \too A^{(0)}_{\!\alpha\dalpha},\cr
&\lambda_{v\alpha}\too
\chi^{(-1)}_{\beta\alpha},~
\eta^{(-1)},~\lambda^{(+1)}_{i\alpha},\cr
&\bar\lambda^v{}_{\dot\alpha} \too
\psi^{(+1)}_{\!\alpha\dot\alpha},~\zeta^{(-1)}_{i\dot\alpha},\cr
&\phi_{uv}\too B^{(-2)},~C^{(+2)},~G^{(0)}_{\! i\alpha},\cr}
\eqn\cien
$$
where  we have indicated the ghost number  carried by the
fields after the twisting by a superscript. Some of the definitions in \cien\
need clarification. Our choices for the anticommuting fields are:
$$
\eqalign{
&\lambda_{v\alpha}= \cases{\lambda_{(v=1,2)\alpha}\to
\lambda_{\beta\alpha}\to\cases{\chi_{
\beta\alpha}=\lambda_{(\beta\alpha)},\cr
\eta=2\lambda_{[\beta\alpha]},\cr}\cr
\lambda_{(v=3,4)\alpha}\to
\lambda_{i\alpha},\cr}\cr
&\bar\lambda^v{}_{\dot\alpha} =
\cases{\bar\lambda^{v=1,2}{}_{\!\dot\alpha}\to
\psi^{\alpha}{}_{\!\dot\alpha},\cr
\bar\lambda^{v=3,4}{}_{\dot\alpha}\to
\zeta^{i}{}
_{\!\dot\alpha},\cr}\cr
}
\eqn\ccuno
$$
whereas for the commuting ones:
$$
\eqalign{B&=\phi_{12},\cr
G_{i\alpha}&=\cases{G_{(i=1)1}=\phi_{13},\cr
G_{(i=1)2}=\phi_{23},\cr}\cr}\qquad\qquad
\eqalign{
C&=\phi_{34},\cr
G_{(i=2)1}&=
\phi_{14},\cr
G_{(i=2)2}&=\phi_{24}.\cr}
\eqn\cctres
$$

In terms of the twisted fields, the action for the theory (on flat $\Erre
^4$) takes the form:
$$
\eqalign{ {\cal S}^{(0)}&= {1\over e^2}\int d^4 x\, \tr\,
\bigl\{\,\half\nabla_{\!\alpha\dalpha} B\nabla^{\dalpha\alpha}C
-{1\over4}\nabla_{\!\alpha\dalpha} G_{i\beta}\nabla^{\dalpha\alpha}
G^{i\beta}  -i\psi^{\beta}{}_{\dot\alpha}\nabla^{\dot\alpha\alpha}
\chi_{\alpha\beta}\cr& 
-{i\over 2}\psi_{\alpha\dalpha}\nabla^{\dalpha\alpha}
\eta
-i\zeta^i{}_{\dot\alpha}\nabla^{\dot\alpha\alpha}\lambda_{i\alpha}-
{1\over4}F_{mn} F^{mn}-
{i\over\raiz}\,\chi^{\alpha\beta}[\chi_{\alpha\beta},C]
\cr&-{i\over\raiz}\,\lambda^{i\alpha}[\lambda_{i\alpha},B] 
+i\raiz\,\chi^{\alpha\beta}[\lambda_{i\alpha},G^i{}_\beta]
+{i\over\raiz}\,\eta[\lambda_i{}^\alpha,G^i{}_\alpha]
-{i\over{2\raiz}}\,\eta[\eta,C]
\cr&+{i\over\raiz}\,\psi_{\alpha\dalpha}[\psi^{\dalpha\alpha},B]
+i\raiz\,\psi^\alpha{}_{\dalpha}[\zeta^{i\dalpha},G_{i\alpha}]+
{i\over\raiz}\,\zeta_{i\dot\alpha}[\zeta^{i\dot\alpha},C]-
\half[B,C]^2\cr& 
-[B,G_{i\alpha}][C,G^{i\alpha}] 
+{1\over 4}[G_{i\alpha},G_{j\beta}]
[G^{i\alpha},G^{j\beta}]\,\bigr\}
-{i\theta\over 32\pi^2}\int d^4 x\,\tr\,\bigl\{\, *  F_{mn}F^{mn}
\,\bigr\}
.
\cr
}
\eqn\cccuatro
$$ 

To obtain the corresponding topological symmetry we proceed as follows.
First of all, we recall that the $N=4$ supersymmetry transformations
\Vian\ are generated by $\xi_v{}^\alpha Q^v{}_\alpha + 
\bar\xi^v{}_\dalpha \bar Q_v{}^\dalpha$. According to our conventions, to
obtain the $Q$-transformations we must set $\bar\xi^v{}_\dalpha=0$ and
make the replacement:  
$$
\xi_{v\alpha}=\cases{\xi_{(v=1,2)\alpha}\to\epsilon C_{\beta\alpha},\cr
\xi_{(v=3,4)\alpha}\to 0.\cr}
\eqn\ndos
$$
The resulting transformations turn out to be:
$$
\eqalign{
&\delta A_{\alpha\dalpha} =
2i\epsilon\psi_{\alpha\dalpha},\cr
&\delta\psi_{\alpha\dot\alpha} = 
-i{\raiz}\epsilon\nabla_{\alpha\dot\alpha}C,\cr
&\delta C=0,\cr
&\delta\chi_{\alpha\beta} = 
-i\epsilon F^{+}_{\alpha\beta}
-i\epsilon[G_{i\alpha},G^i{}_{\beta}],\cr 
&\delta\zeta^j{}_{\dot\alpha}=-i{\raiz}\epsilon\nabla_{\alpha\dot\alpha}
G^{j\alpha},\cr}\qquad\qquad
\eqalign{
&\delta G_{j\alpha}=-\raiz\epsilon\lambda_{j\alpha},\cr
&\delta\lambda_{j\alpha}=-2i\epsilon [G_{j\alpha},C],\cr
&\delta F^{+}_{\alpha\beta}=2\epsilon\nabla_{(\alpha}{}^{\dot\alpha}
\psi{}_{\beta)\dot\alpha},\cr
&\delta B=\raiz\epsilon\eta,\cr
&\delta\eta =2i\epsilon[B,C].\cr 
}
\eqn\cccinco
$$
The BRST generator $Q$ associated to the transformations \cccinco\
satisfies the on-shell algebra 
$\{Q,Q\}=\delta_g(C)$
where by $\delta_g(C)$ we mean a non-abelian gauge transformation
generated by $C$.  
It is possible to realize this algebra off-shell, \ie, without the input
of the equations of motion for some of the fields in the theory. A minimal
off-shell formulation can be constructed by introducing in the theory the
auxiliary fields $N_{\alpha\beta}$ (symmetric in its spinor indices) and
$P^i{}_{\!\alpha}$ (both with ghost number $0$).  The off-shell BRST
transformations which correspond to the enlarged topological multiplet can be
cast in the form:
$$
\eqalign{
&[Q, A_{\alpha\dalpha}] =
2i\psi_{\alpha\dalpha},\cr
&[Q, F^{+}_{\alpha\beta}]=2\nabla_{(\alpha}{}^{\dot\alpha}
\psi{}_{\beta)\dot\alpha},\cr
&[Q, G_{j\alpha}]=-\raiz\lambda_{j\alpha},\cr
&\{Q,\chi_{\alpha\beta}\} =  N_{\alpha\beta},\cr 
&\{Q,\zeta^j{}_{\dot\alpha}\}= P^j{}_{\!\dalpha}
,\cr
&[Q, B]=\raiz\eta,\cr
}\qquad\qquad
\eqalign{
&\{Q,\psi_{\alpha\dot\alpha}\} = 
-i{\raiz}\nabla_{\alpha\dot\alpha}C,\cr
&[Q, C]=0,\cr
&\{Q,\lambda_{j\alpha}\}=-2i [G_{j\alpha},C],\cr
&[Q,N_{\alpha\beta}]=2\raiz i\,[\chi_{\alpha\beta},C],\cr
&[Q,P^i{}_{\!\dalpha}]=2\raiz i\,[\zeta^i{}_{\!\dalpha},C],\cr
&\{Q,\eta\,\} =2i[B,C].\cr  }
\eqn\Medeiros
$$

After some
suitable manipulations [\tesina], the off-shell action which corresponds
to the topopological symmetry \Medeiros\ is:
$$
\eqalign{ 
{\cal S}^{(1)}&={1\over e^2}\int d^4 x\, \tr\,
\bigl\{\,\half\nabla_{\!\alpha\dalpha} B\nabla^{\dalpha\alpha}C
+{1\over4}P_i{}^{\dalpha}\,\bigl (\, P^i{}_{\!\dalpha}+2\raiz
i\,\nabla_{\!\alpha
\dalpha}G^{i\alpha}\,\bigr ) 
-i\psi^{\beta}{}_{\dot\alpha}\nabla^{\dot\alpha\alpha}
\chi_{\alpha\beta}\cr& 
-{i\over 2}\psi_{\alpha\dot\alpha}\nabla^{\dot\alpha\alpha}
\eta
-i\zeta^j{}_{\dot\alpha}\nabla^{\dot\alpha\alpha}\lambda_{j\alpha}+
{1\over4}N_{\alpha\beta}\,\bigl (\, N^{\alpha\beta}+2i
F^{+\alpha\beta} +2i[G_i{}^{\alpha},G^{i\beta}]\,\bigr )\cr &-
{i\over\raiz}\,\chi^{\alpha\beta}[\chi_{\alpha\beta},C]
-{i\over\raiz}\,\lambda^{i\alpha}[\lambda_{i\alpha},B]+ 
i\raiz\,\chi^{\alpha\beta}[\lambda_{i\alpha},G^i{}_\beta]+
{i\over\raiz}\,\zeta_{i\dot\alpha}[\zeta^{i\dot\alpha},C]\cr 
&+{i\over\raiz}\,\eta[\lambda_i{}^\alpha,G^i{}_\alpha]
-{i\over{2\raiz}}\,\eta[\eta,C]
+{i\over\raiz}\,\psi_{\alpha\dot\alpha}[\psi^{\dalpha\alpha},B]+
i\raiz\,\psi^\alpha{}_{\dot\alpha}[\zeta^{i\dot\alpha},G_{i\alpha}]\cr 
&-\half[B,C]^2  -[B,G_{i\alpha}][C,G^{i\alpha}]\,\bigr\}
-2\pi i\tau{1\over 32\pi^2}\int d^4 x\,\tr\,\bigl\{\, *  F_{mn}F^{mn}
\,\bigr\}.
\cr
}
\eqn\cinza
$$ 
The $\tau$-independent part of the 
topological action above is, as it could be expected, BRST-exact, that
is, it can be written as $\{Q,\Psi\,\}$. The  appropriate gauge
fermion is easily seen to be: 
$$
\eqalign
{\Psi=&{1\over e^2}\int d^4 x\,\tr\,\bigl\{\,\, 
{1\over4}\zeta_i{}^{\!\dot\alpha}\,\bigl (\,P^i{}_{\!\dot\alpha}+
2\raiz i\,\nabla_{\!\alpha\dalpha}G^{i\alpha}\,\bigr )\cr
&+{1\over4}\chi_{\alpha\beta}\,\bigl (\, N^{\alpha\beta}+
2iF^{+\alpha\beta}
+2i[G_{i}{}^{\alpha},G^{i\beta}]\,\bigr )\,\bigr \}\cr
&-{1\over e^2}\int d^4 x\,\tr\,\bigl\{\,\,
{i\over{2\raiz}}\,B\,\bigl (\,
\nabla_{\!\alpha\dalpha}\psi^{\dalpha\alpha}
-\raiz\,[G_{i\alpha},\lambda^{i\alpha}]\,\bigr )\,\bigr \}\cr
&-{1\over e^2}\int d^4 x\,\tr\,\bigl\{\,\,
{i\over 4}\,B[\eta,C]\,\bigr\}.\cr
}
\eqn\cerne
$$ 

The next  step will consist of the coupling the theory to an arbitrary
background metric $g_{\mu\nu}$ of Euclidean signature. To achieve this
goal we make use of the covariantized version of the topological symmetry 
\Medeiros\ (which is trivial to obtain), and of the gauge fermion $\Psi$, 
and then define the topological action to be ${\cal
S}^{(1)}_{c}=\{Q,\Psi\,\}_{\hbox{\sevenrm cov}}-2 \pi i k \tau$. 
The resulting action is:
$$
\eqalign{ 
{\cal S}^{(1)}_{c}&={1\over e^2} \int_X  d^4 x\,\sqrt{g}\,\tr\,
\bigl\{\,\half\deriv_{\alpha\dalpha} B\deriv^{\dalpha\alpha}C
+{1\over4}P_i{}^{\dalpha}\,\bigl (\, P^i{}_{\!\dalpha}+2\raiz i\,\deriv_
{\alpha\dot\alpha}G^{i\alpha}\,\bigr ) 
-i\psi^{\beta}{}_{\!\dot\alpha}\deriv^{\dot\alpha\alpha}
\chi_{\alpha\beta}\cr& 
-{i\over 2}\psi_{\alpha\dot\alpha}\deriv^{\dot\alpha\alpha}
\eta
-i\zeta^j{}_{\dot\alpha}\deriv^{\dot\alpha\alpha}\lambda_{j\alpha}+
{1\over4}N_{\alpha\beta}\,\bigl (\, N^{\alpha\beta}+2i
F^{+\alpha\beta} +2i[G_i{}^{\!\alpha},G^{i\beta}]\,\bigr )\cr &-
{i\over\raiz}\,\chi^{\alpha\beta}[\chi_{\alpha\beta},C]
-{i\over\raiz}\,\lambda^{i\alpha}[\lambda_{i\alpha},B]+ 
i\raiz\,\chi^{\alpha\beta}[\lambda_{i\alpha},G^i{}_\beta]+
{i\over\raiz}\,\zeta_{i\dot\alpha}[\zeta^{i\dot\alpha},C]\cr 
&+{i\over\raiz}\,\eta[\lambda_i{}^\alpha,G^i{}_\alpha]
-{i\over{2\raiz}}\,\eta[\eta,C]
+{i\over\raiz}\,\psi_{\alpha\dot\alpha}[\psi^{\dalpha\alpha},B]+
i\raiz\,\psi^\alpha{}_{\dot\alpha}[\zeta^{i\dot\alpha},G_{i\alpha}]\cr 
&-\half[B,C]^2  -[B,G_{i\alpha}][C,G^{i\alpha}]\,\bigr\}
-2\pi i\tau{1\over 32\pi^2}\int_X  d^4 x\,\tr\,\bigl\{\, *  F_{\mu\nu}
F^{\mu\nu}
\,\bigr\}
.
\cr
}
\eqn\covcinza
$$ 

Up to now we have carefully  studied the ``standard" formulation of the second 
twist, and we have been able to reproduce faithfully previously known 
results [\yamron]. However, we think there are several subtleties that demand 
clarification. Since the twisted theory contains several spinor fields taking 
values in the fundamental representation of the internal $SU(2)_F$ symmetry
group, and these fields are necessarily complex, as they live in complex
representations of the rotation group and of the isospin group, it can be 
seen that the action \covcinza\ is not real. Moreover, there are more fields 
in the twisted theory than in the physical theory. To see this, pick for
example the scalar fields $\phi_{uv}$ in the physical $N=4$ theory. They
are $6$ real fields that after the twisting become the  
scalar fields $B$ and $C$ (which can be safely taken to be real, thus making a
total of $2$ real fields) and  the isospin doublet bosonic spinor field 
$G_{i\alpha}$, which is necessarily complex and thus is built out of $2\times2
\times2=8$ real fields. Thus we see that $6$ real fields in the $N=4$ theory 
give rise to $10$ real fields in the twisted theory. With the anticommuting
fields this overcounting is even worse.  In what follows we will break
$SU(2)_F$  explicitly and rearrange the resulting fields wisely so as to avoid  
these problems. The outcome of this reformulation is that we will 
make contact with the non-abelian monopole theory formulated in 
[\marmon,\marpol,\zzeta]. For a thoroughful and self-contained review of these 
theories  see [\marth].  

We start with the fields $G_{i\alpha}$, which we rearrange in a complex 
commuting 
two-component Weyl spinor $M_{\alpha}\equiv G_{2\alpha}$ and its complex 
conjugate 
$\overline M^{\alpha}\equiv G_{1}{}^{\alpha}$. The constraint $G_{1}{}^{\alpha}
=(G_{2\alpha})^*$  
looks rather natural when considered from the viewpoint of the physical 
$N=4$ theory, 
in terms of which --recall eqn. \Inclan\ and \cctres -- 
$$
G_{1\alpha}=\pmatrix{\phi_{13}\cr\phi_{23}\cr}=\pmatrix{B^{\dag 2}
\cr-B^{\dag 1}\cr},
\qquad 
G_{2\alpha}=\pmatrix{\phi_{14}\cr
\phi_{24}\cr}=\pmatrix{-B_{1}\cr-B^{2}\cr}.
\eqn\sienna
$$
Similarly, for the other isodoublets in the theory we make the rearrangements:
$$
\eqalign{
\lambda_{1\alpha}&=\bar\mu_\alpha,\cr
\lambda_{2\alpha}&=\mu_\alpha,\cr
P_{1\dalpha}&=\bar h_{\dalpha},\cr}
\qquad\qquad
\eqalign{
\zeta_{1\dalpha}&=\bar\nu_\dalpha,\cr
\zeta_{2\alpha}&=\nu_\dalpha,\cr
P_{2\dalpha}&=h_{\dalpha}.\cr}
\eqn\bologna
$$
Finally, after redefining $\psi\to -i\psi$, $C\to\phi$, $B\to\lambda$ and 
$N_{\alpha\beta}\to H_{\alpha\beta}$ ($A$ and $\eta$ remain the same),
the action \covcinza\ becomes:
$$
\eqalign{ 
{\cal S}^{(1)}_c&={1\over e^2}\int_X  d^4 x\,\sqrt{g}\,\tr\,
\bigl\{\,\half\deriv_{\alpha\dalpha}
\phi\deriv^{\dalpha\alpha}\lambda +{1\over4}\bar h^{\dalpha}\bigl (
h_{\dalpha}+2\raiz i\,\deriv_{\alpha\dalpha}M^{\alpha}\bigr )\cr& 
-{1\over4}h^{\dalpha}\bigl (\bar h_{\dalpha}+2\raiz i\,\deriv_{\alpha
\dalpha}\overline M^{\alpha}\bigr )- 
\psi^{\beta}{}_{\!\dalpha}\deriv^{\dalpha\alpha}
\chi_{\alpha\beta} 
-{1\over 2}\psi_{\alpha\dalpha}\deriv^{\dalpha\alpha}
\eta\cr&
-i\nu_{\dalpha}\deriv^{\dalpha\alpha}\bar\mu_{\alpha}+
i\bar\nu_{\dalpha}\deriv^{\dalpha\alpha}\mu_{\alpha}+
{1\over4}H_{\alpha\beta}\bigl ( H^{\alpha\beta}+2i
F^{+\alpha\beta} +4i[\overline M^{(\alpha},M^{\beta)}]\bigr )\cr&-
{i\over\raiz}\,\chi^{\alpha\beta}[\chi_{\alpha\beta},\phi]
+i\raiz\,\bar\mu^{\alpha}[\mu_{\alpha},\lambda]\cr&+ 
 i\raiz\,\chi^{\alpha\beta}[\bar\mu_{\alpha},M_\beta]-
 i\raiz\,\chi^{\alpha\beta}[\mu_{\alpha},\overline M_\beta]+
i\raiz\,\bar\nu_{\dalpha}[\nu^{\dalpha},\phi]\cr 
&-{i\over\raiz}\,\eta[\bar\mu_\alpha,M^\alpha]
+{i\over\raiz}\,\eta[\mu_\alpha,\overline M^\alpha]
-{i\over{2\raiz}}\,\eta[\eta,\phi]
-{i\over\raiz}\,\psi_{\alpha\dalpha}[\psi^{\dalpha\alpha},\lambda]\cr&+
\raiz\,\psi^\alpha{}_{\!\dalpha}[\nu^{\dalpha},\overline M_{\alpha}]-
\raiz\,\psi^\alpha{}_{\!\dalpha}[\bar\nu^{\dalpha},M_{\alpha}]- 
\half[\phi,\lambda]^2  \cr&-
[\lambda,\overline M_{\alpha}][\phi,M^{\alpha}]+
[\lambda,M_{\alpha}][\phi,\overline M^{\alpha}]
\,\bigr\}
-2\pi i\tau{1\over 32\pi^2}\int_X  d^4 x\,\tr\,\sqrt{g}\bigl\{\, *  
F_{\mu\nu}F^{\mu\nu}
\,\bigr\}.
\cr
}
\eqn\coventry
$$

Let us focus now on the bosonic part of the action not containing the
scalar fields $\phi$ and $\lambda$. After integrating out the auxiliary fields,
this  part reads:
$$
\int_X  d^4 x\,\sqrt{g}\,\tr\, \bigl\{\,-\deriv_{\alpha
\dalpha}\overline M^{\alpha}\deriv_{\beta}{}^{\dalpha}M^{\beta}
+{1\over4} 
\,(\,F^{+\alpha\beta} +2[\overline M_{(\alpha},M_{\beta)}]\,)^2\,\bigr\}.
\eqn\Assur
$$
Expanding the squares we obtain the contributions:
$$
\eqalign{
\int_X  d^4 x\,\sqrt{g}\,\tr\,&\bigl\{-g^{\mu\nu}\deriv_{\mu}
\overline M^{\alpha}\deriv_{\nu}M_{\alpha}
-{1\over4}R\,\overline M^{\alpha}M_{\alpha}-{1\over2}
F^{+}_{\mu\nu}F^{+\mu\nu}\cr
&+[\overline M_{(\alpha},M_{\beta)}]
[\overline M^{(\alpha},M^{\beta)}]\bigr\}.
\cr}
\eqn\Dorma
$$ 
%In the derivation of \Dorma\ we have used the identity
%$$
%\tr\,\bigl\{\,[G_{i\alpha},G^i{}_{\!\beta}][G_j{}^{\!\alpha},
%G^{j\beta}]\,\bigr\}=\tr\,\bigl\{\,
%[G_{i\alpha},G_{j\beta}]
%[G^{i\alpha},G^{j\beta}]\,\},
%$$
In the derivation of \Dorma\ we have used the Weitzenb\"ock formula,
$$
\deriv_{\alpha\dalpha}\deriv^{\dalpha\beta}=\half\delta_\alpha{}^{\!\beta}
\,\deriv_{\gamma\dalpha}\deriv^{\gamma\dalpha}+{1\over4}\delta_\alpha{}
^{\!\beta}\,R +F_{\;\alpha}^{+a\beta}T^a
\eqn\weitzenbock
$$
being $R$ the scalar curvature and 
$T^a$, $a=1,\dots,$dim$(G)$, the generators of the 
gauge group in the appropriate
representation.

The corresponding BRST symmetry is readily obtained from \Medeiros :
$$
\eqalign{
&[Q, A_{\alpha\dalpha}] =
2\psi_{\alpha\dalpha},\cr
&[Q, F^{+}_{\alpha\beta}]=2\deriv_{(\alpha}{}^{\dot\alpha}
\psi{}_{\beta)\dot\alpha},\cr
&[Q, M_{\alpha}]=-\raiz\mu_{\alpha},\cr
&\{Q,\chi_{\alpha\beta}\} =  H_{\alpha\beta},\cr 
&\{Q,\nu_{\dalpha}\}= h_{\dalpha},\cr
&[Q, \lambda]=\raiz\eta,\cr
}\qquad\qquad
\eqalign{
&\{Q,\psi_{\alpha\dot\alpha}\} = 
{\raiz}\deriv_{\alpha\dot\alpha}\phi,\cr
&[Q, \phi]=0,\cr
&\{Q,\mu_{\alpha}\}=-2i [M_{\alpha},\phi],\cr
&[Q,H_{\alpha\beta}]=2\raiz i\,[\chi_{\alpha\beta},\phi],\cr
&[Q,h_{\dalpha}]=2\raiz i\,[\nu_{\dalpha},\phi],\cr
&\{Q,\eta\,\} =2i[\lambda,\phi].\cr  }
\eqn\covMedeiros
$$ 

The covariantized gauge fermion \cerne\ takes now the form:
$$
\eqalign
{\Psi&={1\over e^2}\int_X  d^4 x\,\sqrt{g}\,\tr\,\bigl\{\,\, 
{1\over4}\bar\nu^{\dalpha}\bigl (h_{\dalpha}+
2\raiz i\,\deriv_{\alpha\dalpha}M^{\alpha}\bigr )-
{1\over4}\nu^{\dalpha}\bigl (\bar h_{\dalpha}+
2\raiz i\,\deriv_{\alpha\dalpha}\overline M^{\alpha}\bigr )\cr
+&{1\over4}\chi_{\alpha\beta}\,\bigl (\, H^{\alpha\beta}+
2i(F^{+\alpha\beta}
+2[\overline M^{(\alpha},M^{\beta)}])\,\bigr )\,\bigr \}\cr
&-{1\over e^2}\int_X  d^4 x\,\tr\,\bigl\{\,\,
{1\over{2\raiz}}\,\lambda\,\bigl (\,
\deriv_{\alpha\dalpha}\psi^{\dalpha\alpha}
+i\raiz\,[\overline M^{\alpha},\mu_{\alpha}]
-i\raiz\,[\bar\mu^{\alpha},M_{\alpha}]
\,\bigr )\,\bigr \}\cr
&-{1\over e^2}\int_X  d^4 x\,\tr\,\bigl\{\,\,
{i\over 4}\,\lambda[\eta,\phi]\,\bigr\}.\cr
}
\eqn\covcerne
$$ 

The resulting theory is equivalent to the theory of  non-abelian monopoles
discussed at length in [\marmon,\marpol,\marth],  but with the monopole
multiplet in the adjoint representation of the gauge  group. That theory in
turn is a generalization of the abelian monopole  equations proposed in
[\monop].  The reason for this equivalence can be explained as follows.
First recall that  from the viewpoint of  $N=1$ superspace  both, $N=2$
supersymmetric gauge theory  coupled to an $N=2$  hypermultiplet in the
adjoint of the gauge group, and $N=4$ supersymmetric gauge theory,  are built
out of   the same set of $N=1$ superfields, namely a vector superfield and
three chiral  superfields. In the case of $N=4$ supersymmetric gauge theory
we have a  cuadruplet of gauginos in the  ${\bf 4}$ of $SU(4)_I$, which
correspond to a $SU(2)_I$ doublet of gauginos and  an $SU(2)_I$ singlet Dirac
spinor (\ie , two $SU(2)_I$ singlet Weyl spinors) in  the case of the $N=2$
theory. Notice that in the transition the decomposition ${\bf
4}\to {\bf 2}\oplus {\bf 1}\oplus {\bf 1}$ has to be done, which is 
equivalent  to the decomposition defining the second twist of $N=4$.    In this
framework, the $T_3$ subgroup of the former $SU(2)_F$ symmetry  remains in the
theory as an $U(1)$ symmetry which involves the monopole  sector only and which
corresponds to the $N=2$   central charge (trivial in  this case) that remains
after the twisting [\zzeta].

%%%%%%%%%%  %%%%%%%%%%%%%%%%%%%%%%%%%%%%%%%%%%%%%%%%%%%%%%%%%%%%%%%%%%%%%%%%%%%%
%%MARCUS%%
%%%%%%%%%%  %%%%%%%%%%%%%%%%%%%%%%%%%%%%%%%%%%%%%%%%%%%%%%%%%%%%%%%%%%%%%%%%%%%%

\vskip 0.5cm
\subsection{(3) ${\bf 4}\too ({\bf 2},{\bf 2})$ Amphicheiral Theory}

The last theory we will consider was  briefly introduced at the end of
reference [\yamron], and afterwards it was  considered in detail in
[\marcus,\blauthomp].  
It corresponds to the decomposition ${\bf 4}\too ({\bf 2},{\bf
2})$, but it is easier (and equivalent) to start from the second twisted theory 
and replace $SU(2)_R$ with the diagonal sum $SU(2)'_R$ of $SU(2)_R$ itself
and  the remaining isospin group $SU(2)_F$ (this is very much alike to a
conventional $N=2$ twisting). This introduces in the theory a second
BRST-like symmetry, which comes from the $N=4$
spinor supercharges $\bar Q_{v\dalpha}$. As we pointed out at the end of the
introduction, there are several unusual features in this theory that we think 
deserve a detailed analysis. We begin by recalling the fundamentals of
the second twist. The symmetry group
${\cal H}=SU(2)_L\otimes SU(2)_R\otimes SU(4)_I$ of the original $N=4$
supersymmetric gauge theory is twisted to give the symmetry  group 
${\cal H'}=SU(2)'_L\otimes SU(2)_R\otimes SU(2)_F\otimes U(1)$ 
(we will refer to this as the $L$ twist) of the half-twisted theory. The
supersymmetry charges $Q^v_{\!\alpha}$ and $\bar Q_{v\dalpha}$ decompose under
${\cal H'}$ as:
$$
Q^v{}_\alpha \oplus \bar Q_{v\dot\alpha}\too   
Q^{(+1)}\oplus Q^{(+1)}_{(\alpha\beta)}\oplus Q^{(-1)}_{i\alpha} \oplus
\bar Q^{(-1)}_{\alpha\dot\alpha}\oplus \bar Q^{(+1)}_{i\dalpha}.
\eqn\Reeve
$$
But one can also twist with $SU(2)_R$ thus obtaining its corresponding $\tilde
T$ twist with symmetry group  
${\tilde{\cal H'}}=SU(2)_L\otimes SU(2)'_R\otimes SU(2)_F\otimes U(1)$ 
($R$ twist). Both formulations are related \titin\ through an 
orientation reversal and a change of sign in $\theta$. Now we can twist
$SU(2)_F$ away in four different ways. Two of these ($LL$ and $RR$) take us back
to the Vafa-Witten twists $T$ and $\tilde T$. The other two ($LR$ and 
$RL$) should lead us to the twist considered in [\marcus,\blauthomp] and its
corresponding $\tilde T$ twist. The non-trivial result is that either of these
two different choices leads to the  same topological theory. This can be seen
as follows. Pick one of the possibilities,  say, $LR$. After the first twist
we have the half-twisted  theory with symmetry group ${\cal H'}$ and
supersymmetry charges
\Reeve . If we now twist $SU(2)_F$ with $SU(2)_R$ we obtain, from
the last charge in
\Reeve , a second scalar charge  $\tilde Q$ given by:
$$
\bar Q_{i\dalpha}\to \bar Q_{\dbeta\dalpha}\to \tilde Q=
C^{\dbeta\dalpha}\bar Q_{\dbeta\dalpha}.
\eqn\Sontag
$$

Notice that both the anticommuting symmetries, $Q$ and $\tilde Q$, have the
same ghost number, so they are both to be considered either as BRST
or anti-BRST operators. This is in contrast with the situation we 
found
in the first twist where, after explicitly breaking the isospin group
$SU(2)_F$ down to its
$T_3$ subgroup, we were left with two scalar charges $Q^{(+)}$ and
$Q^{(-)}$ with opposite ghost numbers, which were then
interpreted as a BRST-antiBRST system.   

The fields of the new theory can be obtained from those in the half-twisted
theory as follows:
$$
\eqalign{
A_{\alpha\dalpha} &\too A^{(0)}_{\alpha\dalpha}\too
A^{(0)}_{\alpha\dalpha},\cr 
\lambda_{v\alpha}&\too
\chi^{(-1)}_{\beta\alpha},~
\eta^{(-1)},~\lambda^{(+1)}_{i\alpha}\too\chi^{(-1)}_{\beta\alpha},~
\eta^{(-1)},~\tilde\psi^{(+1)}_{\alpha\dot\alpha},\cr
\bar\lambda^v{}_{\!\dot\alpha} &\too
\psi^{(+1)}_{\alpha\dalpha},~\zeta^{(-1)}_{i\dalpha}\too
\psi^{(+1)}_{\alpha\dalpha},~\tilde\eta^{(-1)},~\tilde\chi^{(-1)}
_{\dalpha\dbeta},
\cr
\phi_{uv}&\too B^{(-2)},~C^{(+2)},~G^{(0)}_{i\alpha}\too
B^{(-2)},~C^{(+2)},~V^{(0)}_{\alpha\dalpha},\cr
}
\eqn\pcien
$$
where we have included also the corresponding fields of the $N=4$ theory and
the ghost number carried by the twisted fields. The notation is similar to
that in ref. [\marcus]. Notice that if
we exchange $SU(2)_L$ by $SU(2)_R$ the field content in \pcien\ does not
change. This in turn implies that the $LR$ and $RL$ twists are in fact the
same,
$$
{\cal S}^{LR}_{X}={\cal S}^{RL}_{X},
\eqn\anfiuno
$$
or, in other words, the third twist leads to an amphicheiral topological
quantum field theory (see \amphiuno). Since it is known that the
two twists are related by
$ {\cal S}^{LR}_{X}={\cal S}^{RL}_{\tilde X}\big|_{\tau\rightarrow
-\bar\tau}$ ($\tilde X$ denotes the manifold $X$ with the opposite
orientation), it follows that by reversing the sign of the $\theta$-angle one
can jump from 
$X$ to $\tilde X$:
$$
{\cal S}_{X}={\cal S}_{\tilde X}\big|_{\tau\rightarrow - \bar\tau}.
\eqn\anfidos
$$
We will see in a moment that this information is encoded in the conjugation 
discrete symmetry introduced in [\marcus].

The definitions in \pcien\ are almost self-evident. The only ones
which need clarification are those corresponding to $\tilde \eta$ and
$\tilde\chi_{\dalpha\dbeta}$. Our conventions are:
$$
\zeta^{i}{}_{\!\dot\alpha}\to \zeta^{\dot\beta}{}_{\!\dot\alpha}\to
\cases{\tilde\eta =-\zeta^{\dot\alpha}{}_{\dot\alpha},\cr
\tilde\chi_{\dot\alpha\dot\beta}=-C_{\dot\gamma(\dot\beta}\zeta^{\dot
\gamma}{}_{\dot\alpha)}.\cr}
\eqn\Moon
$$

In terms of the fields in \pcien , the on-shell action \cccuatro\ takes the
form:
$$
\eqalign{ {\cal S}^{(0)}&={1\over e^2}\int d^4 x\,
\tr\,\bigl\{\,\half\nabla_{\!\alpha\dalpha}
 B\nabla^{\dalpha\alpha}C -{1\over4}\nabla_{\!\beta\dbeta}
V_{\alpha\dalpha}\nabla^{\dbeta\beta}V^{\dalpha\alpha} 
-i\psi^{\beta}{}_{\dalpha}\nabla^{\dalpha\alpha}
\chi_{\alpha\beta}\cr& 
-{i\over 2}\psi_{\alpha\dalpha}\nabla^{\dalpha\alpha}
\eta
+{i\over 2}\tilde\eta\nabla^{\dalpha\alpha}\tilde\psi_{\alpha\dalpha}+
i\tilde\chi_{\dalpha\dbeta}\nabla^{\dalpha\alpha}\tilde\psi
^\dbeta{}_{\alpha}-{1\over 4}F_{mn}
F^{mn}\cr&-{i\over\raiz}\,\chi^{\alpha\beta}[\chi_{\alpha\beta},C]
-{i\over\raiz}\,\tilde\psi^{\dalpha\alpha}[\tilde\psi_{\alpha\dalpha},B]+ 
i\raiz\,\chi^{\alpha\beta}[\tilde\psi_{\alpha\dalpha},V_\beta{}^{\dalpha}]
+{i\over\raiz}\,\eta[\tilde\psi^\alpha{}_{\dalpha},V_\alpha{}^{\dalpha}]
\cr&-{i\over{2\raiz}}\,\eta[\eta,C]
+{i\over\raiz}\,\psi_{\alpha\dalpha}[\psi^{\dalpha\alpha},B]-
{i\over\raiz}\,\tilde\eta[\psi^\alpha{}_{\dalpha},V_\alpha{}^\dalpha]-
i\raiz\,\tilde\chi^{\dalpha\dbeta}[\psi_{\alpha\dalpha},V^\alpha{}_\dbeta]\cr&+
{i\over\raiz}\,\tilde\chi_{\dalpha\dbeta}[\tilde\chi^{\dalpha\dbeta},C]
+{i\over{2\raiz}}\,\tilde\eta[\tilde\eta,C]-\half[B,C]^2 
-[B,V_{\alpha\dalpha}][C,V^{\dalpha\alpha}]\cr&+ 
{1\over 4}[V_{\alpha\dalpha},V_{\beta\dbeta}]
[V^{\dalpha\alpha},V^{\dbeta\beta}]\,\bigr\}
-{i\theta\over 32\pi^2}\int d^4 x\,\tr\,\bigl\{\, *  F_{mn}F^{mn}
\,\bigr\}.
\cr
}
\eqn\Light
$$ 
The next thing to do is to obtain the symmetry transformations which
correspond to the new model. Recall that we have now two fermionic
charges  $Q$ and $\tilde Q$. The transformations generated by $Q$ are
easily obtained from those in the previous twist \cccinco . To obtain
the transformations generated by $\tilde Q$ we must return to the $N=4$
theory. Let us  recall that the $N=4$ supersymmetry transformations are
generated by $\xi_v{}^\alpha Q^v{}_\alpha + 
\bar\xi^v{}_\dalpha \bar Q_v{}^\dalpha$. The transformations corresponding
to $\tilde Q$ are readily extracted by setting 
$\bar\xi^1=\bar\xi^2=0$ and making the replacement
$$
\bar\xi^{\,3,4}{}_\dalpha\to \bar\xi^{\,i}{}_{\!\dalpha}\to
\bar\xi^{\,\dbeta}{}_{\!\dalpha}\to 
\tilde\epsilon\,\delta^\dbeta{}_{\!\dalpha}.
\eqn\pera
$$
In this way one gets the following set of transformations:
$$
\eqalign{
&\delta A_{\alpha\dalpha} = 2i\epsilon\psi_{\alpha\dalpha},\cr 
&\delta F^{+}_{\alpha\beta}=2\epsilon\nabla_{(\alpha}{}^{\dot\alpha}
\psi{}_{\beta)\dot\alpha},\cr
&\delta\psi_{\alpha\dot\alpha} = 
-i{\raiz}\epsilon\nabla_{\alpha\dot\alpha}C,\cr
&\delta\tilde\eta=i{\raiz}\epsilon\nabla_{\alpha\dot\alpha}
V^{\dalpha\alpha},\cr
&\delta\tilde\chi_{\dot\alpha\dot\beta}=-i{\raiz}\epsilon\nabla_{\alpha
(\dot\alpha}V^{\alpha}{}_{\dot\beta)},\cr
&\delta\chi_{\alpha\beta} = 
-i\epsilon F^{+}_{\alpha\beta}
-i\epsilon[V_{\alpha\dot\alpha},V_{\beta}{}^{\dot\alpha}],\cr 
&\delta\eta  =2i\epsilon[B,C],\cr
&\delta\tilde\psi_{\alpha\dot\alpha}=-2i\epsilon [V_{\alpha\dot\alpha},C],
\cr
&\delta B=\raiz\epsilon\eta,\cr
&\delta C=0,\cr
&\delta V_{\alpha\dot\alpha}=-\raiz\epsilon\tilde\psi_{\alpha\dot\alpha},
\cr}
\qquad\qquad
\eqalign{
&\tilde\delta A_{\alpha\dalpha} =
-2i\tilde\epsilon\tilde\psi_{\alpha\dalpha},\cr  
&\tilde\delta
F^{+}_{\alpha\beta}=-2\tilde\epsilon\nabla_{(\alpha}{}^
{\dot\alpha}\tilde\psi{}_{\beta)\dot\alpha},\cr
&\tilde\delta\psi_{\alpha\dot\alpha} = 
-2i\tilde\epsilon[V_{\alpha\dot\alpha},C]\cr
&\tilde\delta\tilde\eta=2i\tilde\epsilon[B,C],\cr
&\tilde\delta\tilde\chi_{\dot\alpha\dot\beta}=i\tilde\epsilon F^{-}_
{\dot\alpha\dot\beta}
-i\tilde\epsilon[V_{\alpha\dot\alpha},V^{\alpha}{}_{\dot\beta}],\cr 
&\tilde\delta\chi_{\alpha\beta} = i{\raiz}\,\tilde\epsilon\,\nabla^{\dot
\alpha}{}_{(\alpha}V_{\beta)\dot\alpha},\cr
&\tilde\delta\eta  =i{\raiz}\,\tilde\epsilon\,\nabla_{\alpha\dot\alpha}
V^{\dalpha\alpha},\cr
&\tilde\delta\tilde\psi_{\alpha\dot\alpha}=-i\raiz\tilde\epsilon\nabla
_{\alpha\dot\alpha}C ,\cr
&\tilde\delta B=-\raiz\tilde\epsilon\tilde\eta,\cr
&\tilde\delta C=0,\cr
&\tilde\delta V_{\alpha\dot\alpha}=\raiz\tilde\epsilon\psi_
{\alpha\dot\alpha}.\cr}
\eqn\melon
$$

%
%notice that $[Q,\{Q,\psi\}]={1\over2}[\{Q,Q\},\psi]$
%or, in terms of the transformations themselves,
%$$
%\eqalign{&[\delta_1,\delta_2]=4\raiz \epsilon_1\epsilon_2[~,C],\cr
%&[\tilde\delta_1,\tilde\delta_2]=-4\raiz
%\tilde\epsilon_1\tilde\epsilon_2[~,C],\cr 
%&[\delta,\tilde\delta]=0,\cr}
%\eqn\uruguay
%$$

Since there are no half-integer spin fields in the theory it is preferable to
convert the spinor indices into vector indices. To do this we make the 
following definitions:
$$
\pmatrix{V\cr\psi\cr\tilde\psi\cr}_{\!\alpha\dalpha}\equiv
\sigma^m{}_{\!\alpha\dalpha}\pmatrix{V\cr\psi\cr\tilde\psi\cr}_{\!m},\qquad
\chi_{\alpha\beta}=\sigma^{mn}{}_{\!\alpha\beta}\chi^{+}_{mn},\qquad
\tilde\chi_{\dalpha\dbeta}=\bar\sigma^{mn}{}_{\!\dalpha\dbeta}\chi^{-}
_{mn}
\eqn\uva
$$

where $\chi^{\pm}{}_{\!mn}=(1/2)\{\chi_{mn}\pm(1/2)\epsilon_{mn
pq}\chi^{pq}\}$. In order to extract a manifestly real action we also make
the replacements $\psi\to -i\psi$, $\chi^{+}\to i\chi^{+}$, $\tilde\eta\to 
i\tilde\eta$ and $\tilde Q\to i\tilde Q$.   The
resulting action is:
$$
\eqalign
{ {\cal S}^{(0)}&={1\over e^2}\int d^4 x\, \tr\, \bigl\{\,-\nabla_m
B\nabla^m C -\nabla_m
V_n\nabla^m V^n 
+4\psi^m\nabla^n\chi^{+}_{mn} 
+\psi^m\nabla_m\eta\cr&
+\tilde \psi^m\nabla_m\tilde\eta+
4\tilde\psi^m\nabla^n\chi^{-}_{mn}-{1\over 4}F_{mn}
F^{mn}- i\raiz\,\chi^{+mn}[\chi^{+}_{mn},C]\cr
&+i\raiz\,\tilde\psi^m[\tilde\psi_m,B]- 
4i\raiz\,\chi^{+}_{mn}[\tilde\psi^m,V^n]
+i\raiz\,\eta[\tilde\psi_m,V^m]
-{i\over{2\raiz}}\,\eta[\eta,C]\cr
&+i\raiz\,\psi_m[\psi^m,B]-
i\raiz\,\tilde\eta[\psi_m,V^m]+
4\raiz\,i\chi^{-}_{mn}[\psi^m,V^n]-
i\raiz\,\chi^{-}_{mn}[\chi^{-mn},C]\cr
&-{i\over{2\raiz}}\,\tilde\eta[\tilde\eta,C]-\half[B,C]^2 
+2[B,V_m][C,V^m]+ 
[V_m,V_n]
[V^m,V^n]\,\bigr\}\cr
&-{i\theta\over 32\pi^2}\int d^4 x\,\tr\,\bigl\{\, *  F_{mn}F^{mn}
\,\bigr\},\cr}
\eqn\pLight
$$ 
and the corresponding transformations become:
$$
\eqalign{
&\delta A_m = 2\epsilon\psi_m,\cr 
&\delta\psi_m = 
{\raiz}\epsilon\nabla_m C,\cr
&\delta\tilde\eta=-2{\raiz}\epsilon\nabla_m V^m,\cr
&\delta\chi^{-}_{mn}=2{\raiz}\epsilon (\nabla_{[m}V_{n]})^{-},\cr
&\delta\chi^{+}_{mn} = 
-\epsilon F^{+}_{mn}
+2i\epsilon ([V_m,V_n])^{+},\cr 
&\delta\eta  =2i\epsilon[B,C],\cr
&\delta\tilde\psi_m=-2i\epsilon [V_m,C],\cr
&\delta B=\raiz\epsilon\eta,\cr
&\delta C=0,\cr
&\delta V_m=-\raiz\epsilon\tilde\psi_m ,\cr}
\qquad\qquad
\eqalign{
&\tilde\delta A_m = -2\tilde\epsilon\tilde\psi_m,\cr 
&\tilde\delta\psi_m = 
-2i\tilde\epsilon[V_m,C]\cr
&\tilde\delta\tilde\eta=-2i\tilde\epsilon[B,C],\cr
&\tilde\delta\chi^{-}_{mn}=\tilde\epsilon F^{-}_
{mn}
-2i\tilde\epsilon ([V_m,V_n])^{-},\cr 
&\tilde\delta\chi^{+}_{mn} =
2{\raiz}\tilde\epsilon (\nabla_{[m}V_{n]})^{+},\cr
&\tilde\delta\eta  =-2{\raiz}\tilde\epsilon\nabla_m
V^m,\cr
&\tilde\delta\tilde\psi_m=-\raiz\tilde\epsilon\nabla
_m C ,\cr
&\tilde\delta B=-\raiz\tilde\epsilon\tilde\eta,\cr
&\tilde\delta C=0,\cr
&\tilde\delta V_m =-\raiz\tilde\epsilon\psi_
m,\cr}
\eqn\melones
$$
where $(X_{[mn]})^{\pm}\equiv \half(X_{[mn]}\pm *X_{[mn]})$, and
$X_{[mn]}\equiv \half(X_{mn}-X_{nm})$.
The generators $Q$ and $\tilde Q$ satisfy the on-shell algebra:
$$
\eqalign{
\{\,Q,Q\,\}&=\delta_g (C),\cr
\{\,\tilde Q,\tilde Q\,\}&=\delta_g (C),\cr
\{\,Q,\tilde Q\,\}&=0.\cr
}
\eqn\aceituna
$$

Now consider the following discrete transformations acting on the fields of
the theory:
$$
\eqalign{&B\too B,\qquad\qquad C\too C,\cr
&A\too A,\qquad\qquad\quad V\too -V,\cr
&\eta\too -\tilde\eta,\qquad\qquad\quad\,\,\psi\too -\tilde\psi,\cr
&\tilde\eta\too -\eta,\qquad\qquad\quad\,\,\tilde\psi\too -\psi,\cr\cr 
&\chi^{+}\longleftrightarrow -\chi^{-} \Rightarrow \cases{\chi\to -\chi,\cr
*\chi\to *\chi,\cr}\cr\cr
&F^{+}\longleftrightarrow F^{-} \Rightarrow \cases{F\to F,\cr
*F\to -*F.\cr}\cr
}
\eqn\melocoton
$$
Notice that these transformations involve the simultaneous replacement
$\epsilon_{mnpq}\to
-\epsilon_{mnpq}\,$, which is equivalent to a reversal of the orientation
of the four-manifold $X$. Because of this orientation reversal, the sign of
the $\theta$-term  in \pLight\ is also reversed. Thus the ${\Bbb
Z}_2$-like transformations \melocoton\ map the action on $X$ to the same   
action on $\tilde X$ but with  $-\theta$. This is precisely the
information encoded in \anfidos .   

It is also noteworthy that the transformations 
\melocoton\
exchange the BRST generators $Q$ and $\tilde Q$ (one can realize this by
looking at \melones ). Indeed, had we not known about the existence of one
of the topological symmetries, say $\tilde Q$, we would have discovered it
immediately with the aid of the symmetry \melocoton . In addition to this,
one can readily see that the replacements dictated by \melocoton\
preserve the ghost number assignments of the fields. In what
follows, we will usually refer to the transformations 
\melocoton\  by ${\Bbb Z}_2$, but the reader must be aware of this abuse of
notation.

Several things remain to be done. It would be desirable to obtain an
off-shell formulation of the model. Besides, it would be interesting to
find out whether the off-shell action (provided that it exists) can be
written as a $Q$- (or $\tilde Q$, or both) commutator, and write down the
explicit expression for the corresponding gauge fermion. And finally, the
theory should be generalized to any arbitrary four-manifold of euclidean
signature. 

We have found a complete off-shell 
formulation involving both  BRST symmetries simultaneously such that 
the action \pLight\ is (up to  appropriate theta-terms) $Q$ and $\tilde
Q$-exact. 
%In brief, the whole picture can be summarized as follows:
%
%$$
%\cases{{\cal S}=\{Q,\Psi^{+}\}-2\pi ik\tau =\{\tilde Q,\Psi^{-}\}
%-2\pi ik\bar\tau\,\quad  [Q, {\cal S}]=[\tilde Q,{\cal S}]=0,\cr
%\Psi^{+}\mapboth{{\Bbb Z}_2}\Psi^{-}\cr
%}
%\eqn\pomme
%$$
%
Let us examine these results  in more detail. The on-shell algebra \aceituna\ 
can be extended off-shell by introducing the auxiliary fields 
$N^{+}_{mn}$, $N^{-}_{mn}$ and $P$, which have zero ghost numbers and 
are taken to transform under ${\Bbb Z}_2$ as $N^{+}\leftrightarrow 
-N^{-}$, $P\to -P$. In terms of these fields, the transformations \melones\ 
are modified as follows:
$$
\eqalign{
&\delta\tilde\eta=-2{\raiz}\epsilon\nabla_m V^m +\epsilon P,\cr
&\delta P=-4\epsilon\nabla_m\tilde\psi^m +4\raiz i\epsilon [\psi^m,V_m]+
2\raiz i\epsilon[\tilde \eta,C],\cr
&\delta\chi^{-}_{mn}=2{\raiz}\epsilon (\nabla_{[m}V_{n]})^{-}
+\epsilon N^{-}_{mn},\cr
&\delta N^{-}_{mn}=4\epsilon\nabla_{[m}\tilde\psi^{-}_{n]}-
4\raiz i\epsilon [\psi_{[m},V_{n]}]^{-}
+2\raiz i\epsilon[\chi^{-}_{mn},C],\cr
&\delta\chi^{+}_{mn} = 
-\epsilon F^{+}_{mn}+
2i\epsilon ([V_m,V_n])^{+}+\epsilon N^{+}_{mn},\cr 
&\delta N^{+}_{mn}=4\epsilon\nabla_{[m}\psi^{+}_{n]}
+4\raiz i\epsilon [\tilde\psi_{[m},V_{n]}]^{+}+
2\raiz i\epsilon[\chi^{+}_{mn},C].\cr}
\eqn\Basler
$$
The other transformations in \melones\ remain the same. 
Equivalent formulas hold for $\tilde Q$ and are related to those in 
\Basler\ through the ${\Bbb Z}_2$ transformation. In this off-shell
realization the auxiliary fields appear in the action only quadratically, that 
is, 
$${\cal S}^{(1)}={\cal
S}^{(0)}+\int
\tr\{\,\half(N^{+})^2+
\half(N^{-})^2+{1\over 8}P^2\}.
\eqn\meca
$$
%
%${\cal S}_{\hbox{\sevenrm off-shell}}={\cal
%S}_{\hbox{\sevenrm on-shell}}+\int
%\tr\{-\half(N^{+})^2-
%\half(N^{-})^2-{1\over 8}P^2\}$. 
%
The action ${\cal S}^{(1)}$ can be written either
as a $Q$ commutator or as a $\tilde Q$ commutator and is invariant under both, 
$Q$ and $\tilde Q$, that is,
$$
{\cal S}^{(1)}=\{Q,\hat \Psi^{+}\}-2\pi ik\tau =\{\tilde Q,\hat \Psi^{-}\}
-2\pi ik\bar\tau\,;\quad  [Q, {\cal S}^{(1)}]=[\tilde Q,{\cal S}^{(1)}]=0
\eqn\boomerang
$$
where the gauge fermions $\hat\Psi^{\pm}$ are not equal but are formally
interchanged by the ${\Bbb Z}_2$ transformation
and $k$ is the instanton number \knumber.
It is possible to redefine the
auxiliary fields to cast either the $Q$ or the $\tilde Q$ transformations 
(but not both simultaneously) in the standard form,
$$
\eqalign{&\{Q,{\hbox{\sevenrm ANTIGHOST}}
\}={\hbox{\sevenrm AUXILIARY FIELD}}, \cr
&[Q,{\hbox{\sevenrm AUXILIARY FIELD}}]=
\delta^{gauge}{\hbox{\sevenrm ANTIGHOST}},\cr}
$$
which is essential to make contact with the Mathai-Quillen interpretation. 
Performing the shifts,
$$
\eqalign{
P&\too P+2{\raiz}\nabla_m V^m,\cr
N^{-}_{mn}&\too N^{-}_{mn} -2{\raiz} (\nabla_{[m}V_{n]})^{-},\cr
N^{+}_{mn}&\too N^{+}_{mn}+ F^{+}_{mn}
-2i ([V_m,V_n])^{+},\cr}
\eqn\Allou
$$
which can be guessed from \Basler , the $Q$ transformations take the 
simple form:
$$
\eqalign{ 
&\delta A_m = 2\epsilon\psi_m,\cr
&\delta V_m =-\raiz\epsilon\tilde\psi_m,\cr
&\delta C=0,\cr
&\delta B=\raiz\epsilon\eta,\cr
&\delta\tilde\eta=\epsilon P,\cr
&\delta\chi^{+}_{mn}=\epsilon N^{+}_{mn},\cr
&\delta\chi^{-}_{mn}=\epsilon N^{-}_{mn},\cr
}\qquad\qquad
\eqalign{
&\delta\psi_m=\raiz\,\epsilon\nabla_m C,\cr
&\delta\tilde\psi_m=-2i\epsilon [V_m,C],\cr\cr
&\delta\eta =2i\epsilon [B,C],\cr
&\delta P=2\raiz i\epsilon [\tilde\eta,C],\cr
&\delta N^{+}_{mn}=2\raiz i\epsilon [\chi^{+}_{mn},C],\cr
&\delta N^{-}_{mn}=2\raiz i\epsilon [\chi^{-}_{mn},C].\cr}
\eqn\butterfly
$$
The point is that if instead of \Allou\  we make the ${\Bbb Z}_2$ conjugate 
shifts, 
$$
\eqalign{
P&\too P+2{\raiz}\nabla_m V^m,\cr
N^{+}_{mn}&\too N^{+}_{mn} -2{\raiz} (\nabla_{[m}V_{n]})^{+},\cr
N^{-}_{mn}&\too N^{-}_{mn}- F^{-}_{mn}
+2i([V_m,V_n])^{-},\cr}
\eqn\Proust
$$
it is $\tilde\delta\equiv\tilde\epsilon\tilde Q$ the one which can be cast in
the  simple form:
$$
\eqalign{ 
&\tilde\delta A_m = -2\tilde\epsilon\tilde\psi_m,\cr
&\tilde\delta V_m =-\raiz\tilde\epsilon\psi_m,\cr
&\tilde\delta C=0,\cr
&\tilde\delta B=-\raiz\tilde\epsilon\tilde\eta,\cr
&\tilde\delta\eta=\tilde\epsilon P,\cr
&\tilde\delta\chi^{+}_{mn}=\tilde\epsilon N^{+}_{mn},\cr
&\tilde\delta\chi^{-}_{mn}=\tilde\epsilon N^{-}_{mn},\cr
}\qquad\qquad
\eqalign{
&\tilde\delta\tilde\psi_m=-\raiz\,\tilde\epsilon\nabla_m C,\cr
&\tilde\delta\psi_m=-2i\tilde\epsilon [V_m,C],\cr\cr
&\tilde\delta\tilde\eta =-2i\tilde\epsilon [B,C],\cr
&\tilde\delta P=2\raiz i\tilde\epsilon [\eta,C],\cr
&\tilde\delta N^{+}_{mn}=2\raiz i\tilde\epsilon [\chi^{+}_{mn},C],\cr
&\tilde\delta N^{-}_{mn}=2\raiz i\tilde\epsilon [\chi^{-}_{mn},C].\cr
}
\eqn\boheme
$$
Notice that since the appropriate shifts are in each case different, the one
which simplifies the $Q$ transformations makes the corresponding $\tilde Q$
transformations (not shown)  much more complicated and conversely, the shift
which simplifies the $\tilde Q$ transformations makes the corresponding $Q$
transformations (not shown) much more complicated.

Keeping these  results in
mind from now on we will focus on the $Q$ formulation, that is, on the
off-shell formulation in which the $Q$ transformations take the form
\butterfly.  The off-shell action which corresponds to this formulation is:
$$
\eqalign
{ {\cal S}^{(2)}&= {1\over e^2}\int d^4 x\, \tr\, \bigl\{\,-\nabla_m
B\nabla^m C +\half N^{+}_{mn}\bigl (\,N^{+mn}+2F^{+mn}-4i[V^m,V^n]^{+}\,
\bigr )\cr  
&+\half N^{-}_{mn}\bigl (\,N^{-mn}-4\raiz (\nabla^{[m}V^{n]})^{-}\,\bigr ) 
+{1\over8}P\bigl (\,P+4\raiz \,\nabla_m V^m\,\bigr)
\cr&+4\psi^m\nabla^n\chi^{+}_{mn} +\psi^m\nabla_m\eta
+\tilde \psi^m\nabla_m\tilde\eta+
4\tilde\psi^m\nabla^n\chi^{-}_{mn}
 \cr &-i\raiz\,\chi^{+mn}[\chi^{+}_{mn},C]+
i\raiz\,\tilde\psi^m[\tilde\psi_m,B]-
4\raiz\,i\chi^{+}_{mn}[\tilde\psi^m,V^n]
\cr&+i\raiz\,\eta[\tilde\psi_m,V^m]
-{i\over{2\raiz}}\,\eta[\eta,C]+
i\raiz\,\psi_m[\psi^m,B]-
i\raiz\,\tilde\eta[\psi_m,V^m]\cr&+
4\raiz\,i\chi^{-}_{mn}[\psi^m,V^n]-
i\raiz\,\chi^{-}_{mn}[\chi^{-mn},C]
-{i\over{2\raiz}}\,\tilde\eta[\tilde\eta,C]-\half[B,C]^2
\cr&+2[B,V_m][C,V^m]
\,\bigr\}
-2\pi i\tau{1\over 32\pi^2}\int d^4 x\,\tr\,\bigl\{\, *  F_{mn}F^{mn}
\,\bigr\},\cr}
\eqn\isinha
$$
and reverts to \pLight\  after integrating 
out the auxiliary fields.   
The $\tau$-independent part of the action \isinha\ is $Q$-exact, that is, 
it can be written as a $Q$-commutator. The appropriate gauge fermion 
is:
$$
\eqalign{
\Psi^{+}=&{1\over e^2}\int d^4 x\,\tr\,\bigl\{\,\,\half\chi^{+}_{mn}\,\bigl 
(\,N^{+mn}+2F^{+mn}-4i[V^m,V^n]^{+}\,
\bigr )\cr
&+{1\over2} \chi^{-}_{mn}\bigl (\,N^{-mn}-4\raiz (\nabla^{[m}V^{n]})^{-}\,
\bigr)
+{1\over8}\tilde\eta\,\bigl (\,P+4\raiz \,\nabla_m V^m\,\bigr )\,\bigr \}
\cr 
&+{1\over e^2}\int d^4 x\,\tr\,\bigl\{\,{1\over\raiz}B\,\bigl
 (\,\nabla_m \psi^m 
+i\raiz [\tilde\psi_m,V^m]\,\bigr )\,\bigr\}\cr
&+{1\over e^2}\int d^4 x\,\tr\,\bigr\{\,{i\over4}\eta [B,C]\,\bigr\}.\cr}
\eqn\sesamo
$$
notice that $\Psi^-$ would correspond to the ${\Bbb Z}_2$-transformed of
$\Psi^+$. The gauge fermions $\hat\Psi^+$ and $\hat\Psi^-$ in \boomerang\ are
easily obtained after undoing the shifts \Allou\ and \Proust, respectively.

%We want to point out   
% that the action \isinha\ is $\tilde Q$-invariant and can be rewriten as a 
%$\tilde Q$-exact 
%$F$ plus  the same theta term as in \isinha , that is,
%$$
%{\cal S}=\{Q,\Psi^{+}\}-2\pi ik\tau =\{\tilde Q,\tilde\Psi\}
%-2\pi ik\tau
%\eqn\Rigoletto
%$$
%Of course, $\Psi^{+}$ and $\tilde\Psi$ are not related by ${\Bbb Z}_2$
%anymore. The form of the 
%$\tilde Q$ transformations and of the gauge fermion, however, are much more
%complicated  than \butterfly\ and \sesamo\ and are not shown.    

Now we switch on an arbitrary background metric $g_{\mu\nu}$ of euclidean 
signature. This is straightforward once we have expressed the model in the 
form of eqs. \butterfly\ and \sesamo . The covariantized transformations  are 
the following: 
$$
\eqalign{
&\delta A_\mu = 2\epsilon\psi_\mu,\cr
&\delta V_\mu =-\raiz\epsilon\tilde\psi_\mu,\cr
&\delta C=0,\cr
&\delta B=\raiz\epsilon\eta,\cr
&\delta\tilde\eta=\epsilon P,\cr
&\delta\chi^{+}_{\mu\nu}=\epsilon N^{+}_{\mu\nu},\cr
&\delta\chi^{-}_{\mu\nu}=\epsilon N^{-}_{\mu\nu},\cr
}\qquad\qquad
\eqalign{
&\delta\psi_\mu=\raiz\,\epsilon\deriv_\mu C,\cr
&\delta\tilde\psi_\mu=-2i\epsilon [V_\mu,C],\cr\cr
&\delta\eta =2i\epsilon [B,C],\cr
&\delta P=2\raiz i\epsilon [\tilde\eta,C],\cr
&\delta N^{+}_{\mu\nu}=2\raiz i\epsilon [\chi^{+}_{\mu\nu},C],\cr
&\delta N^{-}_{\mu\nu}=2\raiz i\epsilon [\chi^{-}_{\mu\nu},C],\cr
}
\eqn\butter
$$ and the action for the model is defined to be ${\cal
S}^{(2)}_c=\{Q,\Psi^{+}_c\}-
2\pi i k \tau$, with the gauge fermion (appropriately covariantized):
$$
\eqalign{
\Psi^{+}_c=&{1\over e^2}\int_X d^4 x\,\sqrt{g}\,\tr\,\bigl\{\,\,\half\chi^{+}_
{\mu\nu}
\,
\bigl (\,N^{+\mu\nu}+2F^{+\mu\nu}-4i[V^\mu,V^\nu]^{+}\,
\bigr )\cr &+{1\over2}\chi^{-}_{\mu\nu}\bigl
(\,N^{-\mu\nu}-4\raiz (\deriv^{[\mu}V^{\nu]}) ^{-}\,
\bigr ) +{1\over8}\tilde\eta\,\bigl (\,P+4\raiz \,\deriv_\mu V^\mu\,\bigr
)\,
\bigr \}
\cr 
&+{1\over e^2}\int_X d^4 x\,\sqrt{g}\,\tr\,\bigl\{\,{1\over\raiz}B\,\bigl 
(\,\deriv_\mu \psi^\mu +i\raiz [\tilde\psi_\mu,V^\mu]\,\bigr )\,\bigr\}\cr
&+{1\over e^2}\int_X d^4 x\,\sqrt{g}\,
\tr\,\bigr\{\,{i\over4}\eta [B,C]\,\bigr\}.
\cr }
\eqn\epi
$$ 
The resulting action reads:
$$
\eqalign { {\cal S}^{(2)}_c&= {1\over e^2}\int_X d^4 x\,\sqrt{g}\, \tr\, 
\bigl\{\,-\deriv_\mu
B\deriv^\mu C +\half N^{+}_{\mu\nu}\bigl (\,N^{+\mu\nu}+2F^{+\mu\nu}-
4i[V^\mu,V^\nu]^{+}\,
\bigr )\cr
&+\half N^{-}_{\mu\nu}\bigl (\,N^{-\mu\nu}-4\raiz (\deriv^{[\mu}V^{\nu]})
^{-}\,\bigr )
+{1\over8}P\bigl (\,P+4\raiz \,\deriv_\mu V^\mu\,\bigr )
\cr&+4\psi^\mu\deriv^\nu\chi^{+}_{\mu\nu}
+\psi^\mu\deriv_\mu\eta
+\tilde \psi^\mu\deriv_\mu\tilde\eta+
4\tilde\psi^\mu\deriv^\nu\chi^{-}_{\mu\nu}
 \cr &-i\raiz\,\chi^{+\mu\nu}[\chi^{+}_{\mu\nu},C]+
i\raiz\,\tilde\psi^\mu[\tilde\psi_\mu,B]-
4\raiz\,i\chi^{+}_{\mu\nu}[\tilde\psi^\mu,V^\nu]
+i\raiz\,\eta[\tilde\psi_\mu,V^\mu]\cr&
-{i\over{2\raiz}}\,\eta[\eta,C]+
i\raiz\,\psi_\mu[\psi^\mu,B]-
i\raiz\,\tilde\eta[\psi_\mu,V^\mu]+
4\raiz\,i\chi^{-}_{\mu\nu}[\psi^\mu,V^\nu]\cr&-
i\raiz\,\chi^{-}_{\mu\nu}[\chi^{-\mu\nu},C]
-{i\over{2\raiz}}\,\tilde\eta[\tilde\eta,C]-\half[B,C]^2
+2[B,V_\mu][C,V^\mu]
\,\bigr\}\cr &
-2\pi i\tau{1\over 32\pi^2}\int_X d^4 x\,\sqrt{g}\tr\,\bigl\{\, *  F_{\mu\nu}
F^{\mu\nu}\,\bigr\}.
\cr
}\eqn\mitocondria
$$
If we integrate out the auxiliary fields in \mitocondria\ we recover
the action \pLight. Some important issues relative to this theory 
will be addressed in sect. 4.

\endpage

%%%%%%%%%%%%%%%%%%%%%%%%%%%%%%%%%%%%%%%%%%%%%%%%%%%%%%%%%%%%%%%%%%%%%%%%%%%%%

%%%%%%%%%%%%%%%%%%%%%%%%%%%%%%%%%%%%%%%%%%%%%%%%%%%%%%%%%%%%%%%%%%%%%%%%%%%%%

\chapter{The Topological Actions in the Mathai-Quillen Approach}

In the first part of this paper we have reviewed in great detail 
the four dimensional topological field theories that can be obtained 
by twisting the symmetry group of the $N=4$ supersymmetric gauge theory. 
The twisting 
procedure has been repeatedly shown to be a very powerful technique 
for the construction of topological quantum field theories. However, 
it suffers from serious drawbacks, the main one being that it is not 
possible to identify from the very beginning the underlying 
geometrical structure that is involved. Rather, in most of the cases 
the underlying geometrical scenario is unveiled only after a careful 
 analysis with techniques borrowed from conventional quantum field 
theory is carried out [\tqft]. In what follows, we will change our 
scope and try to concentrate on the geometrical formulation of these 
theories. We will make use of the Mathai-Quillen formalism (see  
\REF\atiy{M. F. Atiyah, L. Jeffrey\journal\jgp &7(90)119.}
\REF\mathaiq{V. Mathai, D. Quillen\journal\topo &25(86)85.}
\REF\blau{M. Blau\journal\jgp&11(93)95.}
\REF\moore{S. Cordes, G. Moore, S. Rangoolam, ``Lectures on 2D Yang-Mills 
theory, equivariant cohomology and topological field theory", 
hep-th/9411210.}
\REF\coho{E. Witten\journal\ijmp ~A &6(91)2775.}
[\atiy-\coho] and references therein), which is very well suited 
for our purposes. Let us recall briefly the fundamentals of this 
approach. In the framework of topological quantum field theories of 
cohomological type [\coho], one deals with a certain set of fields 
(the field space, ${\cal M}$), on which  
the action of a  symmetry group, ${\cal G}$, which is usually a 
local symmetry group, is defined. An appropriate  set of  basic equations
 imposed on 
the fields single out a certain subset (the moduli space) of
${\cal M}/{\cal G}$. The topological quantum field theory associated to  
this moduli problem studies intersection theory on the corresponding 
moduli space. In this context, the Mathai-Quillen formalism involves 
the following steps. Given the field space ${\cal M}$, the basic 
equations of the problem are introduced as sections of an appropriate 
vector bundle ${\cal V}\to {\cal M}$, in such a way that the zero locus  
of these sections is precisely  the  relevant moduli space. The   
Mathai-Quillen  
 formalism  allows the computation of a certain representative 
of the Thom class of the vector 
bundle ${\cal V}$, which turns out to be the exponential of the action of the 
topological  field 
theory under consideration. The integration on ${\cal M}$ of the pullback 
under the sections of 
the Thom class of ${\cal V}$ gives the Euler characteristic of the bundle, 
which is the basic topological invariant associated to the moduli problem.

\section{The Vafa-Witten Problem}

In [\vafa], Vafa and Witten studied the partition function of the 
first of the twisted $N=4$ supersymmetric gauge theories we have
considered, namely that corresponding to the  defining embedding ${\bf 4}\to
({\bf 2},{\bf 1})\oplus ({\bf 2},{\bf 1})$  (see section $2$). They were able
to show that, in favourable conditions, the  partition function is the Euler
characteristic of instanton moduli space, and  then computed it on several
$4$-manifolds in order to make some non-trivial tests of S-duality. 
 The analysis starts from two  basic equations
involving the self-dual part of the gauge connection $F^{+}$, a
certain scalar field $C$ and a bosonic self-dual two-form $B^{+}$, all
taking values in the adjoint representation of some compact finite dimensional 
Lie group $G$. These equations are:   
$$
\cases{\deriv_\mu C+\raiz \deriv^\nu B^{+}_{\nu\mu}=0,\cr
F^{+}_{\mu\nu}-{i\over2}[B^{+}_{\mu\tau},B^{+\tau}{}_{\!\nu}]-{i\over\raiz}
[B^{+}_{\mu\nu},C]=0.\cr}
\eqn\equations
$$  
One can consider  the equations above as defining a certain moduli
problem, and our aim is to construct the topological quantum field theory
which corresponds to it  within the framework of the Mathai-Quillen
formalism. Our analysis will follow closely that in 
\REF\marlag{J. M. F. Labastida, M. Mari\~no\journal\pl&B351(95)146.}
[\marmon,\marth,\marlag].
Recently, this formalism has been applied to the twist  under consideration in
\REF\wang{Pei Wang\journal\pl&B378(96)147.} [\wang]. The construction
presented in  that work differs from ours in the role assigned to the field 
$C$.

\subsection{The topological framework}

The geometrical setting is a certain oriented, compact Riemannian 
four-manifold $X$, and the 
field space is ${\cal M}={\cal A}\times\Omega^0(X,\ad P)\times\Omega^{2,+}
(X,\ad P)$, where ${\cal A}$ is the space of connections on a principal 
$G$-bundle $P\to X$, and the second and third factors denote, respectively,
the
$0$-forms and self-dual differential forms of degree two on $X$ taking 
values in the Lie algebra of $G$. $\ad P$ denotes the adjoint bundle of 
$P$, $P\times_{\ad}\Lie$. The space of sections of this bundle,
$\Omega^0(X,\ad P)$, is the Lie algebra of the group 
${\cal G}$ of gauge transformations (vertical automorphisms)  of the bundle
$P$, whose action on the field space is given locally by:
$$
\eqalign{ g^{*}(A)&=i(dg)g^{-1}+gAg^{-1},\cr g^{*}(C)&=gCg^{-1},\cr
g^{*}(B^{+})&=gB^{+}g^{-1},\cr }
\eqn\jauje
$$  where $C\in \Omega^0(X,\ad P)$ and $B^{+}\in \Omega^{2,+}(X,\ad P)$. In
terms of the covariant derivative, $d_A =d+i[A,~]$, the infinitesimal form
of the transformations \jauje , with $g={\hbox{\rm exp}}(-i\phi)$ and
$\phi\in \Omega^0(X,\ad P)$, takes the form:
$$
\eqalign{
\delta_g(\phi)A&=d_A\phi,\cr
\delta_g(\phi)C&=i[C,\phi],\cr
\delta_g(\phi)B^{+}&=i[B^{+},\phi].\cr }
\eqn\jauja
$$  
The tangent space to the field space at the point $(A,C,B^{+})$ is the
vector space $T_{(A,C,B^{+})}{\cal M}=\Omega^1(X,\ad P)\oplus\Omega^0(X,\ad
P)\oplus\Omega^{2,+} (X,\ad P)$. On
$T_{(A,C,B^{+})}{\cal M}$ we can define a gauge-invariant Riemannian metric
(inherited from that on $X$) as follows:
$$
\langle (\psi,\zeta,\tilde\psi^{+}),(\theta,\xi,\tilde\omega^{+})\rangle
=\int_X \tr(\psi\wedge *\theta)+\int_X \tr(\zeta\wedge *\xi)+
\int_X \tr(\tilde\psi^{+}\wedge *\tilde\omega^{+})
\eqn\metrica
$$  
where $\psi,\theta\in\Omega^1(X,\ad P)$, $\zeta,\xi\in\Omega^0(X,\ad P)$
and $\tilde\psi^{+},\tilde\omega^{+}\in \Omega^{2,+} (X,\ad P)$. 

To introduce the basic equations \equations\ in this framework we proceed
as follows. On the field space
$\mani$ we build a trivial vector bundle ${\cal V}=
\mani\times{\cal F}$, where the fibre is in this case 
 ${\cal F}= \Omega^1(X,\ad P)\oplus\Omega^{2,+} (X,\ad P)$. The basic
equations \equations\ can then  be identified to be a section $s:{\cal
M}\to {\cal V}$ of the vector bundle ${\cal V}$. In our case, the section
reads, with a certain choice  of normalization that makes easier the
comparison with the results in  sect. $2$:
$$  s(A,C,B^{+}) =\bigl (\raiz(\deriv_\mu C+\raiz \deriv^\nu
B^{+}_{\nu\mu}),~-2(
F^{+}_{\mu\nu}-{i\over2}[B^{+}_{\mu\tau},B^{+\tau}{}_{\!\nu}]-{i\over\raiz}
[B^{+}_{\mu\nu},C])\bigr ).
\eqn\seccion
$$  
Notice that this section is gauge ad-equivariant, and the zero locus of
the associated section  
${\tilde s}:{\cal M}/{\cal G}\to {\cal V}/{\cal G}$ gives precisely the 
moduli space of the topological theory. It would be desirable to compute 
the dimension of this moduli space. The best we can do is to build the 
corresponding deformation complex whose index is known to compute, under
certain assumptions, the dimension of the tangent space to the moduli
space. This index provides what is called the virtual dimension of the moduli
space. The deformation complex that corresponds to our moduli space is the
following:
$$
\eqalign{ 0\too\Omega^0(X,\ad P)&\mapright{{\cal C}}\Omega^1(X,\ad
P)\oplus\Omega^0(X,\ad P)\oplus\Omega^{2,+} (X,\ad P)\cr
&\mapright{ds}\Omega^1(X,\ad P)\oplus\Omega^{2,+}(X,\ad P)\too 0.\cr
}\eqn\complejo
$$  The map ${\cal C}:\Omega^0(X,\ad P)\too T{\cal M}$, given by (recall
\jauja):
$$  {\cal C}(\phi)=(d_A\phi,i[C,\phi],i[B^{+},\phi]),\quad
\phi\in\Omega^0(X,\ad P),
\eqn\butraguenho
$$
defines the vertical tangent space (gauge orbits) to the principal 
${\cal G}$-bundle. The map $ds:T_{(A,C,B^{+})}{\cal M}\too {\cal
F}$ is  given by the linearization of the basic equations \equations :
$$
\eqalign{ ds(\psi,\zeta,\tilde\psi^{+})=\biggl (\raiz(\deriv_\mu
\zeta+i[\psi_\mu,C]+\raiz
\deriv^\nu \tilde\psi^{+}_{\nu\mu}+ i\raiz[\psi^\nu,B^{+}_{\nu\mu}]),\cr
-2\bigl(\,2(\deriv_{[\mu}\psi_{\nu]})^{+}+i[\tilde\psi^{+}_{\tau[\mu},
B^{+\tau}{}_{\!\nu]}]-{i\over\raiz}[\tilde\psi^{+}_{\mu\nu},C]
-{i\over\raiz}[B^{+}_{\mu\nu},\zeta]\,\bigr)\,\biggr ).\cr}
\eqn\lordaeron
$$  
Under suitable conditions (which happen to be the same vanishing
theorems   discussed in [\vafa]), the index of the complex \complejo\
computes de  dimension of ${\hbox{\rm Ker}}(ds)/{\hbox{\rm Im}}({\cal C})$,
that is, the dimension of the moduli space under consideration. To calculate
its index, the complex \complejo\ can be split up into the
Atiyah-Hitchin-Singer instanton deformation complex  
\REF\atiyah{M. F. Atiyah, N. J. Hitchin, I. M. Singer\journal\prsl ~A 
&362(78)425.} 
[\atiyah] for
anti-self-dual (ASD) connections,
$$  
(1)~0\too\Omega^0(X,\ad P)\mapright{d_{\!A}}\Omega^1(X,\ad P)
\mapright{p^{+}\!d_{\!A}}\Omega^{2,+}(X,\ad P)\too 0,
\eqn\ahs
$$ 
and the complex associated to the operator,
$$  
(2)~D=p^{+}\!d^{*}_A+d_A:\Omega^0(X,\ad P)\oplus\Omega^{2,+} (X,\ad P)
\too\Omega^1(X,\ad P),
\eqn\fanny
$$ 
which is also the ASD instanton deformation complex. They contribute with
opposite signs and therefore the net contribution to the index is zero, leaving
us with the result that the virtual dimension of the moduli space is zero.
\vskip 1cm
\subsection{The topological action}

We now proceed to construct the topological action associated to this
moduli problem, and we will do it within the Mathai-Quillen formalism.  The
Mathai-Quillen form gives a representative of the Thom class of the  bundle
${\cal E}={\cal M}\times_{\cal G} {\cal F}$, and the integration over
${\cal M}/{\cal G}$ of the pullback of this Thom class under the section
$\tilde s:{\cal M}/{\cal G}\to {\cal E}={\cal M}\times_{\cal G} {\cal F}$
gives the (generalized) Euler characteristic of ${\cal E}$, which is to be
identified, from the field-theory point of view, with the partition
function of the associated topological quantum field theory. 

As a first step to construct the topological theory which corresponds to
the moduli problem defined by the basic equations \equations , we have to
give explicitly the field content and the BRST symmetry of the theory. This
will help to clarify the structure of the topological multiplet we
introduced in sect. $2$. In the field space ${\cal M}= {\cal
A}\times\Omega^0(X,\ad P)\times\Omega^{2,+} (X,\ad P)$ we have the gauge
connection $A_\mu$, the scalar field $C$ and the self-dual two-form
$B^{+}_{\mu\nu}$, all with ghost number $0$. In the (co)tangent space
$T_{(A,C,B^{+})}{\cal M}=\Omega^1(X,\ad P)\oplus\Omega^0(X,\ad
P)\oplus\Omega^{2,+} (X,\ad P)$ we have the anticommuting fields
$\psi_\mu$, $\zeta$ and $\tilde\psi^{+}_{\mu\nu}$, all with ghost number
$1$ and which are to be interpreted as differential forms on the moduli
space. In the fibre ${\cal F}= \Omega^1(X,\ad P)\oplus\Omega^{2,+} (X,\ad
P)$ we have anticommuting fields with the quantum numbers of the equations,
namely a one-form 
$\tilde\chi_\mu$ and a self-dual two-form $\chi^{+}_{\mu\nu}$, both with
ghost number $-1$, and their superpartners, a commuting one-form $\tilde 
H_\nu$ and a commuting self-dual two-form $H^{+}_{\mu\nu}$, both with ghost
number $0$ and which appear as auxiliary fields in the associated  field
theory. And finally, associated to the gauge symmetry, we have  a
commuting scalar field $\phi\in\Omega^{0}(X,\ad P)$ with ghost number
$+2$ [\coho], and a multiplet of scalar fields $\bar\phi$ (commuting and
with ghost number $-2$) and $\eta$ (anticommuting and with ghost number
$-1$), both also in $\Omega^{0}(X,\ad P)$ and which enforce the horizontal
projection ${\cal M}\to {\cal M}/{\cal G}$ [\moore]. The BRST symmetry of
the model is given by:
$$
\eqalign{  [Q, A_\mu] &=\psi_\mu,\cr [Q,C]&=\zeta,\cr
[Q,B^{+}_{\mu\nu}]&=\tilde\psi^{+}_{\mu\nu},\cr [Q,\phi]&=0,\cr
\{Q,\tilde\chi_{\mu}\} &= \tilde H_{\mu},\cr
\{Q,\chi^{+}_{\mu\nu}\} &= H^{+}_{\mu\nu},\cr
[Q,\bar\phi]&=\eta,\cr}\qquad\qquad
\eqalign{
\{Q,\psi_{\mu}\} &= \deriv_{\mu}\phi, \cr
\{Q,\zeta\,\} &=i\,[C,\phi],\cr
\{Q,\tilde\psi^{+}_{\mu\nu}\}&=i\,[B^{+}_{\mu\nu},\phi],\cr\cr
[Q,\tilde H_{\mu}]&=i\,[\tilde\chi_{\mu},\phi],\cr  
[Q,H^{+}_{\mu\nu}]&=i\,[\chi^{+}_{\mu\nu},\phi],\cr 
\{Q,\eta\,\} &=i\,[\bar\phi,\phi].\cr}
\eqn\macarena
$$  The BRST generator $Q$ satisfies the algebra 
$\{Q,Q\}=\delta_g(\phi)$,  and can be seen to correspond to the Cartan
model for the ${\cal G
}$-equivariant cohomology of ${\cal M}$ .

We are now ready to write the action for the topological field theory under
consideration. Instead of writing the full expression for the
Mathai-Quillen form, we define the action to be $\{Q,\Psi\}$ for some
appropriate gauge invariant gauge fermion $\Psi$ [\moore]. The use of gauge
fermions was introduced in the context of topological quantum field theory in
\REF\perni{J. M. F. Labastida, M. Pernici\journal\pl&B212(88)56.}
[\perni] (see [\phyrep] for a review). As it is explained in detail in
[\moore], the gauge fermion consists of two basic pieces,  a localization 
gauge fermion, which essentially involves the equations defining the moduli
problem and which in our case takes the form:
$$
\eqalign{
\Psi_{\hbox{\sevenrm loc}}&=\langle(\tilde\chi,\chi^{+}),s(A,C,B^{+})
\rangle +\langle (\tilde\chi,\chi^{+}),(\tilde H,H^{+})\rangle=\cr &\int_X
\sqrt{g}\,\tr\,\bigl\{\,\,\half\chi^{+}_{\mu\nu}\bigl (\,H^{+\mu\nu} -2(
F^{+\mu\nu}-{i\over2}[B^{+\mu\tau},B^{+}_\tau{}^{\!\nu}]-{i\over\raiz}
[B^{+\mu\nu},C])\bigr )\cr &+\tilde\chi_\mu\bigl (\,\tilde H^\mu+\raiz
\,(\deriv^\mu C+\raiz\deriv _\nu B^{+\nu\mu})\,\bigr )\,\bigr\},\cr
}
\eqn\localiz
$$  
and a projection gauge fermion, which enforces the horizontal
projection,
 and which can be written as:
$$
\Psi_{\hbox{\sevenrm proj}}=\langle\bar\phi,{\cal C}^{\dag}(\psi,
\zeta,\tilde\psi)\rangle_{\hbox{\bf g}},
\eqn\project
$$   
where $\langle,\rangle_{\hbox{\bf g}}$ denotes the gauge invariant 
metric in $\Omega^0(X,\ad P)$, and the map 
${\cal C}^{\dag}:T{\cal M}\to \Omega^0(X,\ad P)$ is the adjoint of  the map
${\cal C}$ \butraguenho\ with respect to the Riemannian  metrics
\metrica\ in $T{\cal M}$ and $ \Omega^0(X,\ad P)$. Since 
${\cal C}(\phi),~\phi\in  \Omega^0(X,\ad P)$, is given by 
\butraguenho , its adjoint is readily computed to be:
$$ 
 {\cal C}^{+}(\psi,\zeta,\tilde\psi)=-\deriv_\mu \psi^\mu 
+{i\over2}[\tilde\psi^{+}_{\mu\nu},B^{+\mu\nu}]+i[\zeta,C],
\eqn\adjoint
$$ 
  where $(\psi,\zeta,\tilde\psi)\in T_{(A,C,B^{+})}{\cal M}$. This 
leaves for the projection fermion \project\ the expression:
$$
\Psi_{\hbox{\sevenrm proj}}=\int_X \sqrt{g}\,\tr\,\bigl\{\,
\bar\phi\bigl (\,-\deriv_\mu\psi^\mu
+{i\over2}[\tilde\psi^{+}_{\mu\nu}, B^{+\mu\nu}]+i[\zeta,C]\,\bigr )\,\bigr\}.
\eqn\projectn
$$

In the Mathai-Quillen formalism the action is built out of the terms
\localiz\ and \projectn.  However, as in the case of the Mathai-Quillen
formulation of Donaldson-Witten  theory  [\atiy], one must add another piece to
the gauge fermion to make full contact with the corresponding twisted
supersymmetric theory. In  our case, this extra term is:
$$
\Psi_{\hbox{\sevenrm{extra}}}=-\int_X \sqrt{g}\,\tr\,\bigl\{\,
{i\over2}\eta[\phi,\bar\phi]\,\bigr\}.
\eqn\extra
$$

It is now straightforward to see that after the rescalings
$$
\eqalign{ A'&=A,\cr
\psi'&=-{1\over2}\psi,\cr
\phi'&={1\over{2\raiz}}\phi,\cr
\bar\phi'&=-2\raiz \bar\phi,\cr
\eta'&=-2\eta,\cr}
\qquad\qquad
\eqalign{  
C'&=-{1\over\raiz}C,\cr
\zeta'&=-\zeta,\cr  
B^{+'}&=\half B^{+},\cr
\tilde\psi^{+'}&={1\over{2\raiz}}\tilde\psi^{+},\cr  
}\qquad\qquad
\eqalign{
\tilde\chi'&=\raiz\tilde\chi,\cr
\tilde H'&=\raiz H,\cr
\chi^{+'}&=\chi^{+},\cr  
H^{+'}&=H^{+},\cr}
\eqn\redef
$$ 
one recovers, in terms of the primed fields,  the twisted model 
we analyzed in sect. $2$ and that is encoded in  
\papagayo\ and \mermelada.

\section{Adjoint Non-Abelian Monopoles}

As we saw before, the model arising from the second twist is 
equivalent to the theory of non-abelian monopoles discussed at length 
in [\marmon-\zzeta].   
The relevant basic equations for this  model  involve the self-dual part of
the gauge connection $F^{+}$ and a certain complex spinor field $M$ 
taking values in the adjoint representation of some compact finite dimensional 
Lie group $G$:  
$$
\cases{ F^{+}_{\alpha\beta}+[\overline M_{(\alpha},M_{\beta)}]=0,\cr
\deriv_{\alpha\dalpha}M^{\alpha}=0,\cr
}
\eqn\marinho
$$ 
where $\overline M$ is the complex conjugate of $M$. 

\subsection{The topological framework}

The geometrical setting is a certain oriented, closed Riemannian 
four-manifold $X$, that we will also assume to be spin. We will denote  the
positive and negative chirality spin bundles by $S^{+}$ and $S^{-}$ 
respectively. The   field space is ${\cal M}={\cal
A}\times\Gamma(X,S^{+}\otimes\ad P)$,   where ${\cal A}$ is the space of
connections on a principal 
$G$-bundle $P\to X$, and the second factor denotes the space of sections of
the product bundle $S^{+}\otimes \ad P$, that is,  positive chirality spinors
taking values in the Lie algebra of  the gauge group.  The group
${\cal G}$ of gauge transformations  of the bundle $P$ has an action on the
field space which is given locally by:
$$
\eqalign{ g^{*}(A)&=i(dg)g^{-1}+gAg^{-1},\cr  
g^{*}(M)&=gMg^{-1},\cr }
\eqn\mahler
$$  
where $M\in \Gamma(X,S^{+}\otimes\ad P)$ and $A$ is the gauge connection. 
In
terms of the covariant derivative $d_A =d+i[A,~]$, the infinitesimal form
of the transformations \mahler , with $g={\hbox{\rm exp}}(-i\phi)$ and
$\phi\in \Omega^0(X,\ad P)$, takes the form:
$$
\eqalign{
\delta_g(\phi)A&=d_A \phi,\cr
\delta_g(\phi)M&=i[M,\phi].\cr }
\eqn\mohler
$$  
The tangent space to the field space at the point $(A,M)$ is the vector
space $T_{(A,M)}{\cal M}=\Omega^1(X,\ad P)\oplus\Gamma(X,S^{+}\otimes\ad
P)$. On
$T_{(A,M)}{\cal M}$ we can define a gauge-invariant Riemannian metric
given by:
$$
\langle (\psi,\mu),(\theta,\omega)\rangle =\int_X
\tr\,(\psi\wedge *\theta)+\half\int_X \tr\,(\bar\mu^\alpha\omega_\alpha+\bar
\omega^\alpha\mu_\alpha),
\eqn\mmetrica
$$  
where $\psi,\theta\in\Omega^1(X,\ad P)$ and
$\mu,\omega\in \Gamma(X,S^{+}\otimes\ad P)$. 

The basic equations \marinho\ are introduced in this framework  as
sections of the trivial vector bundle 
${\cal V}=
\mani\times{\cal F}$, where the fibre is in this case 
 ${\cal F}= \Omega^{2,+}(X,\ad P)\oplus\Gamma (X,S^{-}\otimes\ad P)$. 
Taking into account the form of the basic  equations,
the section reads, up to some harmless normalization factors that we 
introduce for reasons that will become apparent soon:
$$  
s(A,M) =\bigl ( -2(F^{+}_{\alpha\beta}+[\overline M_{(\alpha},M_
{\beta)}]),~
 \raiz\deriv_{\alpha\dalpha}M^{\alpha}\,
\bigr ).
\eqn\secante
$$  
The section \secante\ can be alternatively seen as a gauge ad-equivariant 
map from the principal ${\cal G}$-bundle ${\cal M}\to {\cal M}/{\cal G}$ to  
the vector space ${\cal F}$, and in this way it descends naturally to a 
section 
${\tilde s}$ of the associated vector bundle ${\cal M}\times_{\cal G} {\cal
F}$, whose zero locus  gives precisely the  moduli space of the 
topological
theory. It would be desirable to compute  the dimension of this moduli
space. The relevant deformation complex (which allows one to compute, in a 
general situation, the virtual dimension of the moduli space) is the
following:
$$
\eqalign{  0&\too\Omega^0(X,\ad P)\mapright{{\cal C}}\Omega^1(X,\ad
P)\oplus\Gamma(X,S^{+}\otimes\ad P)\cr &\mapright{ds}\Omega^{2,+}(X,\ad
P)\oplus\Gamma(X,S^{-}\otimes\ad P).\cr}
\eqn\complex
$$  
The map ${\cal C}:\Omega^0(X,\ad P)\too T{\cal M}$ is given by:
$$  
{\cal C}(\phi)=(d_A \phi,i[M,\phi]),\quad \phi\in\Omega^0(X,\ad P),
\eqn\barbarella
$$ 
while the map $ds:T_{(A,M)}{\cal M}\too {\cal F}$ is  provided by the
linearization of the basic equations 
\marinho:
$$
\eqalign{ ds(\psi,\mu)=\bigl
(-4\sigma^{\mu\nu}_{\!\alpha\beta}\deriv_{\mu}\psi_{\nu}
-2[\bar\mu_{(\alpha}, M_{\beta)}]-2[\overline M_{(\alpha},\mu_{\beta)}],\cr
\raiz\deriv_{\alpha\dalpha}\mu^{\alpha}+\raiz
[\psi_{\alpha\dalpha}, M^{\alpha}]\bigr ).
\cr}
\eqn\mozarella
$$  
Under suitable conditions, the index of the complex \complex\ computes
de  dimension of ${\hbox{\rm Ker}}(ds)/{\hbox{\rm Im}}({\cal C})$.
 To calculate the index,
the complex \complex\ can be split up into the ASD-instanton deformation
complex:
$$  
(1)~0\too\Omega^0(X,\ad P)\mapright{d_{\!A}}\Omega^1(X,\ad P)
\mapright{p^{+}\!d_{\!A}}\Omega^{2,+}(X,\ad P)\too 0,
\eqn\hitchin
$$  
whose index is $p_1(\ad P)+{\hbox{\rm dim}}(G)(\chi+\sigma)/2$, 
being $p_1(\ad P)$ the first Pontryagin class of the adjoint bundle 
$\ad P$,
and the complex associated to the twisted Dirac operator  
$$  
(2)~D:\Gamma(X,S^{+}\otimes\ad P)\too
\Gamma(X,S^{-}\otimes\ad P),
\eqn\francis
$$  
whose index is $p_1(\ad P)/2 - {\hbox{\rm dim}}(G)\sigma/8$.
Thus, the index of the total complex 
(which gives minus the virtual dimension of the moduli space) is:
$$
\eqalign{ -&{\hbox{\rm dim}}({\cal M})=\ind(1)-2\times \ind(2)= {\hbox{\rm
dim}}(G){(2\chi+3\sigma)\over4}
\cr }
\eqn\indices
$$ 
where $\chi$ is the Euler characteristic of the $4$-manifold $X$ and 
$\sigma$ is its signature. The factor of two appears in \indices\ since we want 
to compute the real dimension of the moduli space.

\vskip 1cm
\subsection{The topological action}

We now proceed as in the previous case.
To build a topological theory out of the moduli problem defined by  the
equations \marinho\ we need the following   multiplet of fields. For
the field space ${\cal M}= {\cal A}\times\Gamma(X,S^{+}\otimes\ad P)$ we
introduce commuting fields $(A,M)$,  both with ghost number $0$, and their 
corresponding superpartners, 
the anticommuting fields
$\psi$ and $\mu$, both with ghost number
$1$. For
the fibre ${\cal F}= \Omega^{2,+}(X,\ad P)\oplus\Gamma (X,S^{-}\otimes\ad
P)$ we introduce anticommuting fields 
$\chi^{+}$ and  $\nu$ respectively, 
both with ghost number $-1$, and their superpartners, a commuting
self-dual two-form
$H^{+}$ and  a commuting negative chirality spinor 
$h$, both with ghost number
$0$ and which appear as auxiliary fields in the associated  field theory.
And finally, associated to the gauge symmetry, we have  a commuting 
scalar field
$\phi\in\Omega^{0}(X,\ad P)$ with ghost number
$+2$, and a multiplet of scalar fields $\lambda$ (commuting and with
ghost number $-2$) and $\eta$ (anticommuting and with ghost number
$-1$), both also in $\Omega^{0}(X,\ad P)$ and which enforce the horizontal
projection ${\cal M}\to {\cal M}/{\cal G}$. The BRST symmetry of
the model is given by:
$$
\eqalign{  
[Q, A_\mu] &=\psi_\mu,\cr 
[Q,M_{\alpha}]&=\mu_{\alpha},\cr
[Q,\phi]&=0,\cr
\{Q,\chi^{+}_{\alpha\beta}\}&=H^{+}_{\alpha\beta},\cr 
\{Q,\nu_{\dalpha}\} &= h_{\dalpha},\cr
[Q,\lambda]&=\eta,\cr
}\qquad\qquad
\eqalign{
\{Q,\psi_{\mu}\} &= \deriv_{\mu}\phi, \cr
\{Q,\mu_{\alpha}\,\} &=i\,[M_{\alpha},\phi],\cr\cr
[Q,H^{+}_{\alpha\beta}]&=i\,[\chi^{+}_{\alpha\beta},\phi],\cr  
[Q,h_{\dalpha}]&= i\,[\nu_{\dalpha},\phi],\cr  
\{Q,\eta\,\} &=i\,[\lambda,\phi].\cr}
\eqn\darek
$$  
This BRST algebra closes up to a gauge transformation generated  by
$\phi$. 

We have to give now the expressions for the different pieces of the  gauge
fermion. For the localization gauge fermion we have:
$$
\eqalign{
\Psi_{\hbox{\sevenrm loc}}&=i\langle(\chi^{+},\nu),s(A,M)
\rangle -\langle (\chi^{+},\nu),( H^{+},h)\rangle= 
\cr &\int_X
\sqrt{g}\,\tr\,\bigl\{\,{1\over4}\chi^{+}_{\alpha\beta}\bigl (\,
H^{+\alpha\beta}
+2i( F^{+\alpha\beta}+[\overline M^{(\alpha},M^{\beta)}])\,\bigr )\cr &
+\half\bar\nu^{\dalpha}\bigl (\,h_{\dalpha}-i\raiz\deriv_{\alpha\dalpha}M^
{\alpha}\bigr )
-\half\nu^{\dalpha}\bigl (\,\bar h_{\dalpha}-i\raiz\deriv_{\alpha\dalpha}
\overline M^{\alpha}\bigr )
\,\bigr \},\cr 
}\eqn\local
$$  
and for the projection gauge fermion, which enforces the horizontal
projection,
$$
\eqalign{
\Psi_{\hbox{\sevenrm proj}}&=\langle\lambda,{\cal C}^{\dag}(\psi,
\mu)\rangle_{\hbox{\bf g}}=\cr &
\int_X \sqrt{g}\,\tr\,\bigl\{\, \lambda\bigl (\,-\deriv_\mu\psi^\mu
+{i\over2}[\bar\mu^{\alpha}, M_{\alpha}]-
{i\over2}[\overline M^{\alpha}, \mu_{\alpha}]
\,\bigr )\,\bigr\}.\cr
}
\eqn\projecting
$$

As in the previous case, it is necessary to add an extra 
 piece to the gauge fermion to make full contact with
the corresponding twisted
supersymmetric theory. In  this case, this extra term is:
$$
\Psi_{\hbox{\sevenrm{extra}}}=-\int_X \sqrt{g}\,\tr\,\bigl\{\,
{i\over2}\lambda[\eta,\phi]\,\bigr\}.
\eqn\stranger
$$

It is now straightforward to see that, after making the following 
redefinitions,
$$
\eqalign{ A'&=A,\cr
\psi'&={1\over2}\psi,\cr 
\phi'&={1\over{2\raiz}}\phi,\cr
\lambda'&=-2\raiz \lambda,\cr
\eta'&=-2\eta,\cr}
\qquad\qquad
\eqalign{  
M'&=M,\cr
\overline M'&=\half \overline M,\cr
\mu'&=-{1\over\raiz}\mu,\cr
\bar\mu'&=-{1\over{2\raiz}}\bar\mu,\cr
\chi'^{+}&= \chi^{+},\cr 
}\qquad\qquad
\eqalign{
H'^{+}&=H^{+},\cr  
\nu'&=-2\nu,\cr
\bar\nu'&=-\bar\nu,\cr
h'&=2h,\cr 
\bar h'&=-\bar h,\cr
}
\eqn\dolphin
$$ 
one recovers, in terms of the primed fields,  the twisted model 
summarized in \coventry\ and \covMedeiros .

%\section{What is actually behind the third theory?}

\section{The Amphicheiral Theory}  

The relevant basic equations for this  model 
involve the self-dual part of the gauge connection $F^{+}$ and a
certain real vector field $V_\mu$  taking values in the adjoint 
representation
of some finite dimensional  compact Lie group
$G$:
$$
\cases{ F^{+}_{\mu\nu}-i[V_{\mu},V_{\nu}]^{+}=0,\cr
\bigl (\,\deriv_{[\mu}V_{\nu]}\,\bigr )^{-}=0,\cr
\deriv_\mu V^\mu=0.\cr}
\eqn\morequations
$$ 
%Despite of what is said in [\marcus], we have not been able to identify 
%clearly the geometrical problem that underlies the basic equations  above.
%However, we shall assume that this equations are given and try  to 
%write down explicitly the  topological field theory that is naturally 
%associated to them  in the context of the Mathai-Quillen formalism. 

\subsection{The topological framework}

The geometrical setting is a certain compact, oriented Riemannian 
four-manifold $X$, and the 
field space is ${\cal M}={\cal A}\times\Omega^1(X,\ad P)$,  where ${\cal
A}$ is the space of connections on a principal 
$G$-bundle $P\to X$, and the second factor denotes, as we have  already
seen before, 
$1$-forms on $X$ taking  values in the Lie algebra of $G$.   The group
${\cal G}$ of gauge transformations  of the bundle $P$ has an action on the
field space which is given locally by:
$$
\eqalign{ g^{*}(A)&=i(dg)g^{-1}+gAg^{-1},\cr  g^{*}(V)&=gVg^{-1},\cr }
\eqn\mjauje
$$  where $V\in \Omega^1(X,\ad P)$ and $A$ is the gauge connection. In
terms of the covariant derivative $d_A =d+i[A,~]$, the infinitesimal form
of the transformations \mjauje , with $g={\hbox{\rm exp}}(-iC)$ and
$C\in \Omega^0(X,\ad P)$, takes the form:
$$
\eqalign{
\delta_g(C)A&=d_A C,\cr
\delta_g(C)V&=i[V,C].\cr }
\eqn\mjauja
$$  The tangent space to the field space at the point $(A,V)$ is the vector
space $T_{(A,V)}{\cal M}=\Omega^1_{(A)}(X,\ad P)\oplus\Omega^1_{(V)}(X,\ad
P)$, where $\Omega^1_{(A)}(X,\ad P)$ denotes the tangent space to ${\cal
A}$ at $A$, and 
$\Omega^1_{(V)}(X,\ad P)$ denotes the tangent space to 
$\Omega^1(X,\ad P)$ at $V$. On
$T_{(A,V)}{\cal M}$, the  gauge-invariant Riemannian metric
(inherited from that on $X$) is defined as:
$$
\langle (\psi,\tilde\psi),(\theta,\tilde\omega)\rangle =\int_X
\tr(\psi\wedge *\theta)+\int_X \tr (\tilde\psi\wedge *\tilde\omega)
\eqn\mmetrica
$$  where $\psi,\theta\in\Omega^1_{(A)}(X,\ad P)$ and
$\tilde\psi,\tilde\omega\in \Omega^{1}_{(V)} (X,\ad P)$. 

The basic equations \morequations\ are introduced in this framework  as
sections of the trivial vector bundle 
${\cal V}=
\mani\times{\cal F}$, where the fibre is in this case 
 ${\cal F}= \Omega^{2,+}(X,\ad P)\oplus\Omega^{2,-} (X,\ad P)\oplus
\Omega^0(X,\ad P)$. Taking into account the form of the basic  equations,
the section reads:
$$  
s(A,V) =\bigl ( -2(F^{+}_{\mu\nu}-i[V_{\mu},V_{\nu}]^{+}),~
 4(\,\deriv_{[\mu}V_{\nu]}\,)^{-},
\raiz\deriv_\mu V^\mu
\bigr ).
\eqn\mseccion
$$  
The section \mseccion, being gauge ad-equivariant, descends to a 
section
${\tilde s}$ of the associated vector bundle ${\cal M}\times_{\cal G} {\cal
F}$ whose zero locus  gives precisely the  moduli space of the topological
theory. It would be desirable to compute  the dimension of this moduli
space. The relevant deformation complex   is the following:
$$
\eqalign{  0&\too\Omega^0(X,\ad P)\mapright{{\cal C}}\Omega^1_{(A)}(X,\ad
P)\oplus\Omega^1_{(V)}(X,\ad P)\cr &\mapright{ds}\Omega^{2,+}(X,\ad
P)\oplus\Omega^{2,-}(X,\ad P)\oplus
\Omega^0(X,\ad P)\too 0.\cr }
\eqn\complexo
$$  
The map ${\cal C}:\Omega^0(X,\ad P)\too T{\cal M}$ is given by:
$$  
{\cal C}(C)=(d_A C,i[V,C]),\quad C\in\Omega^0(X,\ad P),
\eqn\barbacoa
$$ 
while the map $ds:T_{(A,V)}{\cal M}\too {\cal F}$ is  given by the
linearization of the basic equations 
\morequations:
$$
\eqalign{ &ds(\psi,\tilde\psi)=\bigl (-4(\deriv_{[\mu}\psi_{\nu]})^{+}
+4i[\tilde\psi_{[\mu}, V_{\nu]}]^{+},\cr&
4(\deriv_{[\mu}\tilde\psi_{\nu]})^{-} +4i[\psi_{[\mu},
V_{\nu]}]^{-},\raiz\deriv_\mu\tilde\psi^\mu +\raiz i[\psi_\mu ,V^\mu]\bigr ).
\cr}
\eqn\casper
$$  
Under suitable conditions, the index of the complex \complexo\ computes
the  dimension of ${\hbox{\rm Ker}}(ds)/{\hbox{\rm Im}}({\cal C})$. To
calculate its index, the complex \complejo\ can be split up into the
ASD-instanton deformation complex:
$$  
(1)~0\too\Omega^0(X,\ad P)\mapright{d_{\!A}}\Omega^1(X,\ad P)
\mapright{p^{+}\!d_{\!A}}\Omega^{2,+}(X,\ad P)\too 0,
\eqn\athisng
$$ 
 and the complex associated to the operator 
$$  
(2)~D=p^{-}\!d_A+d^{*}_{\!A}:\Omega^1(X,\ad P)\too
\Omega^0(X,\ad P)\oplus\Omega^{2,-} (X,\ad P),
\eqn\kouba
$$  
which is easily seen to correspond to the instanton deformation complex 
for self-dual (SD) connections. Thus, the index of the total complex 
(which gives minus the virtual dimension of the moduli space) is:
$$
\eqalign{ -&{\hbox{\rm dim}}({\cal M})=\ind(1)-\ind(2)=\ind({\hbox{\rm ASD}})+ 
\ind({\hbox{\rm SD}})=\cr  &= p_1(\ad P)+\half{\hbox{\rm dim}}(G)(\chi
+\sigma) - p_1(\ad P)+\half{\hbox{\rm dim}}(G)(\chi-\sigma)= {\hbox{\rm
dim}}(G)\chi,
\cr }
\eqn\indice
$$ 
where $p_1(\ad P)$ is the first Pontryagin class of the adjoint bundle 
$\ad P$, $\chi$ is the Euler characteristic of the $4$-manifold $X$ and 
$\sigma$ is its signature. 

\vskip 1cm
\subsection{The topological action}

We now proceed as in the previous cases.
To build a topological theory out of the moduli problem defined by  the
equations \morequations , we need the following   multiplet of fields. For
the field space ${\cal M}= {\cal A}\times\Omega^1(X,\ad P)$ we introduce the
gauge connection $A_\mu$ and the one-form
$V_{\mu}$, both commuting and with ghost number $0$. For the (co)tangent 
space
$T_{(A,V)}{\cal M}=\Omega^1_{(A)}(X,\ad P)\oplus\Omega^1_{(V)}(X,\ad P)$ we
introduce the anticommuting fields
$\psi_\mu$ and $\tilde\psi_{\mu}$, both with ghost number
$1$ and which can  be seen as differential forms on the moduli space. For
the fibre ${\cal F}= \Omega^{2,+}(X,\ad P)\oplus\Omega^{2,-} (X,\ad
P)\oplus\Omega^0(X,\ad P)$ we have anticommuting fields with the quantum
numbers  of the equations, namely a  self-dual two-form
$\chi^{+}_{\mu\nu}$,  an  anti-self-dual two-form $\chi^{-}_{\mu\nu}$ and a
$0$-form $\tilde
\eta$, all with ghost number $-1$, and their superpartners, a commuting
self-dual two-form
$N^{+}_{\mu\nu}$,  a commuting anti-self-dual two-form $N^{-}_{\mu\nu}$ and
a commuting $0$-form $P$, all with ghost number $0$ and which appear as
auxiliary fields in the associated  field theory. And finally, associated
to the gauge symmetry, we introduce  a commuting scalar field
$C\in\Omega^{0}(X,\ad P)$ with ghost number
$+2$, and a multiplet of scalar fields $B$ (commuting and with
ghost number $-2$) and $\eta$ (anticommuting and with ghost number
$-1$), both also in $\Omega^{0}(X,\ad P)$ and which enforce the horizontal
projection ${\cal M}\to {\cal M}/{\cal G}$ [\moore]. The BRST symmetry of
the model is given by:
$$
\eqalign{  [Q, A_\mu] &=\psi_\mu,\cr [Q,V_\mu]&=\tilde\psi_\mu,\cr
[Q,C]&=0,\cr
\{Q,\chi^{+}_{\mu\nu}\}&=N^{+}_{\mu\nu},\cr 
\{Q,\chi^{-}_{\mu\nu}\} &= N^{-}_{\mu\nu},\cr
\{Q,\tilde\eta\} &= P,\cr [Q,B]&=\eta,\cr}\qquad\qquad
\eqalign{
\{Q,\psi_{\mu}\} &= \deriv_{\mu}C, \cr
\{Q,\tilde\psi_\mu\,\} &=i\,[V_\mu,C],\cr\cr
 [Q,N^{+}_{\mu\nu}]&=i\,[\chi^{+}_{\mu\nu},C],\cr   [Q,N^{-}_{\mu\nu}]&=
i\,[\chi^{-}_{\mu\nu},C],\cr  [Q, P]&=i\,[\tilde\eta,C],\cr
\{Q,\eta\,\} &=i\,[B,C].\cr}
\eqn\marek
$$  This BRST algebra closes up to a gauge transformation generated  by
$C$. 

We have to give now the expressions for the different pieces of the  gauge
fermion. For the localization gauge fermion we have:
$$
\eqalign{
\Psi_{\hbox{\sevenrm loc}}&=\langle(\chi^{+},\chi^{-},\tilde\eta),s(A,V)
\rangle +\langle (\chi^{+},\chi^{-},\tilde\eta),( N^{+},N^{-},P)\rangle=
\cr &\int_X
\sqrt{g}\,\tr\,\bigl\{\,\,\half\chi^{+}_{\mu\nu}\bigl (\,N^{+\mu\nu} -2
F^{+\mu\nu}+2i[V^{\mu},V^{\nu}]^{+}\,\bigr )\cr &
+\half\chi^{-}_{\mu\nu}\bigl (\,N^{-\mu\nu}+4(\deriv^{[\mu}V^{\nu]})
^{-}\,\bigr )+\tilde\eta\bigl (\,P+\raiz \deriv_\mu V^\mu\,\bigr )
\,\bigr \},\cr }
\eqn\localizac
$$  
while for the projection gauge fermion, which enforces the horizontal
projection, we have:
$$
\Psi_{\hbox{\sevenrm proj}}=\langle B,{\cal
C}^{\dag}(\psi,\tilde\psi)\rangle_{\hbox{\bf g}}=
\int_X \sqrt{g}\,\tr\,\bigl\{\, B\bigl (\,-\deriv_\mu\psi^\mu
+i[\tilde\psi_{\mu}, V^{\mu}]\,\bigr )\,\bigr\}.
\eqn\projection
$$

As in the other cases we have studied, it is necessary to add an extra 
piece to the gauge fermion to make full contact with the corresponding twisted
supersymmetric theory. In  this case, this extra term is:
$$
\Psi_{\hbox{\sevenrm{extra}}}=\int_X \sqrt{g}\,\tr\,\bigl\{\,\,
{i\over2}\eta[B,C]\,\bigr\}.
\eqn\extrah
$$

It is now straightforward to see that, with the redefinitions
$$
\eqalign{ 
A'&=A,\cr
\psi'&={1\over2}\psi,\cr 
C'&={1\over{2\raiz}}C,\cr 
B'&=-2\raiz B,\cr
\eta'&=-2\eta,\cr}
\qquad\qquad
\eqalign{  V'&=-{1\over\raiz}V,\cr
\tilde\psi'&=\half\tilde\psi,\cr
\chi'^{+}&= -\chi^{+},
\cr N'^{+}&=-N^{+},\cr  
}\qquad\qquad
\eqalign{
\tilde\eta'&=-2\raiz\tilde\eta,\cr 
P'&=-2\raiz P,\cr
\chi'^{-}&=\chi^{-},\cr  
N'^{-}&=N^{-},\cr}
\eqn\redefin
$$ 
one recovers, in terms of the primed fields,  the twisted model
summarized in \butter\ and  \epi, 
which corresponds to the topological symmetry $Q$.

It is worth to remark that one could  also consider the ``dual" problem built
out of the  basic equations:
$$
\cases{ F^{-}_{\mu\nu}-i[V_{\mu},V_{\nu}]^{-}=0,\cr
\bigl (\,\deriv_{[\mu}V_{\nu]}\,\bigr )^{+}=0,\cr
\deriv_\mu V^\mu=0.\cr}
\eqn\reversequations
$$ 
The resulting theory corresponds precisely to the second type of theory
obtained in the previous section in our discussion of the third twist. The
corresponding action has the form $\{\tilde Q, \psi^- \}$ where $\tilde Q$ is
given in \boheme\   and $\Psi^-$ is the result of performing a   
${\Bbb Z}_2$-transformation  (see \melocoton) on
the gauge fermion $\Psi^+$ in \sesamo.

\endpage

\chapter{Observables}

In this section we will analyze the structure of the observables for  each of
the three twists. Observables are operators which are $Q$-invariant but are
not $Q$-exact. A quick look at the $Q$-transformations  which hold in each
twist shows that the observables are basically the same as in ordinary
Donaldson-Witten theory. Indeed, from \papagayo\ or \macarena\ one finds that
the trio, $A_\mu$, $\psi_\mu$ and $\phi$, which is present in the first twist
transform adequately so that the operators, 
$$
\eqalign{
W_0 =& \tr(\phi^2), \cr
W_2 =& \tr(\half \psi \wedge \psi + {1\over\raiz}\phi\wedge F), \cr}
\qquad\qquad
\eqalign{
W_1 =& -\raiz\tr(\phi \wedge \psi), \cr
W_3 =& -{1\over2}\tr(\psi \wedge  F), \cr}
\eqn\guta
$$
satisfy the descent equations,
$$
\delta W_i = d W_{i-1},
\eqn\descent
$$
which imply that,
$$
{\cal O}^{(\gamma_j)} = \int_{\gamma_j} W_j,
\eqn\noguta
$$
being $\gamma_j$ homology cycles of $X$, are observables. Of course, as usual,
this set can be enlarged for gauge groups possessing other independent
Casimirs besides the quadratic one. The transformations \covMedeiros\ or
\darek\ for the second twist, and \butter\ or \marek\ for the third (after
replacing $C$ by $\phi$) show that these other twists possess a similar set of
observables.

Topological invariants are obtained considering the vacuum expectation value of
arbitrary products of observables:
$$
\langle \prod_{\gamma_j} {\cal O}^{(\gamma_j)} \rangle.
\eqn\leoondos
$$
The general form of this vacuum expectation value is,
$$
\langle \prod_{\gamma_j} {\cal O}^{(\gamma_j)} \rangle
= \sum_{k} 
\langle \prod_{\gamma_j} {\cal O}^{(\gamma_j)} \rangle_k
\ex^{-2\pi i k \tau},
\eqn\pumba
$$
where $k$ is the instanton number and 
$\langle \prod_{\gamma_j} {\cal O}^{(\gamma_j)} \rangle_k$
is the vacuum expectation value computed 
at a fixed value of $k$ with an action which is $Q$-exact,
$$
\langle \prod_{\gamma_j} {\cal O}^{(\gamma_j)} \rangle_k
=\int [df]_k \ex^{\{Q,\Psi\}} 
\prod_{\gamma_j} {\cal O}^{(\gamma_j)}.
\eqn\timon
$$
In this equation $[df]_k$ denotes collectively the measure indicating that only
gauge configurations of instanton number $k$  enter in the functional
integral. These quantities are independent of the coupling constant $e$. When
analyzed in the weak coupling limit the contributions to the functional
integral come from field configurations which are solutions to the equations
which define the moduli problems which we have associated to each twist in
the previous section. All the dependence of the observables on $\tau$ is
contained in the sum \pumba.

The $Q$-symmetry of the theory impossess a selection rule for the products
entering \leoondos\ which could lead to a possibly non-vanishing result: the
ghost number of  \leoondos\
must be equal to the virtual dimension of the corresponding moduli space. For
the first twist this implies that the only observable is the partition
function of the theory. In fact, this is the quantity computed by Vafa and
Witten in [\vafa] for some specific situations to obtain a test of duality.
The resulting partition functions $Z(\tau)$ turn out to transform as modular
forms under $Sl(2,{\Bbb Z})$ transformations.

For the other two twists the virtual dimension is not zero but it is
independent of the instanton number $k$. This means that, as in the case of
the first twist, one could obtain contributions from many values of $k$.
Possibly non-trivial topological invariants for these cases correspond to
products of operators \leoondos\ such that their ghost number matches the 
virtual
dimension dim$(G)(2\chi+3\sigma)/4$ for the case of the second twist, or
dim$(G)\chi$ for the case of the third. One important question is whether
or not the vacuum expectation values of these observables have good modular
properties under $Sl(2,{\Bbb Z})$ transformations. We will show in the rest of
this section that in the case of the third twist the vacuum expectation values
are actually independent of $\tau$. Thus, further non-trivial duality tests
can be addressed only in the second twist. We will not consider this
issue in this paper.

As indicated in the introduction and proved in sect. 2, the third twist
leads to a topological quantum field theory which is amphicheiral. We will
show now that in addition this theory possesses the property that the vacuum
expectation values of products of its observables are independent of $\tau$.
Thus, in some sense the invariance under $Sl(2,{\Bbb Z})$ is trivially realized
in this case.

In order to study the vacuum expectation values of products of observables in
the third twist we are going to consider the action \boomerang\ (in its
covariantized form) in which the auxiliary fields appear quadratically. The
bosonic part of this action involving only the field strength $F_{\mu\nu}$ and
the vector field $V_\mu$ can be written in three equivalent forms.
The form of the action,
${\cal S}= \{Q,\hat\Psi^+\} - 2 \pi i k \tau $, leads to,
$$
\eqalign{
-\int_X d^4 x\,\sqrt{g}\, \tr\, &\bigl\{\,\,
 {1\over2e^2}\bigl (\,F^{+\mu\nu}-2i[V^\mu,V^\nu]^{+}\,
\bigr )^2
 +{4\over e^2}\bigl (\,(\deriv^{[\mu}V^{\nu]})
^{-}\,\bigr )^2\cr&+{1\over e^2}
\bigl (\,\deriv_\mu V^\mu\,\bigr )^2\,\bigr\}
-2\pi i\tau{1\over 32\pi^2}\int_X d^4 x\,\sqrt{g}\tr\,\bigl\{\, *  F_{\mu\nu}
F^{\mu\nu}\,\bigr\},\cr}
\eqn\golgi
$$
the form, ${\cal S}= \{\tilde Q,\hat\Psi^-\} - 2 \pi i k \bar\tau $, to,
$$
\eqalign{
-\int_X d^4 x\,\sqrt{g}\, \tr\, &\bigl\{\,\,
 {1\over2e^2}\bigl (\,F^{-\mu\nu}-2i[V^\mu,V^\nu]^{-}\,
\bigr )^2
 +{4\over e^2}\bigl (\,(\deriv^{[\mu}V^{\nu]})
^{+}\,\bigr )^2\cr&+{1\over e^2}
\bigl (\,\deriv_\mu V^\mu\,\bigr )^2\,\bigr\}
-2\pi i\bar\tau{1\over 32\pi^2}\int_X d^4 x\,\sqrt{g}\tr\,\bigl\{\, * 
F_{\mu\nu} F^{\mu\nu}\,\bigr\},
\cr}
\eqn\otrogolgi
$$
and, finally, the form,
${\cal S}= \half \{Q,\hat\Psi^+\} + \half \{\tilde Q,\hat\Psi^-\} 
- 2 \pi i k {\hbox{\rm Re}}(\tau) $, to,
$$
\eqalign{
-\int_X\tr\, &\bigl\{\,\,
 {1\over4}\bigl (\,F^{\mu\nu}-2i[V^\mu,V^\nu]\,
\bigr )^2
 +2(\deriv^{[\mu}V^{\nu]})^2+
\bigl (\,\deriv_\mu V^\mu\,\bigr )^2\,\bigr\}\cr&
-2\pi i{\hbox{\rm Re}}(\tau)
{1\over 32\pi^2}\int_X d^4 x\,\sqrt{g}\tr\,\bigl\{\, * 
F_{\mu\nu} F^{\mu\nu}\,\bigr\}.\cr}
\eqn\ateate
$$
Standard arguments in topological quantum field theory show that the weak
coupling limit is exact. In the first case this limit implies that the
contributions to the functional integral correspond to the moduli space
defined by the equations \morequations. Notice that the normalization factor
for $V_\mu$ in \redefin\ has to be taken into account since \golgi\ correspond
to the action resulting after twisting. Similarly,
in the second case the weak coupling limit contributions correspond to the
moduli space defined by the equations \reversequations. In the third case,
however, the contributions correspond to the solution of the following set of
equations:
$$
\cases{ F_{\mu\nu}-2i[V_{\mu},V_{\nu}]=0,\cr
\deriv_{[\mu}V_{\nu]}=0,\cr
\deriv_\mu V^\mu=0,\cr}
\eqn\jontas
$$ 
which define a moduli space which is the intersection of the other two. This
is the moduli space which appears in the formulation of the third twist
presented in [\marcus,\blauthomp]. Notice that the three points of view lead
to three different types of dependence on $\tau$. The first one implies that
vacuum expectation values are holomorphic in $\tau$, the second that they are
antiholomorphic, and the third that they depend only on the real part of
$\tau$. We will solve this puzzle showing that actually the vacuum expectation
values are just real numbers and not functions of $\tau$.

We first prove that any solution of $\jontas$ must involve a gauge connection
whose instanton number is zero. Indeed, from the identity,
$$
\eqalign{
\int_X d^4 x\,\sqrt{g}\, \tr\, &\bigl\{\,
  *  F_{\mu\nu}\bigl (\,F^{\mu\nu}-2i[V^\mu,V^\nu]\,
\bigr )-4 * \deriv^{[\mu}V^{\nu]}\deriv_{[\mu}V_{\nu]}\,\bigr\}
\cr &=
\int_X d^4 x\,\sqrt{g}\,\tr\,\bigl\{\, *  F_{\mu\nu}F^{\mu\nu}\,\bigr\},\cr
}
\eqn\gadget
$$
follows that any solution of $\jontas$ must have $k=0$. This implies that only
configurations with vanishing instanton number contribute and therefore:
$$
\langle \prod_{\gamma_j} {\cal O}^{(\gamma_j)} \rangle
=  
\langle \prod_{\gamma_j} {\cal O}^{(\gamma_j)} \rangle_{k=0},
\eqn\mana
$$
which is clearly independent of $\tau$. From \golgi\ and \otrogolgi\
follows that for $k=0$ a solution to the equations of the first moduli space
\morequations\ is also a solution to the ones of the second
\reversequations\ and therefore also to the ones of the third \jontas. For
$k\neq 0$, however, one can have solutions to the equations of the first
moduli space which are not solutions to the equations of the second and
therefore neither to the ones of the third. For $k\neq 0$ the quantities
$\langle \prod_{\gamma_j} {\cal O}^{(\gamma_j)} \rangle_{k}$ are different in
each point of view.  They clearly vanish in the third case. On the other hand,
there is no reason why they should also vanish in the other two cases. Our
results, however, suggest that they do vanish.

We will end this section discussing a vanishing theorem which tells us when
the third moduli space  \jontas\ reduces to the moduli space of flat
connections. The equations \jontas\ have the immediate solution $V=0$, $F=0$,
that is, the moduli space of flat connections is contained in  the moduli
space defined by the equations \jontas. We will show that under certain
conditions both moduli spaces are in fact the same. 
To see this note that since,  
$$
\eqalign{
&\int_X d^4 x\,\sqrt{g}\, \tr\, \bigl\{\,
 {1\over4}\bigl (\,F^{\mu\nu}-2i[V^\mu,V^\nu]\,
\bigr )^2+2\bigl(\deriv^{[\mu}V^{\nu]}\bigr )^2+
\bigl (\,\deriv_\mu V^\mu\,\bigr )^2\,\bigr\}=\cr
&=
\int_X d^4 x\,\sqrt{g}\,\tr\,\bigl\{\,\deriv_\mu V_\nu \deriv^\mu 
V^\nu +R_{\mu\nu}V^\mu V^\nu+{1\over4}F_{\mu\nu}F^{\mu\nu}-
([V_\mu,V_\nu])^2\,\},\cr
}
\eqn\axon
$$
it follows
that if the Ricci tensor is such that 
$$
R_{\mu\nu}V^\mu V^\nu>0\quad {\hbox{\rm for}}~ V\not=0,
\eqn\vanishth
$$
the solutions to the equations \jontas\ are necessarily of the form $V=0$, 
$F=0$, and thus the moduli space is the space of flat gauge connections on 
$X$. 

\endpage

\chapter{Concluding Remarks}

In this paper we have analyzed in full detail the three non-equivalent 
twists of $N=4$ supersymmetric gauge theory. The first twist leads to a
topological quantum field theory whose observables transform as modular forms
under $Sl(2,{\Bbb Z})$ transformations [\vafa]. The second twist leads to the
theory of non-abelian monopoles in the adjoint representation of the gauge
group. In this theory, as in the previous one,  there is a non-trivial
dependence on $\tau$ and one expects that its observables have good
transformation properties under  $Sl(2,{\Bbb Z})$. This is an important issue
that certainly should be addressed. The third twist leads to a topological
quantum field theory which is amphicheiral. We have shown that in this theory
the vacuum expectation values of products of its observables do not depend on
$k$. Hence, barring possible anomalous dependences in $\tau$ like the ones
explicitly unveiled in [\vafa], the theory is trivially invariant under
$Sl(2,{\Bbb Z})$ transformations.

The moduli spaces which, from the point of view of the Mathai-Quillen
formalism, correspond to each twisted theory have been identified. In the
third twist, due to the amphicheiral character of the topological quantum
field theory one finds three different moduli spaces defined by the equations 
\morequations, \reversequations\ and \jontas. These moduli spaces coincide
when the integral of the Chern class of the gauge field vanishes. We have
shown that only the $k=0$ sector contributes to the functional integral,
leading to topological invariants which, therefore, do not depend on $\tau$.

Of the three topological quantum field theories, the one corresponding to the
second twist is not valid on arbitrary oriented four-manifolds. This theory
contains spinors and therefore it only exists for spin manifolds. The
generalization of this theory to arbitrary oriented four-manifolds can be
easily done introducing a Spin$^c$ structure following the construction
recently presented in \REF\baryon{J. M. F. Labastida 
and M. Mari\~no, ``Twisted Baryon Number in
$N=2$ Supersymmetric QCD", hep-th/9702054.} [\baryon]. In this construction
the baryon number symmetry of the original physical theory is gauged
introducing a connection which in the twisted theory is identified with the
Spin$^c$ connection.

We finish by making the remark that the topological quantum field theories
originated from $N=4$ supersymmetric gauge theories are not the only ones
which can lead to a theory with a non-trivial dependence on $\tau$. Any
conformally invariant $N=2$ supersymmetric gauge theory would have the same
property. This is for instance the case for an $N=2$ supersymmetric gauge theory
with gauge group $SU(N_c)$ and $2N_c$ hypermultiplets in the fundamental
representation. These theories should be studied along the lines of this paper
and the duality properties of the resulting topological quantum field theory
should be analyzed.

\ack
We would like to thank M. Mari\~no for important remarks concerning the
twisting of $N=4$ supersymmetric theories, and for many enlightening 
discussions concerning other aspects of this work. This work was supported in
part by DGICYT under grant PB93-0344.

\appendix

We will now summarize the conventions used in this paper. 
Basically
we will describe the elements of the positive and
negative chirality spin bundles $S^+$ and $S^-$ on a four-dimensional spin
manifold $X$ endowed with a vierbein $e^{m\mu}$ and a spin connection
$\omega_{\mu}^{m n}$. The spaces of sections of the spin bundles $S^+$ 
and $S^-$ correspond, from the field-theory point of view, to the 
set of two-component Weyl spinors defined on the manifold $X$. These are 
the simplest irreducible representations of the holonomy group $SO(4)$. 
We will denote positive-chirality (or negative-chirality) spinors by indices 
$\alpha,\beta,\ldots=1,2$ (or $\dalpha,\dbeta,\ldots=1,2$). Spinor indices 
are raised and lowered with the $SU(2)$ invariant tensor $C_{\alpha\beta}$ 
(or $C_{\dalpha\dbeta}$) and its inverse $C^{\alpha\beta}$ 
(or $C^{\dalpha\dbeta}$), with the conventions $C_{21}=C^{12}=+1$, 
so that, 
$$
\eqalign{&C_{\alpha\beta}C^{\beta\gamma}=\delta_{\alpha}{}^{\gamma},\cr
&C_{\dalpha\dbeta}C^{\dbeta\dgamma}=\delta_{\dalpha}{}^{\dgamma},\cr}\qquad
\qquad
\eqalign{&C_{\alpha\beta}C^{\gamma\delta}=\delta_\alpha{}^\delta
\delta_\beta{}^\gamma-\delta_\alpha{}^\gamma \delta_\beta{}^\delta,\cr 
&C_{\dalpha\dbeta}C^{\dgamma\ddelta}=\delta_\dalpha{}^\ddelta
\delta_\dbeta{}^\dgamma-\delta_\dalpha{}^\dgamma
\delta_\dbeta{}^\ddelta.\cr}
\eqn\setas
$$ 

The 
spinor representations and the vector representation associated to 
$S^+\!\times S^-$ are related by the Clebsch-Gordan 
$\sigma^m{}_{\!\alpha\dalpha}=(i{\bf 1}, \vec\tau)$ and 
$\bar\sigma^{m\dalpha\alpha}=
(i{\bf 1}, -\vec\tau)$, where ${\bf 1}$ is the $2\times2$ unit matrix and 
$\vec\tau=(\tau^1,\tau^2,\tau^3)$ are the Pauli matrices,  
$$
\tau^1 = \left(\matrix{0&{1}\cr
                         {1}&0\cr}\right),\,\,\,\,\,\,\,\,\,\,\,
\tau^2 = \left(\matrix{0&{-i}\cr
                         {i}&0\cr}\right),\,\,\,\,\,\,\,\,\,\,\,
\tau^3 = \left(\matrix{1&{0}\cr
                         {0}&-1\cr}\right).
\eqn\lados
$$
The Pauli matrices satisfy: $$
\tau_a\tau_b = i\epsilon_{abc}\tau_c+\delta_{ab}{\bf 1},
\eqn\latres
$$
where $\epsilon_{abc}$ is the totally antisymmetric tensor with
$\epsilon_{123}=1$.

Under an infinitesimal $SO(4)$ rotation a Weyl spinor $M_\alpha$,
$\alpha=1,2$, associated to $S^+$, transforms as:
$$
\delta M_\alpha = {i\over 2} \epsilon_{mn}
(\sigma^{mn})_\alpha{}^\beta M_\beta,
\eqn\laocho
$$
where $\epsilon_{mn}=-\epsilon_{nm}$ are the infinitesimal parameters
of the transformation. On the other hand, a Weyl spinor $N^{\dot\alpha}$,
$\dot\alpha=1,2$, associated to $S^-$, transforms as,
$$
\delta N^{\dot\alpha} = {i\over 2} \epsilon_{mn}
(\bar\sigma_{mn})^{\dot\alpha}{}_{\dot\beta} N^{\dot\beta}.
\eqn\lanueve
$$
The matrices $\sigma^{mn}$ and $\bar\sigma^{mn}$ are antisymmetric in $m$ 
and $n$ and are defined as follows:
$$
\sigma^{mn}{}_{\!\alpha}{}^{\!\beta}={i\over
2}\sigma^{[m}{}_{\!\alpha\dalpha}\bar\sigma
^{n]\dalpha\beta},\qquad\qquad
\bar\sigma^{mn\dalpha}{}_{\!\dbeta}={i\over
2}\bar\sigma^{[m\dalpha\alpha}\sigma^{n]}{}_{\!\alpha\dbeta}.
\eqn\Pollux
$$ 
They satisfy the self-duality properties,
$$
\sigma^{mn} =\half
\epsilon^{mnpq}\sigma_{pq},
\qquad\qquad 
\bar\sigma^{mn} = -\half
\epsilon^{mnpq}
\bar\sigma_{pq},
\eqn\Io
$$
and the $SO(4)$ algebra,
$$
[\sigma_{mn},\sigma_{pq}]=
i(\delta_{mp}\sigma_{nq}-
\delta_{mq}\sigma_{np}-
\delta_{np}\sigma_{mq}+
\delta_{nq}\sigma_{mp}).
\eqn\Ramses
$$
The same algebra is fulfilled by $\bar\sigma^{mn}$.

Let us consider the covariant derivative $ D_\mu$ on the manifold
$X$. Acting on an element of $\Gamma(X,S^+)$ it has the form:
$$
D_\mu M_\alpha = \partial_\mu M_\alpha + {i\over 2}
\omega_{\mu}^{m n} (\sigma_{mn})_\alpha{}^\beta M_\beta,
\eqn\carve
$$
where $\omega_{\mu}^{m n}$ is the spin connection. Defining 
$ D_{\alpha\dot\alpha}$ as,
$$
 D_{\alpha\dot\alpha} = (\sigma_n)_{\alpha\dot\alpha} e^{n\mu} 
D_\mu,
\eqn\cave
$$
where $e^{n\mu}$ is the vierbein on $X$, the Dirac equation for 
$M\in \Gamma(X,S^+)$ and $N\in \Gamma(X,S^-)$ 
can be simply written as,
$$
 D_{\alpha\dot\alpha} M^\alpha=0,
\,\,\,\,\,\,\,\,\,\,\,\,\,
D_{\alpha\dot\alpha} N^{\dot\alpha}=0.
\eqn\dirac
$$

Let us now introduce a principal 
$G$-bundle $P\to X$ with its associated connection one-form $A$, 
and let us consider
that the Weyl spinors $M_\alpha$ realize locally an element of 
$\Gamma(S^+\otimes \ad P)$, \ie, they transform under a $G$ gauge
transformation
in the adjoint representation --indeed,  $\ad P$ is the vector bundle 
associated 
to $P$ through the adjoint representation of the gauge group on 
its Lie algebra: 

$$
\delta M_\alpha^a = i[M_\alpha,\phi]^a=-i(T^c)^{ab}M_\alpha^b \phi^c,
\eqn\gaugetransf
$$
where $(T^a)^{bc}=-if^{abc}$, $a=1,\cdots,{\rm dim}(G)$,  are the
generators of $G$ in the adjoint representation, which are traceless
and chosen to be hermitian and are normalized as follows: $\tr\,
(T^a T^b)=\delta^{ab}$. In \gaugetransf\ $\phi^a$, $a=1,\cdots,
{\rm dim}(G)$,
denote the infinitesimal parameters of the gauge transformation.

In terms of the gauge connection $A$, the covariant derivative 
\carve\ can be promoted to a full covariant derivative acting on  
sections in $\Gamma(X, S^+\otimes\ad P)$,
$$
{\cal D}_\mu M_\alpha = \partial_\mu M_\alpha + {i\over 2}
\omega_{\mu}^{m n} (\sigma_{mn})_\alpha{}^\beta M_\beta
+i[A_\mu,M_\alpha],
\eqn\carvemore
$$
and its analogue in \cave:
$$
{\cal D}_{\alpha\dalpha} = (\sigma_n)_{\alpha\dalpha} e^{n\mu}
 {\cal D}_\mu.
\eqn\cavemore
$$
In terms of the full covariant derivative the Dirac equations \dirac\ 
become:
$$
{\cal D}_{\alpha\dalpha} M^{\alpha}=0,
\,\,\,\,\,\,\,\,\,\,\,\,\,
{\cal D}_{\alpha\dalpha} N^{\dalpha}=0.
\eqn\diraci
$$

Given an element of $\Gamma(X,S^+\otimes \ad P)$, $M_\alpha = (a,b)$ we
define $\overline M^{\alpha} = (a^{*},b^{*})$. In this way, 
given $M,N \in \Gamma(X,
S^+\otimes \ad P)$, the gauge-invariant quantity entering the metric 
%\mmetrica,
$$
\half\tr\,\bigl (\,\overline M^{\alpha} N_{\alpha} +
\overline N^{\alpha} M_{\alpha}\,\bigr ),
\eqn\mmmetric
$$
is positive definite. With similar arguments the corresponding gauge 
invariant metric in the fibre  $\Gamma(X,S^-\otimes \ad P)$, which 
we define as 
$$
\half\tr\,\bigl (\,\overline M_{\dalpha} N^{\dalpha} +
\overline N_{\dalpha} M^{\dalpha}\,\bigr ),
\eqn\mmmetrica
$$
for $M,N\in\Gamma(X,S^-\otimes \ad P)$, can be seen to be 
positive definite, too.
For self-dual two-forms $Y,Z\in \Gamma(X,\Lambda^{2,+}T^{*}X\otimes\ad P)\equiv
\Omega^{2,+}(X,\ad P)$ our definition of the metric is the following:
$$
\langle Y,Z\rangle =\int_X\tr\,\bigl (\,Y\wedge *Z\,\bigr)=\half 
\int_X\tr\,\bigl (\,Y_{\mu\nu}Z^{\mu\nu}\,\bigr )=-{1\over4}
\int_X\tr\,\bigl (\,Y_{\alpha\beta}Z^{\alpha\beta}\,\bigr ),
\eqn\gangan
$$
where $(Y,Z)_{\alpha\beta}=\sigma^{\mu\nu}_{\alpha\beta}(Y,Z)_{\mu\nu}$ and 
we have used the identity 
$(Y,Z)_{\alpha\beta}(Y,Z)^{\alpha\beta }=-2(Y,Z)^+_{\mu\nu}(Y,Z)^{+\mu\nu}$.

Acting on an element of $\Gamma(X,S^+\otimes\ad P)$ the covariant derivatives
satisfy:
$$
[{\cal D}_\mu,{\cal D}_\nu]M_\alpha = i[F_{\mu\nu}, M_\alpha] +
{i\over2} R_{\mu\nu}{}^{mn}(\sigma_{mn})_\alpha{}^\beta M_\beta,
\eqn\commutador
$$
where $F_{\mu\nu}$ are the components of the two-form field strength:
$$
F = d A + iA\wedge A,
\eqn\fuerzacampo
$$
and $R_{\mu\nu}{}^{mn}$ the components of the curvature two-form,
$$
R^{mn} = d\omega^{mn}+\omega^{mp}\wedge \omega^{pn},
\eqn\curvatura
$$
being $\omega^{mn}$ the spin connection one-form. The scalar curvature is
defined as:
$$
R=e^{\mu}_m e^{\nu}_n R_{\mu\nu}{}^{mn},
\eqn\curvescalar
$$
and the Ricci tensor as:
$$ 
R_{\kappa\lambda}=e^{\mu}{}{\!_m} e_{n\lambda} R_{\mu\kappa}{}^{mn}.
\eqn\Ricciten
$$
The components of the curvature two-form \curvatura\ are related to the 
components of the Riemann tensor as follows:
$$
R_{\mu\nu\kappa\lambda}=e_{\kappa m}e_{\lambda n}R_{\mu\nu}{}^{\!mn}.
\eqn\riemann
$$
The Riemann tensor satisfies the following algebraic properties:

(a) Symmetry:
$$
R_{\lambda\mu\nu\kappa}=R_{\nu\kappa\lambda\mu},\eqn\weinbuno
$$

(b) Antisymmetry:
$$
R_{\lambda\mu\nu\kappa}=-R_{\mu\lambda\nu\kappa}=
-R_{\lambda\mu\kappa\nu}=+R_{\mu\lambda\kappa\nu},\eqn\weinbdos
$$

(c) Cyclicity:
$$
R_{\lambda\mu\nu\kappa}+R_{\lambda\kappa\mu\nu}+
R_{\lambda\nu\kappa\mu}=0.\eqn\weinbtres
$$
Notice that \weinbtres\ implies that 
$$
\epsilon^{\mu\nu\kappa\sigma}R_{\lambda\mu\nu\kappa}=0.
\eqn\weinbcuatro
$$ 
This result is essential in the verification of the identity \gadget .
\endpage
\refout
\end

*****************

   LEFT OVERS

*****************

, for the 
bosonic part of the action not containing the scalar fields $B$ and $C$:
$$
\eqalign{
\int d^4 x\,\sqrt{g}\, \tr\, &\bigl\{\,
 -{1\over2e^2}\bigl (\,F^{+\mu\nu}-2i[V^\mu,V^\nu]^{+}\,
\bigr )^2
 -{4\over e^2}\bigl (\,(\deriv^{[\mu}V^{\nu]})
^{-}\,\bigr )^2-\cr&-{1\over e^2}
\bigl (\,\deriv_\mu V^\mu\,\bigr )^2\,\bigr\}
-2\pi i\tau{1\over 32\pi^2}\int d^4 x\,\sqrt{g}\tr\,\bigl\{\, *  F_{\mu\nu}
F^{\mu\nu}\,\bigr\}\cr
}
\eqn\golgi
$$
It can be shown that \golgi\ can be also written in the form:
$$
\eqalign{
\int d^4 x\,\sqrt{g}\, \tr\, &\bigl\{\,
 -{1\over2e^2}\bigl (\,F^{-\mu\nu}-2i[V^\mu,V^\nu]^{-}\,
\bigr )^2
 -{4\over e^2}\bigl (\,(\deriv^{[\mu}V^{\nu]})
^{+}\,\bigr )^2-\cr&-{1\over e^2}
\bigl (\,\deriv_\mu V^\mu\,\bigr )^2\,\bigr\}
-2\pi i\bar\tau{1\over 32\pi^2}\int d^4 x\,\sqrt{g}\tr\,\bigl\{\, * 
F_{\mu\nu} F^{\mu\nu}\,\bigr\}
\cr}
\eqn\otrogolgi
$$
In fact, this is precisely what we would have obtained if we had worked with 
$\tilde Q$ \boheme\ instead of $Q$. This observation has far reaching
consequences. First of all, we shift the auxiliary fields so as to cast the $Q$
transformations as in \Basler . We have seen that in this situation the action 
${\cal S}$ can be simultaneously written as a $Q$ or a $\tilde Q$
(anti)commutator \boomerang : 
$$
{\cal S}=\{Q,\Psi^{+}\}-2\pi ik\tau =\{\tilde Q,\Psi^{-}\}
-2\pi ik\bar\tau\,;\quad  [Q, {\cal S}]=[\tilde Q,{\cal S}]=0, 
\eqn\gang
$$
so it is also true that:
$$
{\cal S}=\half\{Q,\Psi^{+}\}+\half\{\tilde Q,\Psi^{-}\}
-2\pi ik{\hbox{\rm Re}}(\tau).
\eqn\noe
$$
The bosonic part of the action ${\cal S}$ not containing $B$ or $C$ can thus
be written in the form:
$$
\eqalign{
&\half\int d^4 x\,\sqrt{g}\, \tr\, \bigl\{\,
 -{1\over2e^2}\bigl (\,F^{+\mu\nu}-2i[V^\mu,V^\nu]^{+}\,
\bigr )^2-{4\over e^2}\bigl (\,(\deriv^{[\mu}V^{\nu]})
^{-}\,\bigr )^2-\cr&-{1\over e^2}
\bigl (\,\deriv_\mu V^\mu\,\bigr )^2\,\bigr\}+
\half\int d^4 x\,\sqrt{g}\, \tr\, \bigl\{\,
 -{1\over2e^2}\bigl (\,F^{-\mu\nu}-2i[V^\mu,V^\nu]^{-}\,
\bigr )^2
 -{4\over e^2}\bigl (\,(\deriv^{[\mu}V^{\nu]})
^{+}\,\bigr )^2-\cr&-{1\over e^2}
\bigl (\,\deriv_\mu V^\mu\,\bigr )^2\,\bigr\}
-{{i\theta}\over 32\pi^2}\int d^4 x\,\sqrt{g}\tr\,\bigl\{\, * 
F_{\mu\nu} F^{\mu\nu}\,\bigr\}
\cr}
\eqn\siemon
$$
By standard arguments \noe\ implies that the observables do not depend on the 
coupling constant $e$. Thus, one can safely go to the weak coupling limit $e\to
0$, where the Euclidean path integral $Z\sim \int [{\cal D}\Phi]\ex^{\cal S}$
is dominated by the zeroes of the real part of \siemon , that is, by those
field configurations that satisfy the equations:
$$ 
\eqalign{
&F^{\mu\nu}-2i[V^\mu,V^\nu]=0,\cr
&\deriv^{[\mu}V^{\nu]}=0,\cr
&\deriv_\mu V^\mu=0,\cr
}
\eqn\siemens
$$

These are precisely the field equations behind the theories presented in 
[\marcus] and more recently in 
\REF\tonson{M. Blau, G. Thompson, ``Aspects of $N_{T}\geq 2$ Topological 
Gauge Theories and D-Branes", hep-th/9612143.}
[\tonson]. In fact, we can redefine our fields and BRST charges in such a way
that we reproduce the models presented in [\marcus] and [\tonson]. 

\vskip .5cm
\noindent {\sl A Vanishing Theorem}

The equations \siemens\ have the immediate solution $V=0$, $F=0$, that is, the
moduli space of flat connections is contained in (but it is not necessarily
equal to) the moduli space defined by the equations \siemens . However, it
turns out that under certain conditions (when an appropriate vanishing theorem
holds) both moduli spaces are in fact the same. To see this note that since 
$$
\eqalign{
&\half\int d^4 x\,\sqrt{g}\, \tr\, \bigl\{\,
 {1\over4}\bigl (\,F^{\mu\nu}-2i[V^\mu,V^\nu]\,
\bigr )^2+2\bigl(\deriv^{[\mu}V^{\nu]}\bigr )^2+
\bigl (\,\deriv_\mu V^\mu\,\bigr )^2\,\bigr\}=\cr
&=
\int d^4 x\,\sqrt{g}\,\tr\,\bigl\{\,\deriv_\mu V_\nu \deriv^\mu 
V^\nu +R_{\mu\nu}V^\mu V^\nu+{1\over4}F_{\mu\nu}F^{\mu\nu}-
([V_\mu,V_\nu])^2\,\}\cr
}
\eqn\axon
$$
 ($R_{\mu\nu}=R^\lambda{}_{\!\mu\lambda\nu}$ is the Ricci tensor), it follows
that if the Ricci tensor is such that 
$$
R_{\mu\nu}V^\mu V^\nu>0\quad {\hbox{\rm for}}~ V\not=0,
\eqn\vanishth
$$
the solutions to the equations \siemens\ are necessarily of the form $V=0$, 
$F=0$, and thus the moduli space is 
contained in the $k=0$ sector of the space of gauge connections. But this
result is far more general, for one can show that even in the case that  
\vanishth\ does not hold (and hence one can have solutions to the equations 
\siemens\ with $V\not=0$), all the gauge connections in the moduli space have
$k=0$. To see this note that from the identity 
$$
\eqalign{
\int d^4 x\,\sqrt{g}\, \tr\, &\bigl\{\,
  *  F_{\mu\nu}\bigl (\,F^{\mu\nu}-2i[V^\mu,V^\nu]\,
\bigr )^2-4 * \deriv^{[\mu}V^{\nu]}\deriv_{[\mu}V_{\nu]}\,\bigr\}=
\cr &=
\int d^4 x\,\sqrt{g}\,\tr\,\bigl\{\, *  F_{\mu\nu}F^{\mu\nu}\,\bigr\}\cr
}
\eqn\gadget
$$
(you have to integrate by parts once) it follows that within the moduli space
$k=0$. The value $k=0$ is rather special, for in that case the three
apparently different moduli spaces that naively seem to underlie the theory, 
namely   
$$ 
\eqalign{
&(F^{\mu\nu}-2i[V^\mu,V^\nu])^{+}=0,\cr
&(\deriv^{[\mu}V^{\nu]})^{-}=0,\cr
&\deriv_\mu V^\mu=0,\cr
}
\qquad
\eqalign{
&(F^{\mu\nu}-2i[V^\mu,V^\nu])^{-}=0,\cr
&(\deriv^{[\mu}V^{\nu]})^{+}=0,\cr
&\deriv_\mu V^\mu=0,\cr
}
\qquad
\eqalign{
&F^{\mu\nu}-2i[V^\mu,V^\nu]=0,\cr
&\deriv^{[\mu}V^{\nu]}=0,\cr
&\deriv_\mu V^\mu=0,\cr
}
\eqn\rodin
$$
are all equivalent. This result follows from  the identities: 
$$
\eqalign{
\int\tr\, &\bigl\{\,
 -{1\over2}\bigl (\,F^{+\mu\nu}-2i[V^\mu,V^\nu]^{+}\,
\bigr )^2
 -4\bigl (\,(\deriv^{[\mu}V^{\nu]})
^{-}\,\bigr )^2-\cr&-
\bigl (\,\deriv_\mu V^\mu\,\bigr )^2\,\bigr\}+
{1\over4}\int\tr\,\bigl\{\, *  F_{\mu\nu}
F^{\mu\nu}\,\bigr\}=\cr
\int\tr\, &\bigl\{\,
 -{1\over2}\bigl (\,F^{-\mu\nu}-2i[V^\mu,V^\nu]^{-}\,
\bigr )^2
 -4\bigl (\,(\deriv^{[\mu}V^{\nu]})
^{+}\,\bigr )^2-\cr&-
\bigl (\,\deriv_\mu V^\mu\,\bigr )^2\,\bigr\}
-{1\over4}\int\tr\,\bigl\{\, *  F_{\mu\nu}
F^{\mu\nu}\,\bigr\}=\cr
\int\tr\, &\bigl\{\,
 -{1\over4}\bigl (\,F^{\mu\nu}-2i[V^\mu,V^\nu]\,
\bigr )^2
 -2(\deriv^{[\mu}V^{\nu]})^2-
\bigl (\,\deriv_\mu V^\mu\,\bigr )^2\,\bigr\}
\cr
}
\eqn\adios
$$